\documentclass[useAMS,preprint2,usenatbib]{aastex}
\usepackage{graphicx,amssymb,natbib,rotating,longtable,cite,multicol}

\def\apjl{ApJ}
\def\apjs{ApJS}
\def\aap{A\&A}
\shorttitle{Structural studied of BRCs}
\shortauthors{Sharma et al.}

\begin{document}

\title{Structural studies of eight bright rimmed clouds in the  southern hemisphere}

\author {Saurabh Sharma\altaffilmark{1}, A. K. Pandey\altaffilmark{1}, J. Borissova\altaffilmark{2,7}, D. K. Ojha\altaffilmark{3}, V. D. Ivanov\altaffilmark{4},
K. Ogura\altaffilmark{5}, N. Kobayashi\altaffilmark{6}, R. Kurtev\altaffilmark{2,7}, M. Gopinathan\altaffilmark{1} and Ram Kesh Yadav\altaffilmark{8}}

\altaffiltext{1}{Aryabhatta Research Institute of Observational Sciences (ARIES), Manora Peak, Nainital, 263 001, India, saurabh@aries.res.in}
\altaffiltext{2}{Departamento de F\'isica y Astronom\'ia, Universidad de Valpara\'{\i}so, Ave. Gran Breta\~na 1111, Valpara\'iso, Chile}
\altaffiltext{3}{Tata Institute of Fundamental Research, Homi Bhabha Road, Colaba, Mumbai - 400 005, India}
\altaffiltext{4}{European Southern Observatory, Karl-Schwarzschild-Str. 2, 85748 Garching bei M\~unchen, Germany}
\altaffiltext{5}{Kokugakuin University, Higashi, Shibuya-ku, Tokyo 150-8440, Japan}
\altaffiltext{6}{Institute of Astronomy, University of Tokyo, 2-21-1 Osawa, Mitaka, Tokyo 181-0015, Japan}
\altaffiltext{7}{Millennium Institute of Astrophysics, Chile }
\altaffiltext{8}{National Astronomical Research Institute of Thailand, Chiang Mai, Thailand }

\begin{abstract}
We carried out deep and wide-field near- and mid-infrared observations for a sample of 8 
bright-rimmed clouds (BRCs). Supplemented with the $Spitzer$ archival data, we have identified and classified 44 to 433 young stellar 
objects (YSOs) associated with these BRCs.
The Class I sources are generally located towards the places with 
higher extinction and are relatively closer to each other than the Class II sources, 
confirming that the young protostars are usually found in regions having denser 
molecular material. On the other hand the comparatively older population, Class II objects, 
are more randomly found throughout the regions, which can be due to their dynamical 
evolution. Using the minimal sampling tree analyses, we have extracted 13 stellar cores of 
8 or more members, which contains 60\% of the total YSOs. 
The typical core is $\sim$0.6 pc in radii and somewhat elongated (aspect ratio of 1.45), of 
relatively low stellar density (surface density 60 pc$^{-2}$), consisting of a small (35) number of YSOs of relatively young sources (66\% Class I ), 
and partially embedded (median $A_K$ =1.1 mag).
But the cores show a wide range in their mass distribution 
($\sim$20 to 2400 M$_\odot$) with a median value of around 130  M$_\odot$.
We have found the star  formation efficiencies in the cores to be between 3\% and 30\% 
with an average of $\sim$14\%, which agree with the efficiencies needed to link
the core mass function to the initial mass function.
We also found a linear relation between the density of the clouds and the number of YSOs.
The peaked nearest neighbor spacing distributions of the YSOs and the ratio of Jeans lengths to the YSOs separations
indicates a significant degree of non-thermally driven fragmentation in these BRCs.
\end{abstract}

\section{Introduction}

Observations of embedded star-forming regions (SFRs)
show that stellar distributions in SFRs are often elongated, clumpy, or
both \citep[see][]{2003ARAA..41...57L, 2005ApJ...632..397G, 2008ApJ...674..336G, 2006ApJ...636L..45T, 2007prpl.conf..361A, 2008ApJ...688.1142K}
and seems to be correlated with
the distribution of dense gas of the  natal molecular clouds \citep{2008ApJ...674..336G}.
Mapping the spatial distribution of young stellar objects (YSOs) within these SFRs
is an important means by which one can study the star formation scenario in the region 
and understand the physical processes that influence star formation \citep{2008ApJ...688.1142K}.

Surveys of molecular clouds in the nearest 1 kpc show that approximately 75\% of embedded
young stars are in groups and clusters with 10 or more members
\citep{2000ApJS..130..381C,2007prpl.conf..361A,1993prpl.conf..429Z}.
\citet{2009ApJS..184...18G} have isolated several young stellar cores associated  with each of a sample 
of 36  young stellar clusters and have studied their structures.
Quantitative statistical properties of these structures, especially
the sizes, densities and morphologies of young stellar cores could be used to test the theoretical models 
of star formation \citep{2014ApJ...787..107K,2009ApJ...694..367S}.
While stars in some SFRs are centrally concentrated
with a smooth radial density gradient, in other SFRs stars show
signs of fractal sub-clustering. 
How the different structures are connected to the environmental conditions of the molecular 
clouds and how they depend on the evolutionary stage of the cluster are not yet clear.

Bright-rimmed clouds (BRCs), which are small, more or less isolated molecular clouds found at the edges of large HII regions, usually show the presence of a group of young stars 
near the heads of a pillar of gas that
point towards the central O/B stars in the region. 
These clouds may have resulted 
through the compression of pre-existing molecular clumps in the molecular cloud via a 
photoionisation-induced shock and are potential sites of triggered SFRs
\citep[known as radiatively-driven implosion: RDI;][]{1989ApJ...346..735B,1994A&A...289..559L}.
Numerical simulations on the dynamical evolution of a molecular clump illuminated by the ionizing radiation of OB stars 
can be very useful for understanding the RDI process \citep[e.g.,][]{2006MNRAS.369..143M}. 

Relative isolation and simple geometry of BRCs make them an ideal laboratory to observationally test the RDI star formation. 
Sugitani and collaborators \citep{1991ApJS...77...59S, 1994ApJS...92..163S}
compiled a catalog of 89 BRCs (commonly referred to as the SFO catalog) spread throughout the Galaxy; 
44 clouds located in the northern hemisphere and 45 clouds located in the southern hemisphere. 
These clouds were identified by correlating IRAS point sources - having colors consistent with embedded 
protostars - with clouds displaying optically bright rims 
from the Sharpless HII region catalog \citep{1959ApJS....4..257S} and the ESO(R) Southern Hemisphere Atlas. 
Using the submillimetre continuum observations, \citet{2008A&A...477..557M} have demonstrated the presence of 
at least one core in 39 of the 45 BRCs studied by them.
The morphology of these BRCs is, in general, supportive of the scenario seen in RDI
models; a dense core at the head of an elongated column along with  the presence of young stars.
\citet{2009A&A...497..789U} have used CO, mid-infrared (MIR) and radio data to identify 24 of 
the 45 southern BRCs that are undergoing a strong interaction with their HII region and 
have classified them as triggered candidates.  Fourteen of these 24 interacting BRCs were 
found to show active star formation on the basis of them being associated with 
embedded MIR point sources. The remaining BRCs  did not show any 
sign of triggered star formation or they were at an earlier evolutionary stage of star formation.
Whilst some individual clouds from the SFO catalog have been studied in detail 
\citep[e.g.][]{1997A&A...324..249L,1997AJ....114.1106M,2001A&A...376..271C,2004A&A...419..599T,2004A&A...428..723U,2006A&A...450..625U,2007A&A...467.1125U,2009A&A...497..789U, 2008A&A...477..557M}
and have been shown to harbor protostellar cores, the question of whether star formation is a common occurrence
within BRCs is still unresolved  \citep{2008A&A...477..557M}.
It is essential to carry out a census of YSOs in these SFO BRCs to determine the present status
of star formation within them and to relate it to their physical properties and morphologies.
 In particular, a systematic statistical study of the structure  of the resultant stellar systems of 
BRC star forming activities adjacent to HII regions 
under the influence of high mass O/B stars has not been made to date. 

The main aim of the present study is to investigate statistically the star forming activities in  BRCs, especially the structure of the resultant stellar systems or aggregates and its possible origin. 
Therefore, for our study, we have selected eight triggered BRCs  (cf. Table \ref{Tlog})  from the SFO  catalog which are 
currently showing active star formation.
These BRCs are located at heliocentric distances ranging from 0.95-2.80 kpc
and are distributed in the Galactic longitude $(l)$ between 250 and 340 degrees and latitude $|b|$ $<3$ degrees.
Distances of these BRCs have been taken from the available literature
\citep[][and references therein]{2009A&A...497..789U, 1999PASJ...51..791Y}.
These distances were derived mostly from the photometric data of the O type stars which are being assigned as the exciting star(s) 
of the HII regions in which these BRCs are located. 
For bright stars at the distance of 1-3 kpc, one can expect small
photometric errors ($<1-2\%$), and after taking care of the errors associated with 
intrinsic main sequence (MS), reddening values and fitting, we can safely assume 
that the error in the distances can be $\sim$5-15\% \citep{1994ApJS...90...31P}.

In the optical, the challenges in studying BRCs are their association
with high column density molecular clouds.
Wide-field near-infrared (NIR) cameras installed on moderate size telescopes is needed to
 probe BRCs
with sensitivity to detect sub-solar mass objects as well as the angular resolution to resolve high-density groupings or aggregates of stars. 
The wider field-of-view (FOV) is also necessary to observe the distribution of stars over multiparsec distances \citep[cf.][]{2005ApJ...632..397G}.
In the present study, we carrie out deep NIR observations with Infrared Side Port Imager (ISPI) camera on the 4m Blanco telescope 
at Cerro Tololo Inter-American Observatory (CTIO). We also utilized the infra-red (IR) data from the 
$Spitzer$ archive around these BRCs (except SFO 79, which has no $Spitzer$ data)  
to identify deeply embedded YSOs associated with them.
The spatial distribution of these YSOs has been compared to that of the surroundings as a function of their evolutionary status.
According to the simulations shown by \citet{2006MNRAS.369..143M}, the cloud cores of BRCs have sizes between 1-2 pc. 
The range of the distances of the selected BRCs are $\sim$1-3 kpc, 
therefore, the $10\times10$ arcmin square FOV of ISPI camera  will 
correspond to 3-9 pc, which is sufficient to cover the BRC regions necessary for our analyses.
We are using new quantitative techniques for analyzing the spatial structures of the aggregates 
to reveal the presence of compact 
cores and to compare the properties of
these cores with those found in other SFRs. 
These YSO cores' properties (density, size, etc.)  have been used to infer the history of star formation in the parental molecular cloud. 

In this paper, Section 2 describes the observations and data reduction. The YSO identification technique, 
the completeness of the resultant YSOs sample, and their  spatial distribution, 
mainly with respect to the associated molecular clouds, is discussed in Section 3. The methods for finding the aggregate cores and the active regions
are also explained in Section 3. In Section 4, we will discuss our results and will conclude in Section 5.

\section{Observations and data reduction}

Deep NIR broad band  observations of the fields containing BRCs
along with the $Spitzer$ archival data have been used in the present study.

\subsection{Blanco Observations }

\begin{figure*}
\centering \includegraphics[height=7cm,width=7cm,angle=0]{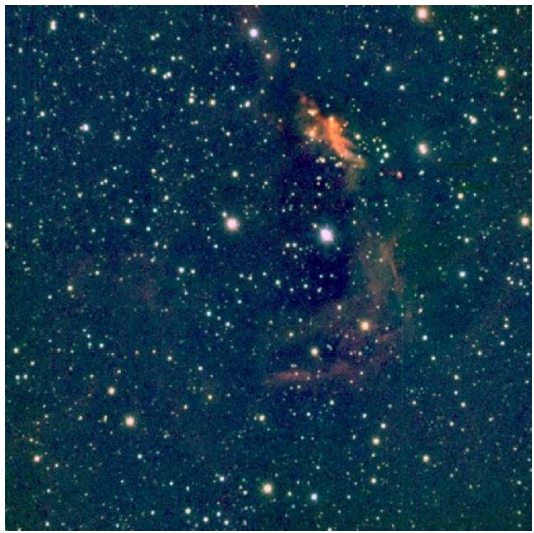}
\caption{\label{Fimage} Color composite image of the SFO 54 region covering $10\times10$ arcmin$^2$ FOV based on the 
ISPI data.
The red, green and blue colors correspond to $K^\prime$, $H$ and $J$ bands, respectively.
  }
\end{figure*}

NIR ($J,H,K^\prime$) data for eight selected BRCs along with two nearby field regions  (cf. Table \ref{Tlog}) were collected with the ISPI camera 
\citep[FOV $\sim$10.5 $\times$ 10.5 arcmin$^2$; scale 0.3 arcsec/pixel;][]{2004SPIE.5492.1582V} on the 4 m Blanco
telescope at CTIO, Chile during the  nights of 2010 March 03 to 04.
The seeing was $\sim$1 arcsec.
The individual exposure times were 60 s per frame for all the filters.
We used a set of $3\times3$ grid pattern, with 1 arcmin step, to
compensate for the cosmetic effects of the detectors, and to create
a sky image for the sky subtraction. 
The total exposure time for the target fields were 540 s 
for each $J,H$ and $K^\prime$ bands. 
Dome flats and dark frames were taken at the beginning and end of each night.
Data reduction followed the usual steps for NIR data:
dark subtraction, flat-fielding, sky subtraction, alignment and averaging of
sky-subtracted frames for each filter separately. 
The dome flats were also used to flag bad pixels. The sky frames
were median combined and subtracted from the science images.
In Fig. \ref{Fimage}, we have shown the color composite image of the SFO 54 region made
by using the clean $J,H$ and $K^\prime$ band images of 10.5 $\times$ 10.5 arcmin$^2$ field.

\begin{figure*}
\centering\includegraphics[height=4cm,width=10cm,angle=0]{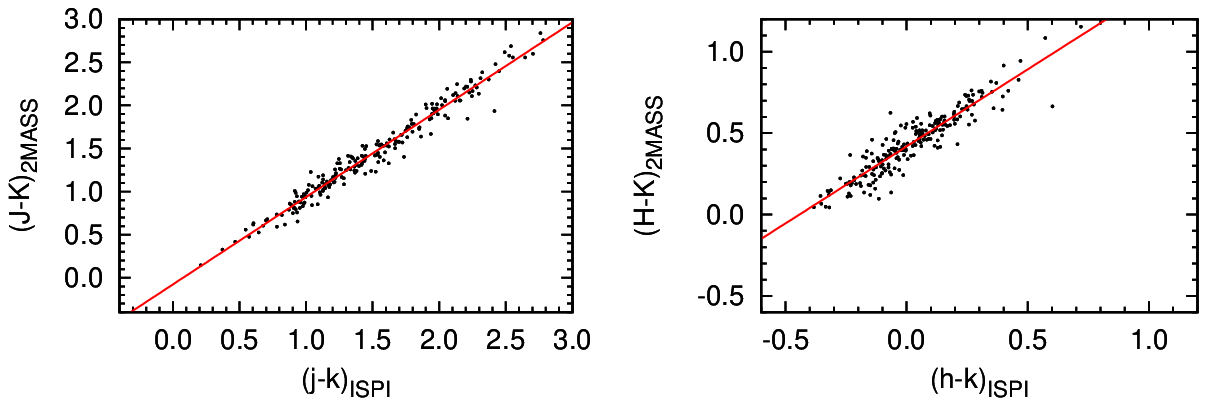}
\centering\includegraphics[height=4cm,width=5cm,angle=0]{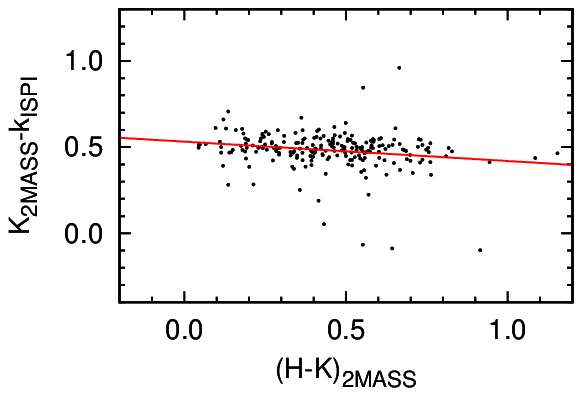}
\caption{\label{Ffit} Fit for the transformation coefficients for ISPI data using 2MASS data for the sources in the SFO 54 region.}
\end{figure*}

\begin{figure*}
\centering\includegraphics[height=7cm,width=16cm,angle=0]{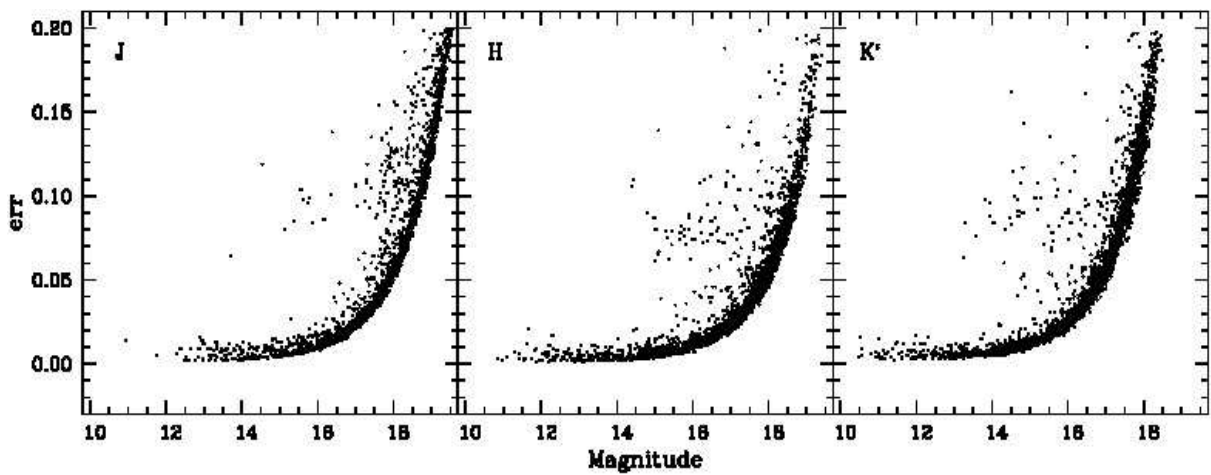}
\caption{\label{Ferr} Photometric errors as a function of magnitudes for all three bands of ISPI data for the sources in the SFO 54 region.}
\end{figure*}

The PSF-fitting stellar photometry on these sky-subtracted and
combined images was carried out by using the $find, phot, psf$ and $allstar$ routines within DAOPHOT \citep{1994PASP..106..250S}. 
The calibration of the photometry to the standard system was done by using the transformation equations:

\begin{equation}
\rm (J-K) = M1\times(j-k) + C1
\end{equation}

\begin{equation}
\rm (H-K) = M2\times(h-k) + C2
\end{equation}

\begin{equation}
\rm (K-k) = M3\times(H-K) + C3
\end{equation}

where the capital $JHK$ are the standard magnitudes of the common stars taken from the 
Two Micron All Sky Survey (2MASS) catalog which provides absolute photometry in the $J$ (1.25 $\mu$m), 
$H$ (1.65 $\mu$m), and $K_s$ (2.17 $\mu$m) bands down to a limiting magnitude
of 15.8, 15.1, and 14.3, respectively, with a signal-to-noise ratio greater than 10.
The small $jhk$ and Ms \& Cs are the instrumental ISPI magnitudes and the transformation coefficients, respectively.
In Fig. \ref{Ffit}, as an example of the fit for the transformation coefficients of the ISPI data 
to the 2MASS data is shown for the sources in the SFO 54 region.
The values of these coefficients are given in Table \ref{Tslopes} along with 
the standard errors ($\sim$0.01-0.02 for the zero point (C)  and less than 0.03 for the color term (M)). 
The typical DAOPHOT errors in magnitude as a
function of corresponding magnitudes are shown in Fig. \ref{Ferr}.
It can be seen that the errors become large (0.2 mag) for stars fainter than $K=18.5$ mag, so
the measurements beyond these magnitudes are not reliable and were not used in this study.

Because of the higher number of detected stars in $K$ and $H$ bands, we have used the $H-K$ color to calibrate $K$ magnitude.
Therefore, in the final catalog, we have only those stars which are detected in at least $K$ and $H$ bands and
found by merging the individual photometric catalog by a search radius of 0.3 arcsec.
We have done the alignment of the individual frames using the IMALIGN task of IRAF\footnote{IRAF is distributed by the National Optical Astronomy Observatory,
which is operated by the Association of Universities for Research
in Astronomy (AURA) under a cooperative agreement with the National
Science Foundation.}
with an accuracy better than 0.1 arcsec; our 0.3 arcsec match threshold should 
match sources for upto 3$\sigma$ positional errors. 
In a few cases when there was a second astrometric match, we adopted the closer match for our final merged catalog. 
We used the $K$ band position for  astrometry of the final catalog. 
Stars brighter than 10 mag in $K$ are saturated in our observations, so we have taken their respective
magnitude from the 2MASS point source catalog.
The final catalog contains 2000-15000 sources per field depending on the crowding/nebulosity in the region.

\subsection{{\it Spitzer} observations \label{obs-spit}}

We have used the archived data taken from the Infrared Array 
Camera \citep[IRAC;][]{2004ApJS..154...10F}  of the space-based 
{\it Spitzer} telescope at 3.6 $\mu$m, 4.5 $\mu$m, 5.8 $\mu$m and 8.0 $\mu$m bands.
We obtained basic calibrated data (BCD) from the {\it Spitzer} data archive for all the BRCs, except for SFO 76. 
The exposure time of each BCD was 10.4 sec and for creating a mosaic, few hundreds of 
BCDs have been used. Mosaicing in each wavelength was performed by the MOPEX software provided by 
{\it Spitzer} Science Center (SSC). All of our mosaics were built at the native instrument resolution of 1.2 arcsec pixel$^{-1}$ with the standard BCDs.
All the mosaics were then aligned and trimmed to make a  $10.5\times10.5$ arcmin$^2$ box size containing the same 
area as observed in the ISPI observations. These trimmed sections have been used for further analyses.
Some of the studied regions were not fully 
observed in the outer parts ($<$10\% of the whole area) in all the four channels of IRAC
but it does not affect our analyses, since our region of interest is always closer to the image center containing  the H II regions.

We used the {\it DAOPHOT} package in IRAF to detect sources and to perform photometry in each of the IRAC mosaics. 
The FWHM of every detection was measured and all detections with a FWHM $>$3.6 arcsec 
were considered resolved and were removed. The detections were also examined visually in each band to remove 
non-stellar objects and false detections. 
In order to avoid source confusion due to crowding, {\it PSF} photometry for all the sources was carried out. 
Aperture photometry for well isolated sources was first done by using an aperture radius of 3.6 arcsec 
with a concentric sky annulus of the inner and outer radii of 3.6 and 8.4 arcsec, respectively. 
 The FWHM of the star's intensity profile were between 2.4 - 3.6 arcsec in different IRAC bands.
As the studied regions are nebulous and sometimes crowded, the inner bright region of the intensity profile 
($\sim1\times$ FWHM: 3.6 arcsec) of stars has been used to derive the aperture magnitude. This will lose some photons from the outer wings of the intensity profile of stars.
Since this profile behaves like a Gaussian, at 3$\times$FWHM radii, around $\sim$99\% of star photons can be considered integrated. 
Therefore, we have applied the aperture correction as a difference in aperture magnitudes at these two radii, 
as has been suggested in the IRAC Handbook for data reduction\footnote{http://irsa.ipac.caltech.edu/data/SPITZER/docs\/irac/\\iracinstrumenthandbook/IRAC$\_$Instrument$\_$Handbook.pdf.}.
We adopted the zero-point magnitudes, for the standard aperture radius of 12 arcsec ($\sim 3 \times FWHM$) and background annulus 
of 12-22.4 arcsec, as 19.670, 18.921, 16.855 and 17.394 in the 3.6 $\mu$m, 4.5 $\mu$m, 5.8 $\mu$m and 8.0 $\mu$m bands, respectively (IRAC data Handbook). 
The necessary aperture corrections for the {\it PSF} photometry were also calculated
as a difference between aperture and {\it PSF} magnitudes of the selected well-isolated sources and
were  applied to the {\it PSF} magnitudes of all the sources. 

The sources with photometric uncertainties $<0.2$ mag in each 
band were considered as good detections and are used in further analyses. 
Around 600-1900 sources have been detected in 3.6 $\mu$m in different regions with fewer detections at longer wavelengths.
This may be because the shorter wavelength channels are more sensitive and less affected by the
bright diffuse emission that dominates the 5.0 $\mu$m and 8.0 $\mu$m observations.
The underlying typical stellar photosphere is also intrinsically fainter at 5.0 $\mu$m and 8.0 $\mu$m  than at 3.6 $\mu$m and 4.5 $\mu$m.
The typical magnitude limits of the data in the 3.6 $\mu$m, 4.5 $\mu$m, 5.8 $\mu$m and 8.0 $\mu$m bands having $S/N > 5$ (error $\le 0.2$ mag) 
were found to be $\sim$ 16.0, 15.5, 13.0 and 12.8 mag, respectively, but they varied from region to region.
The NIR ISPI counterparts of the IRAC sources were then searched for within a radius of 1 arcsec.

\section{Results}

\subsection{Strategy of YSO identification \label{idf}}

The YSOs are usually grouped in an evolutionary sequence representing:
accreting protostars (Class 0), evolved protostars (Class I), classical T-Tauri stars (CTTSs: Class II) and weak-
line T-Tauri stars (WTTSs: Class III) \citep[cf.][]{1999ARAA..37..363F}. 
The YSOs of earlier stages are usually deeply buried inside the molecular clouds hence it is difficult to detect them
 at optical wavelengths.
The most prominent feature of these YSOs is the accreting circumstellar disks.
The radiations from the central YSO are more or less absorbed by these 
circumstellar material and re-emitted in IR.
Therefore, excepting Class III sources these YSOs with disks can  be probed through 
their IR excess (compared to normal stellar photospheres). 

However, observations of YSOs in SFRs such as Taurus Auriga show a wide dispersion ($\sim$1-10 Myr)
in the lifetimes of circumstellar disks \citep{1989AJ.....97.1451S,2001ApJ...553L.153H,2003MNRAS.342.1139A}.
Recently, \citet{2012ApJ...745...19K} studied excess emission in Taurus binaries over  
NIR to millimeter wavelengths
 and concluded that the 
prompt disk dispersal only occurs for a small fraction of single stars, and that $\sim80\%-90\%$ retains their disks for at least $\sim$2-3 Myr (but rarely for more than $\sim$5 Myr).
The YSO census based on IR excesses is not complete since it detects  only
YSOs having an accreting disk. But we are interested in whether our target BRCs have been
showing star forming activities in the last couple of Myr,
this limitation is small where few YSOs have had sufficient time for their disks to disperse 
\citep[cf.][]{2008ApJ...686.1195H,2001ApJ...553L.153H,2009ApJS..184...18G,2011ApJ...739...84G}.

The ISPI data along with the $Spitzer$ IRAC data have been used to identify and classify the YSOs
associated with the BRCs based on their excess emission in IR
by using the following classification schemes.

\subsubsection{Step  1: IRAC four-band YSO classification scheme:}

We have applied Step 1 to all sources which are detected in all four IRAC bands.
We separated out IR excess contaminants such as star forming galaxies, broad-line active galactic nuclei,
unresolved shock emission knots, objects that suffer from polycyclic aromatic hydrocarbon (PAH) emissions etc.,
by applying constraints in various color spaces \citep{2009ApJS..184...18G}. 

To minimize the inclusion of faint extra-galactic contaminants, a brightness limit to IRAC 
magnitudes has been applied. Also to discriminate out brighter contaminants, 
various IRAC color spaces along with magnitude limits have been used. 
Active star forming galaxies have a very strong PAH features yielding very 
red 5.8  and 8.0 $\mu$m colors 
\citep[the 6.2 $\mu$m and 7.7 $\mu$m PAH features  are much stronger than the 3.3 $\mu$m PAH feature:][]{2005ApJ...631..163S}, which can
therefore be very well separated out from YSOs by using 
IRAC two-color diagrams (TCDs). The resultant sample will have negligible residual 
contaminants \citep{2009ApJS..184...18G}. Broad line AGN  having MIR colors 
consistent with YSOs \citep{2005ApJ...631..163S} can also be separated out from  
YSOs simply by applying IRAC color selections. In  [5.8 - 8.0] vs. [3.6 - 4.5] TCD,  
the broad line AGNs lie along a vertical branch because of the lack of strong PAH feature at 5.8 $\mu$m 
and 8.0 $\mu$m and dominating power law emission peaking at 1.6 $\mu$m \citep{2005ApJ...631..163S}. 
Color/magnitude selection criteria based on \citet{2009ApJS..184...18G} has been used 
to eliminate these AGNs. Even after applying these cut-offs, we can still expect $\sim$8 contaminants 
per square degree \citep{2009ApJS..184...18G}. 
But our FOV is $\sim10 \times 10$ arcmin square, therefore the resultant contaminants 
will be very low ($\sim$0.2 contaminant  per field).

\begin{figure*}
\includegraphics[height=4cm,width=4cm,angle=0]{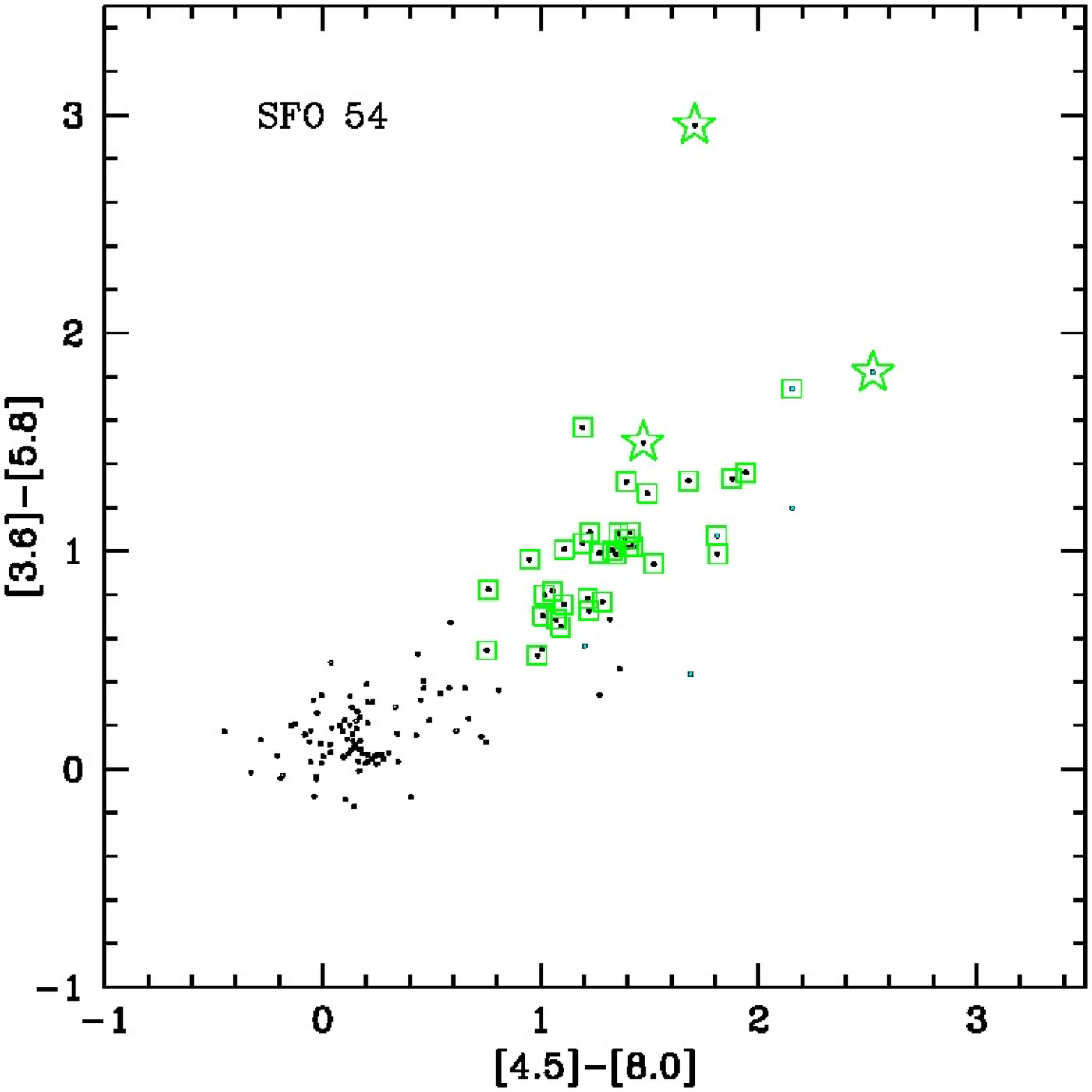}
\includegraphics[height=4cm,width=4cm,angle=0]{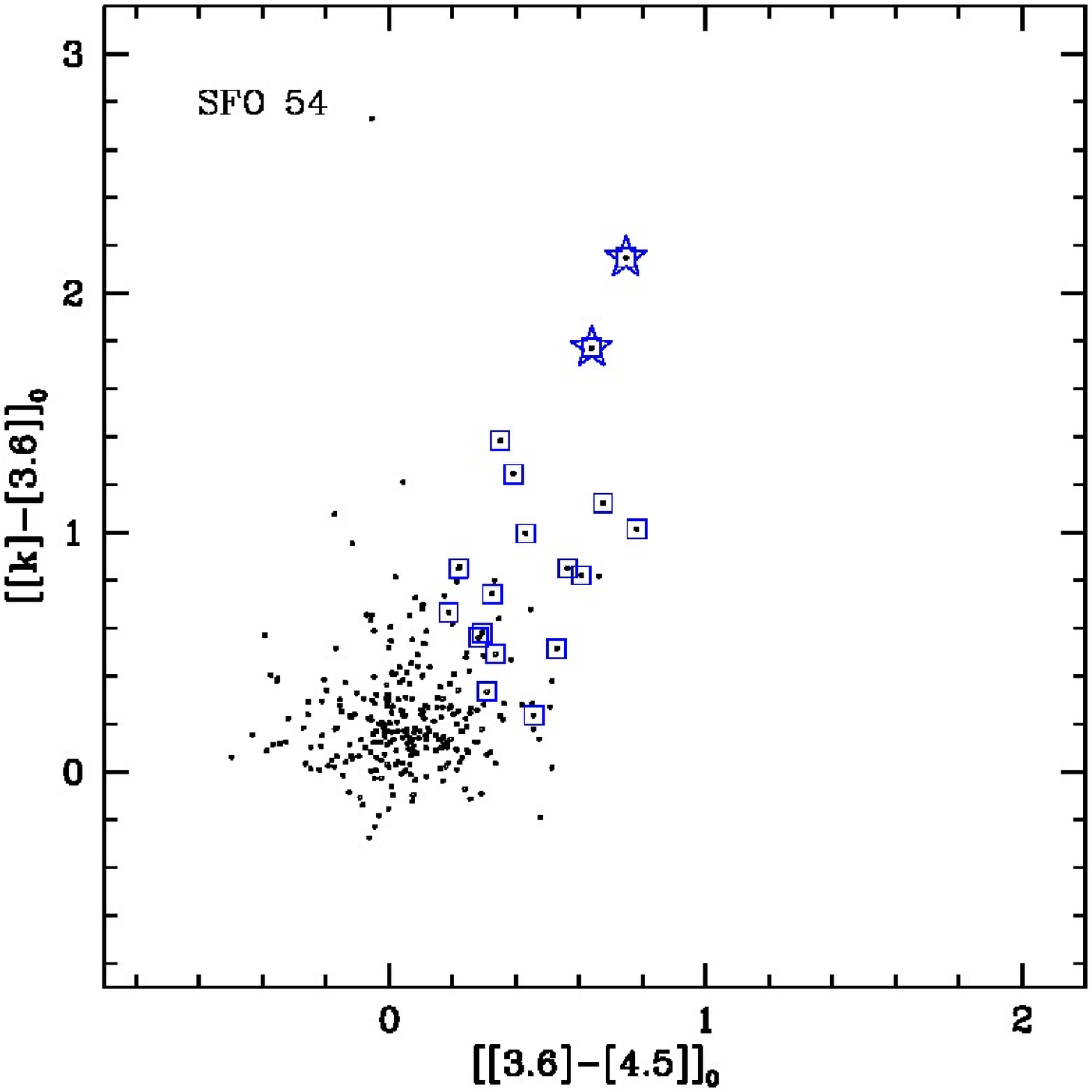}
\includegraphics[height=4cm,width=4cm,angle=0]{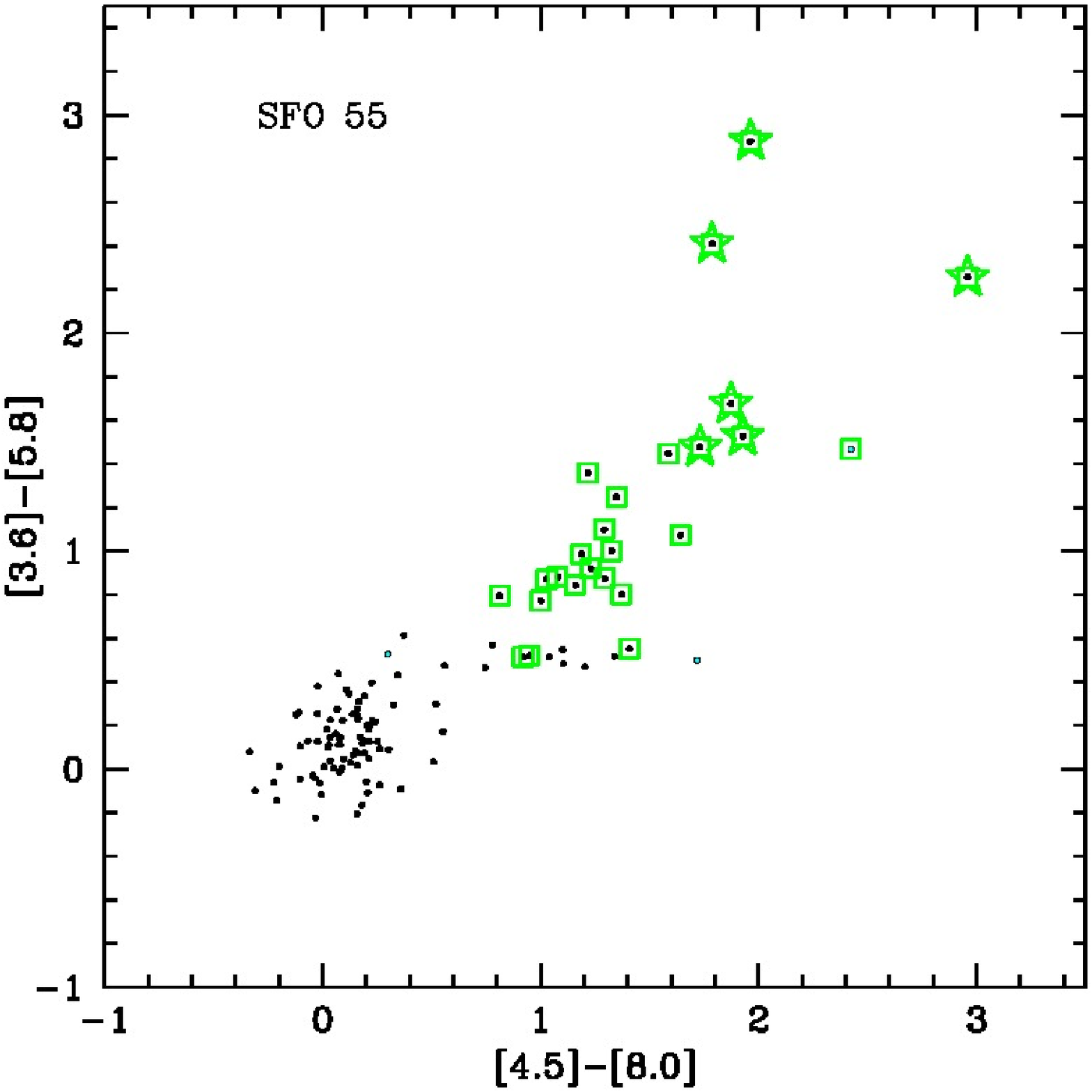}
\includegraphics[height=4cm,width=4cm,angle=0]{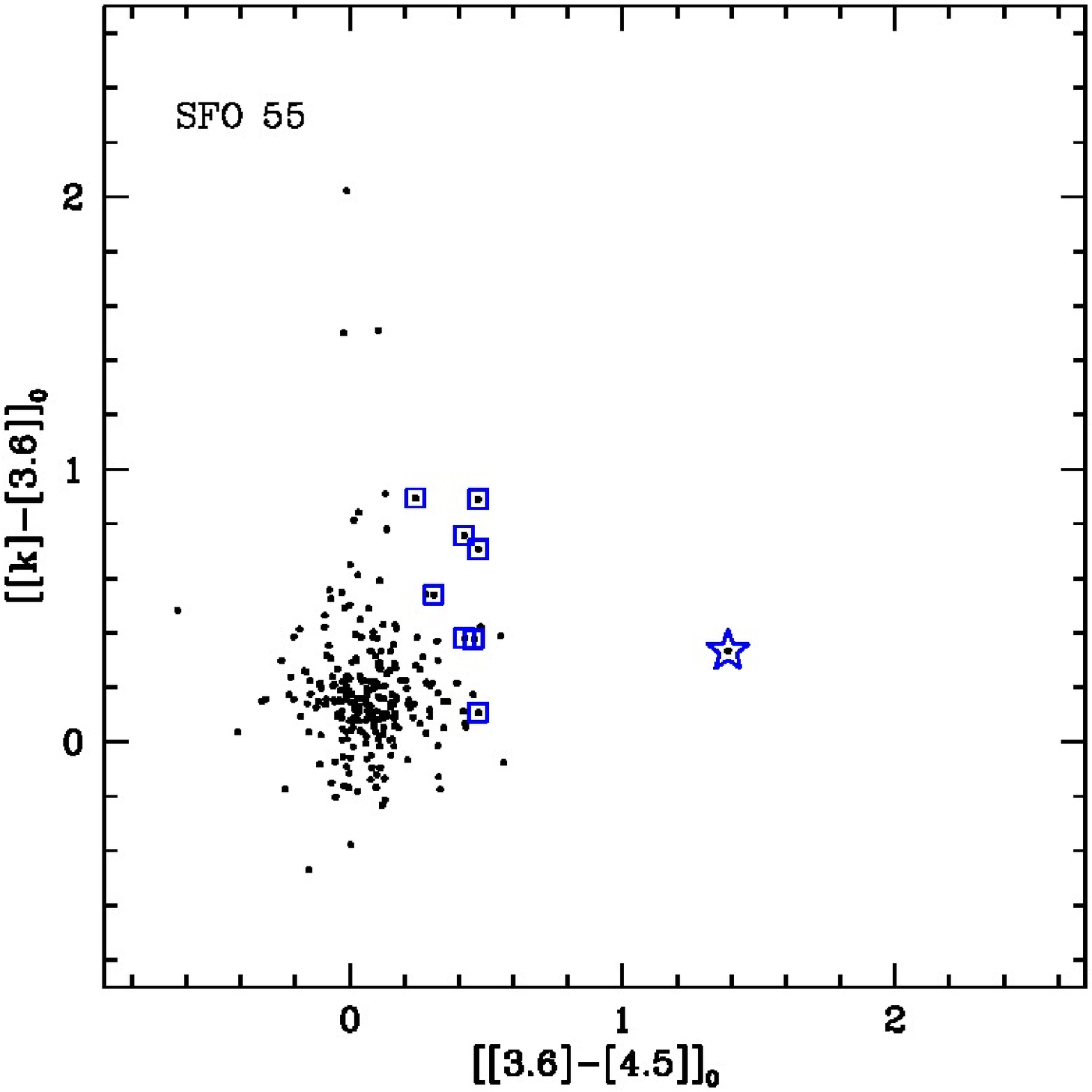}
\includegraphics[height=4cm,width=4cm,angle=0]{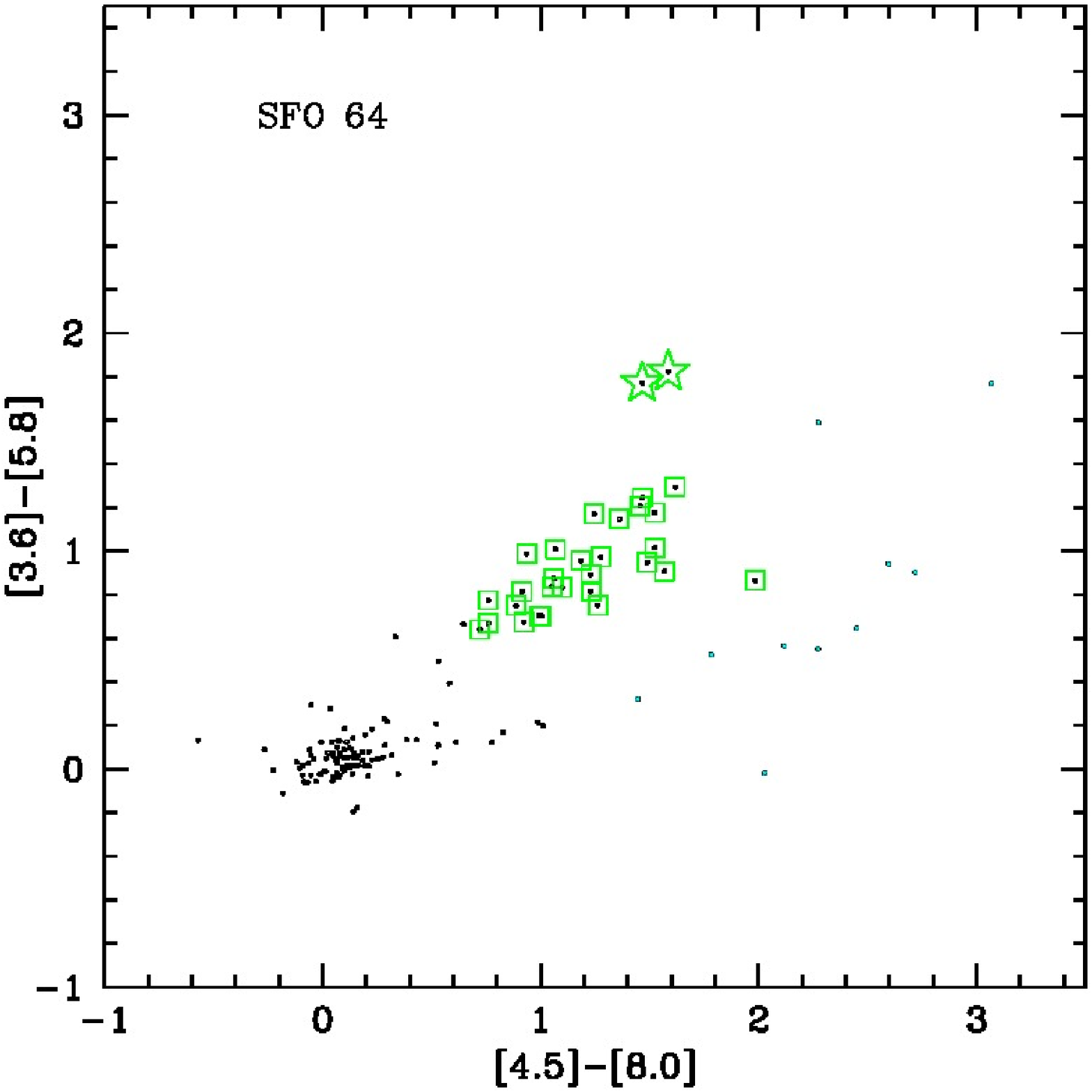}
\includegraphics[height=4cm,width=4cm,angle=0]{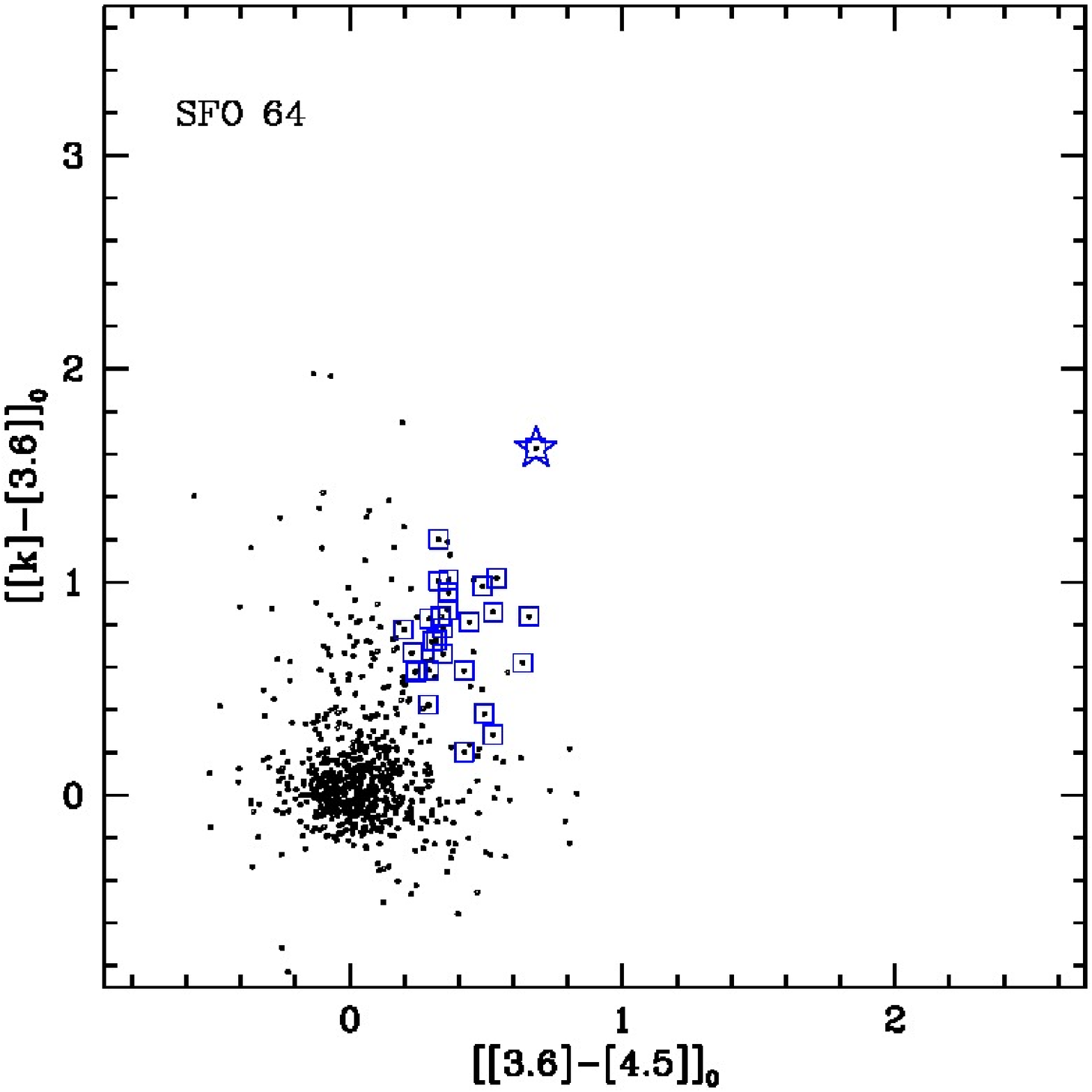}
\includegraphics[height=4cm,width=4cm,angle=0]{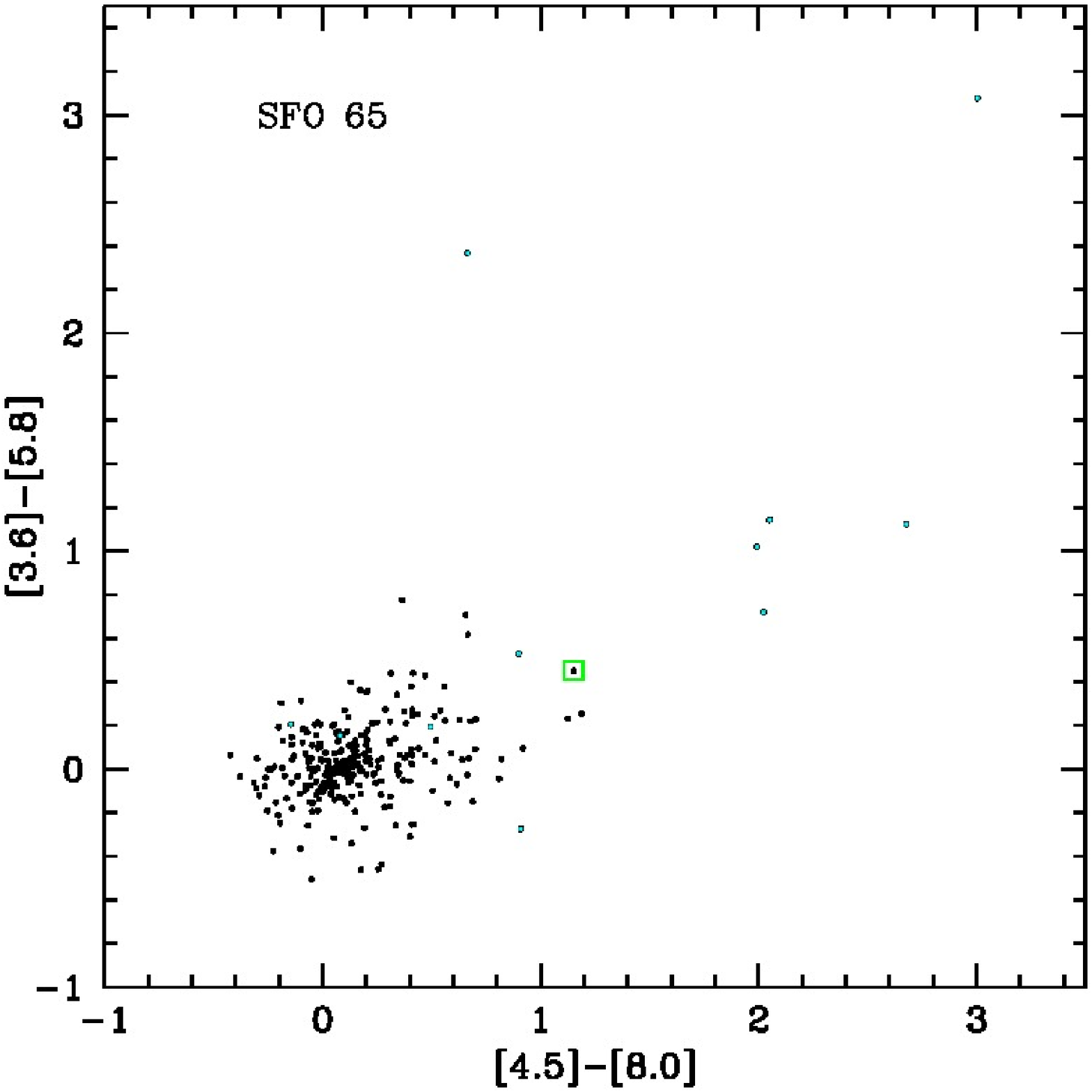}
\includegraphics[height=4cm,width=4cm,angle=0]{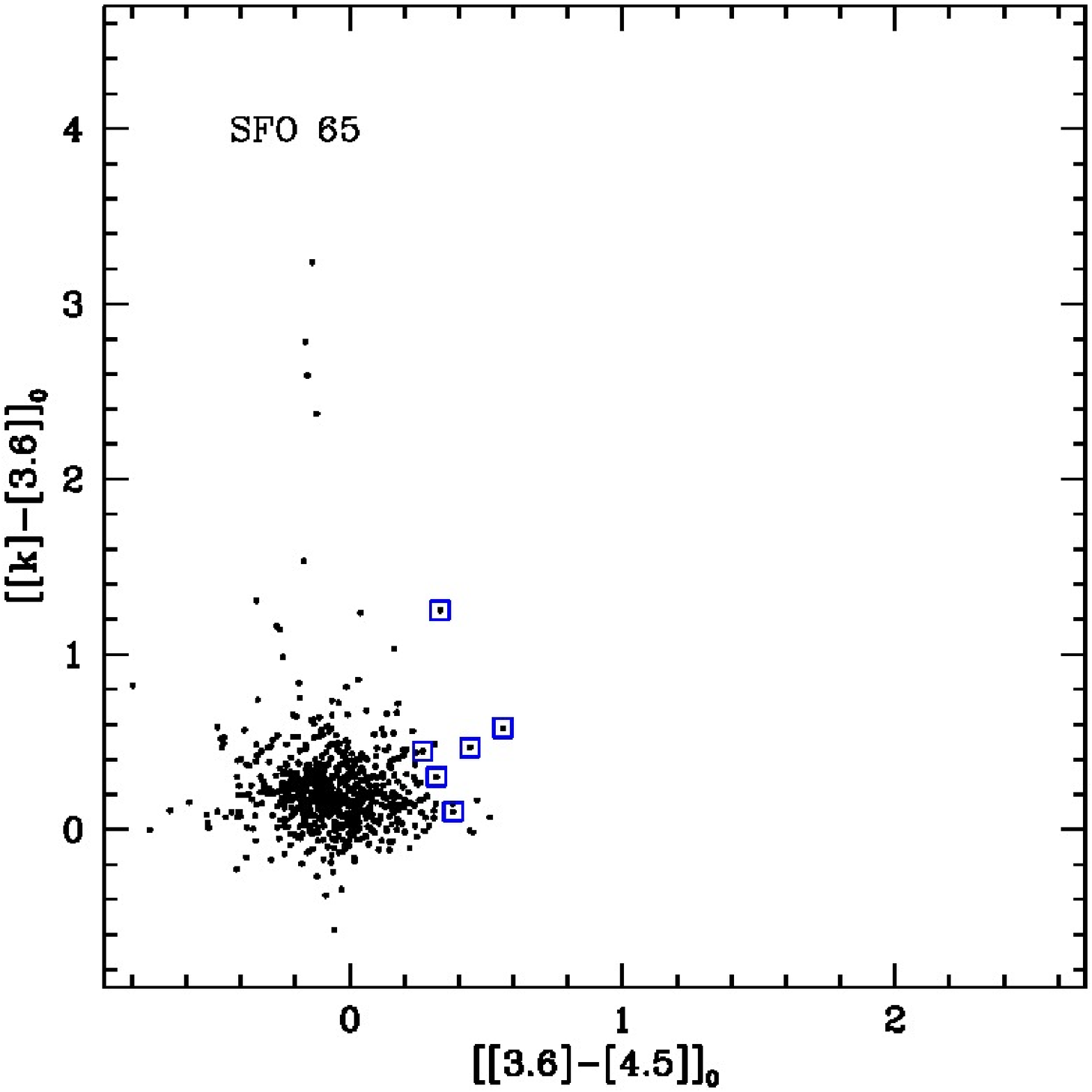}
\includegraphics[height=4cm,width=4cm,angle=0]{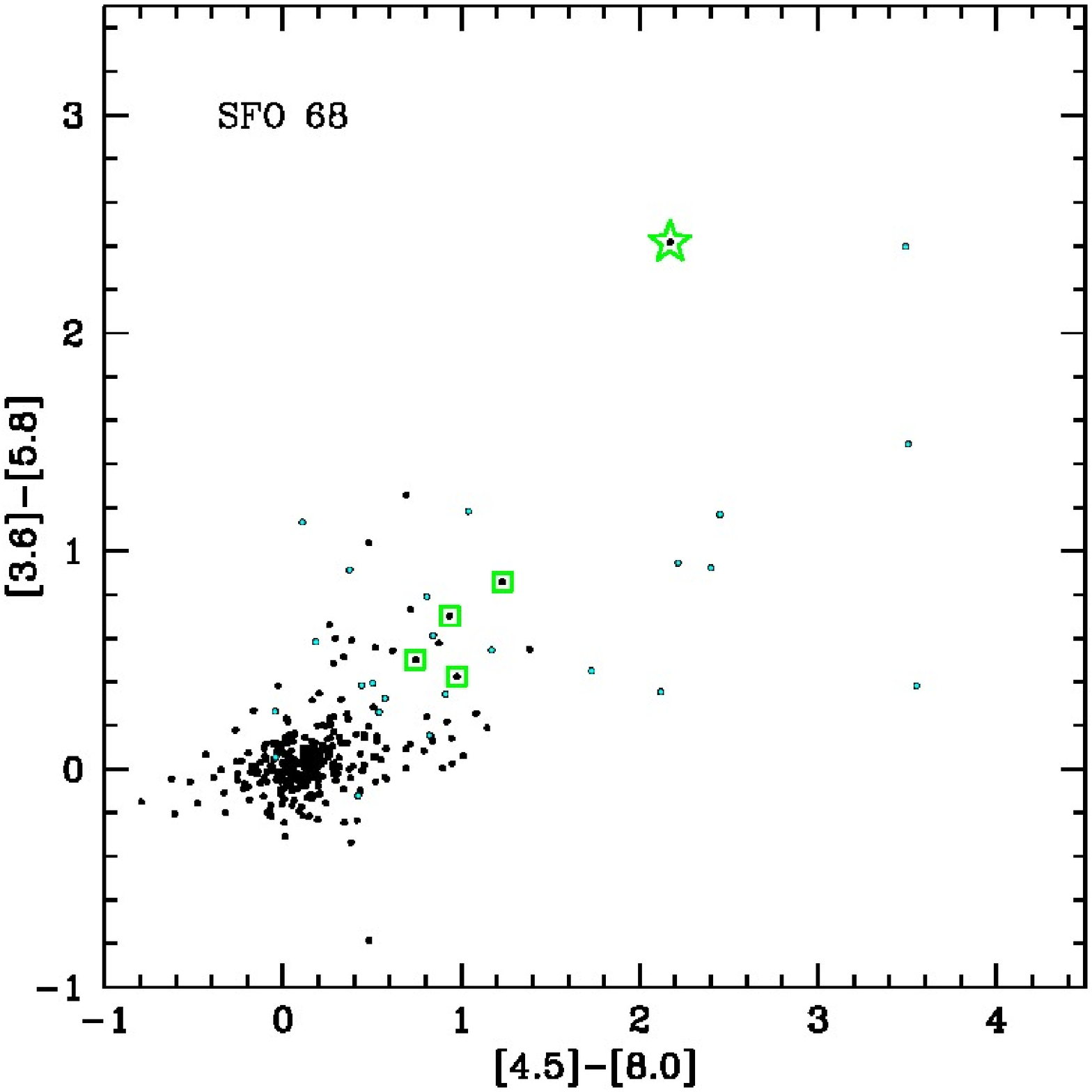}
\includegraphics[height=4cm,width=4cm,angle=0]{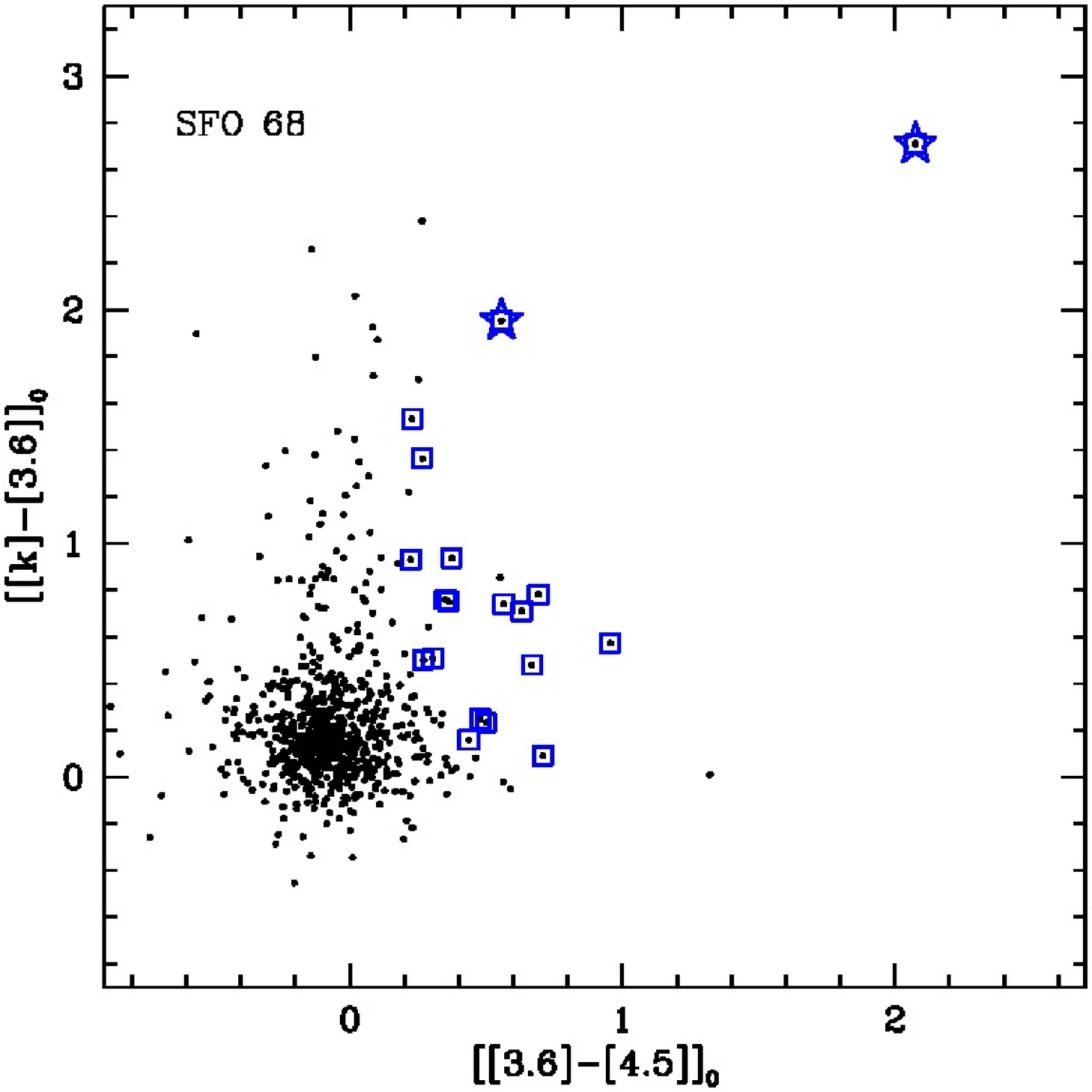}
\includegraphics[height=4cm,width=4cm,angle=0]{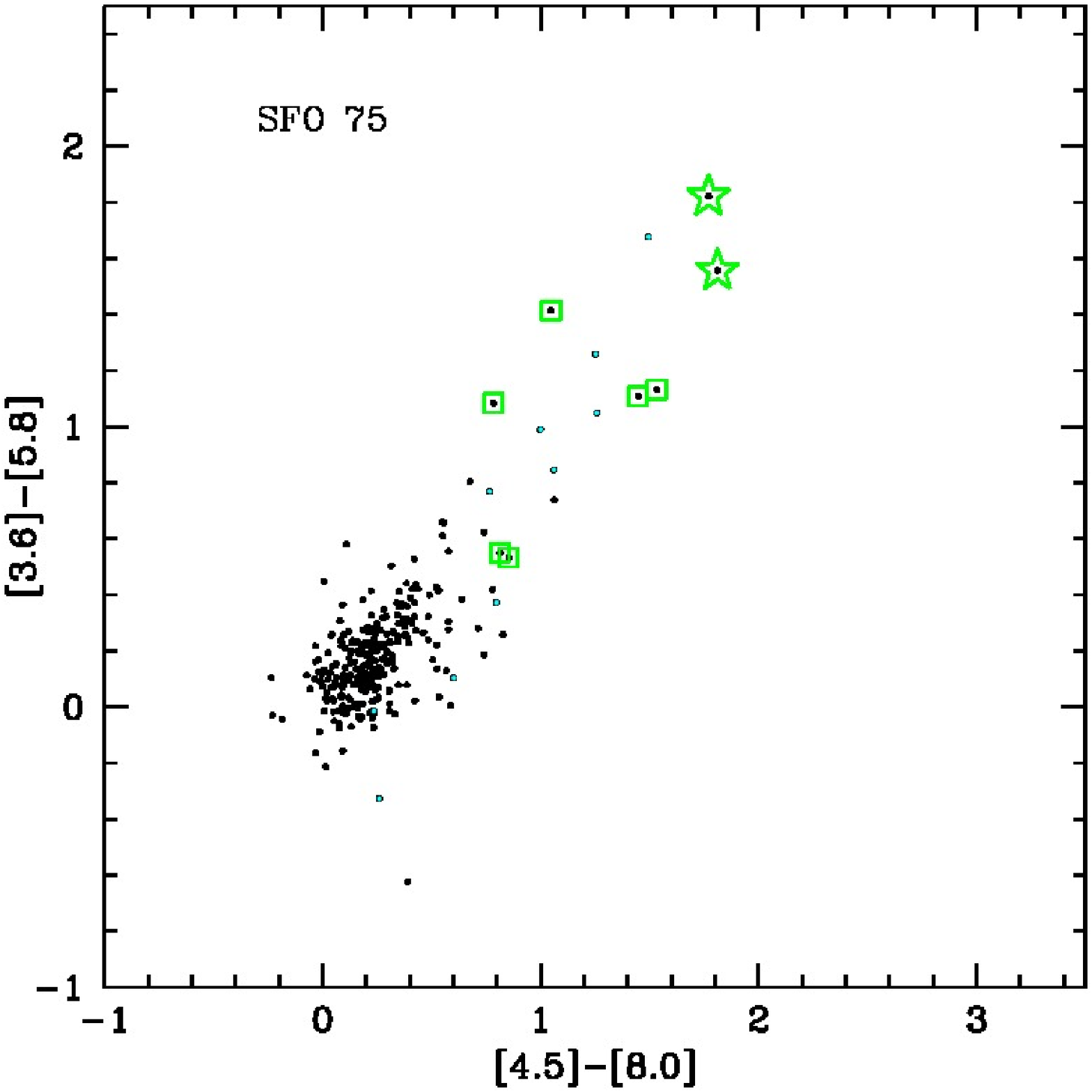}
\includegraphics[height=4cm,width=4cm,angle=0]{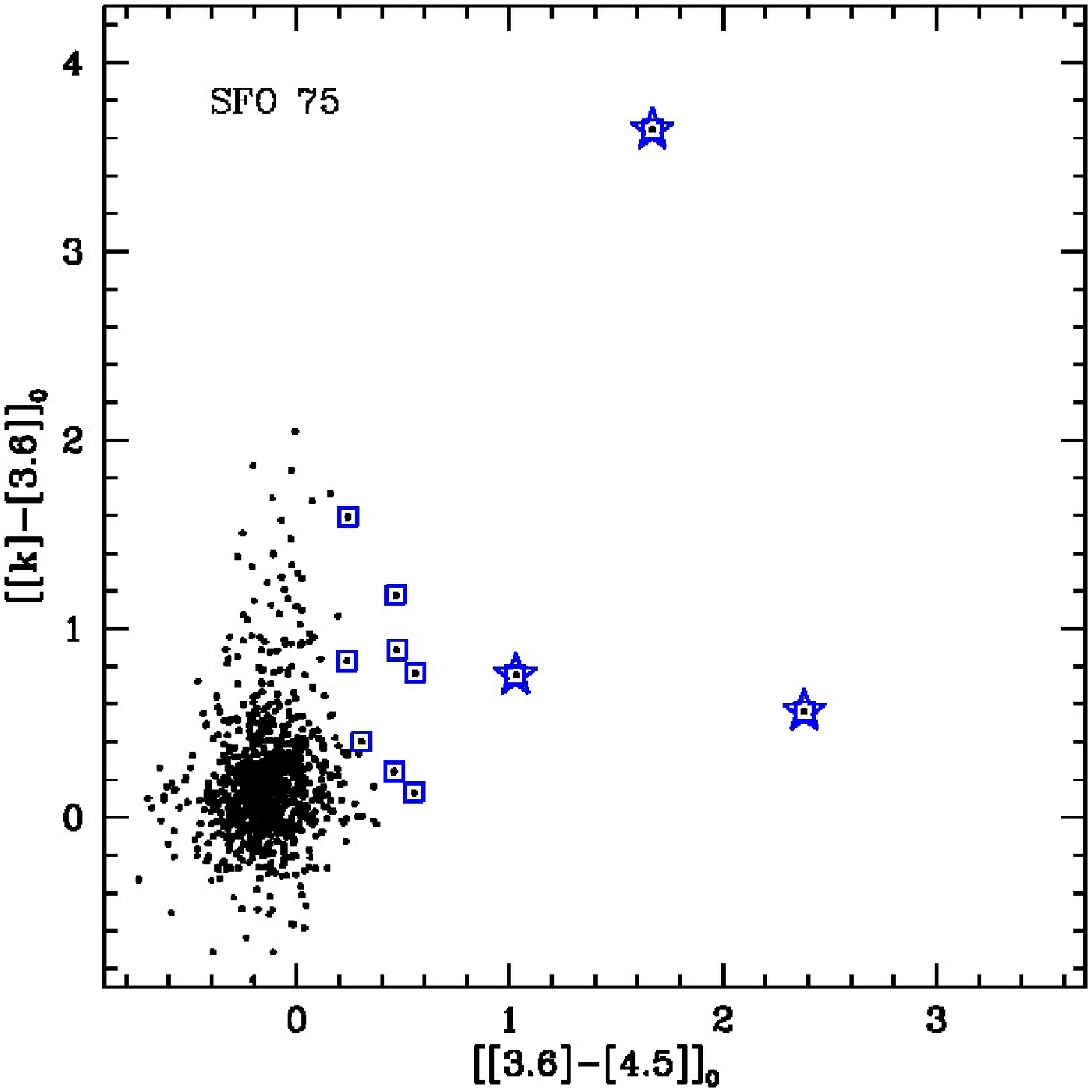}
\includegraphics[height=4cm,width=4cm,angle=0]{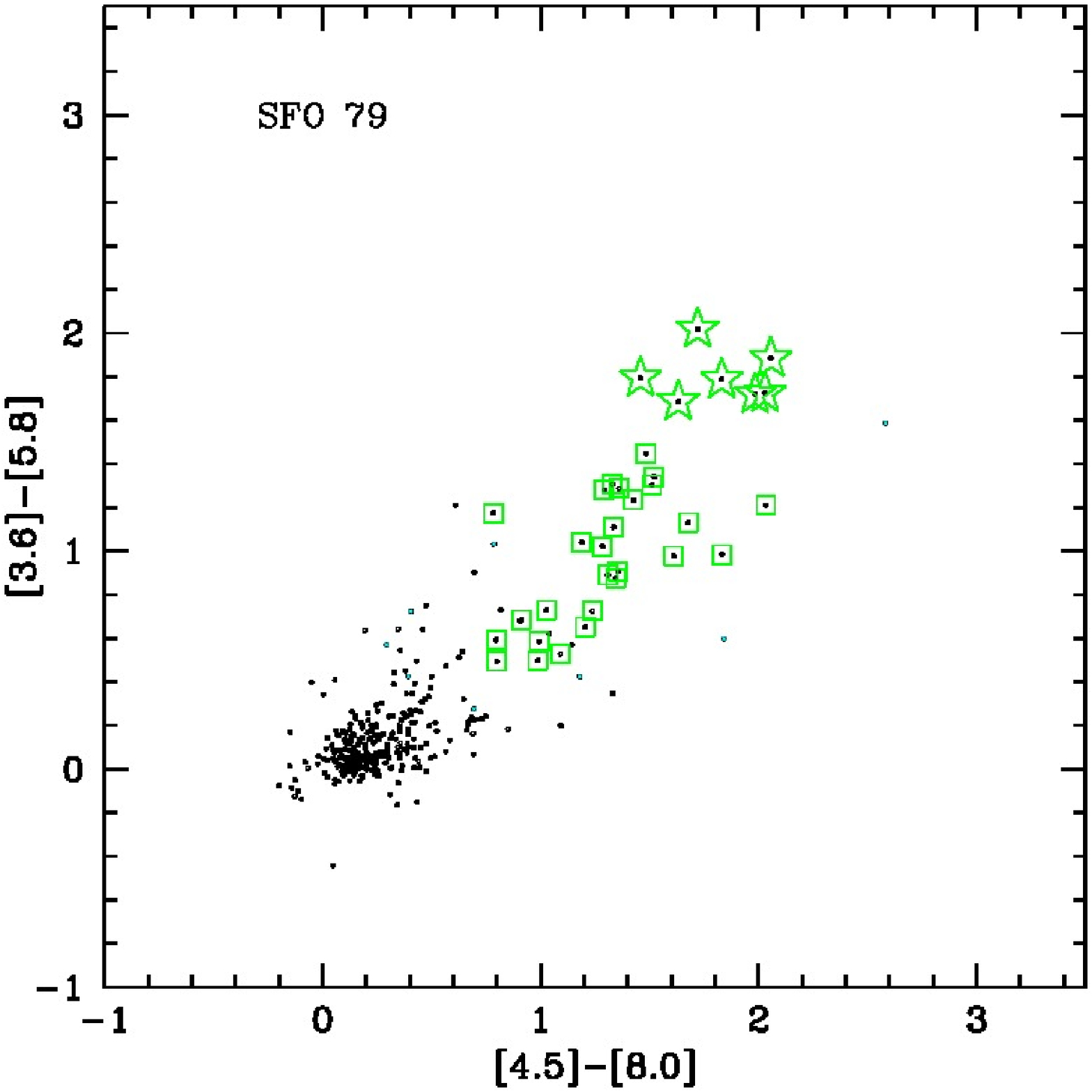}
\includegraphics[height=4cm,width=4cm,angle=0]{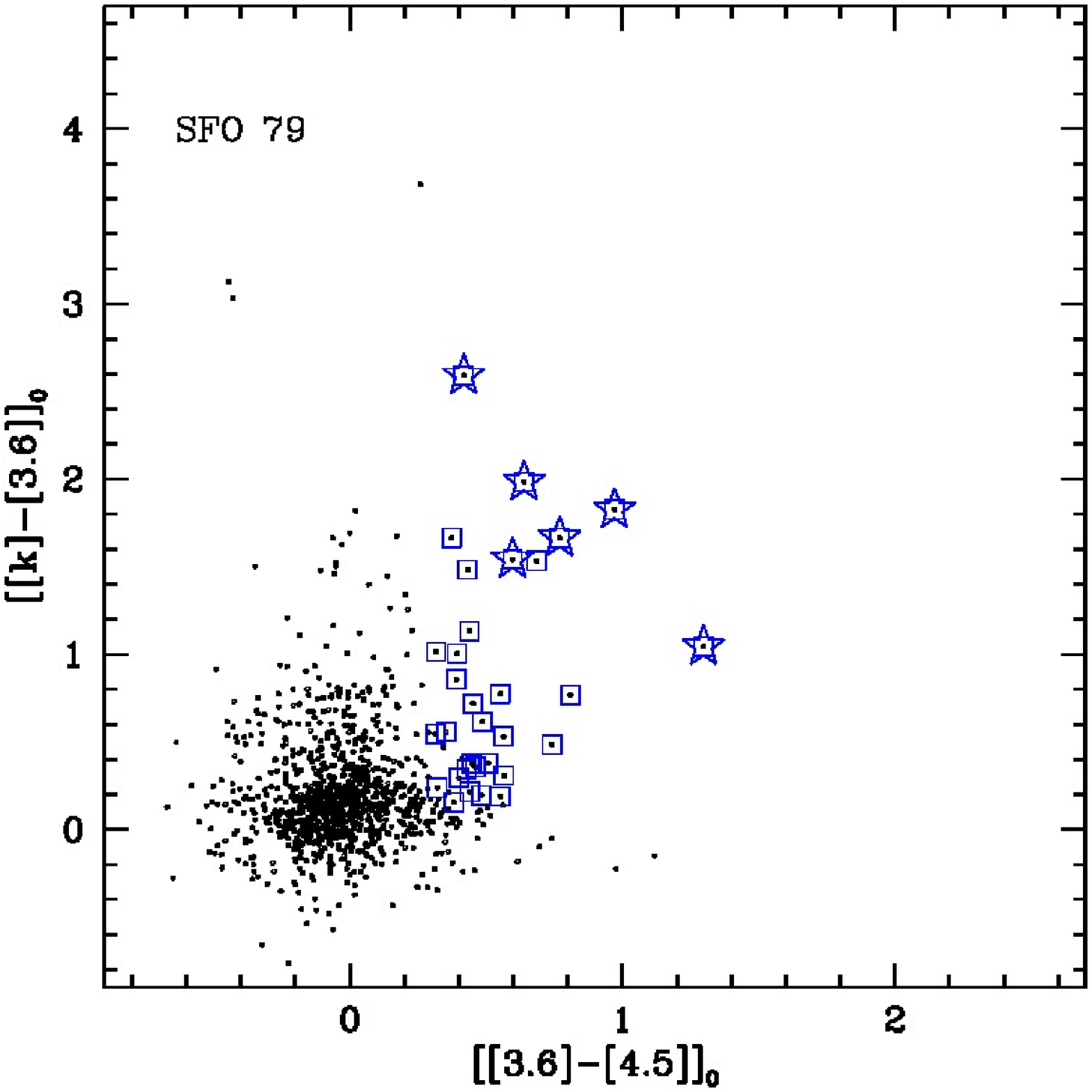}
\caption{\label{Firac} IRAC/2MASS TCDs for the stars in all the regions studied. The
YSOs classified as Class I and Class II, based on the color criteria by \citet{2009ApJS..184...18G}, are marked  using  star and square symbols, respectively. 
  }
\end{figure*}

We, then isolated YSOs with IR excess from those without IR excess and 
classified them as Class I or as Class II YSOs based on their prominence at longer wavelengths.
In Fig. \ref{Firac}, we have shown the [3.6 - 5.8] vs [4.5 - 8.0] TCD for
all the IRAC sources in  all regions studied, where Class I and Class II sources are represented by green star and green open square symbols, respectively. 
The identified contaminating sources are shown by blue dots. 

\subsubsection{Step 2: $K$, [3.6], [4.5] three band YSO classification scheme:}

Since the studied regions are highly nebulous, the YSO selection 
based on the IRAC four band photometry may  not be complete
as many sources falling in these regions could not be detected at 
longer wavelengths due to the saturation of detectors.
Therefore, we apply Step 2 to those sources that lack detection at
either 5.8 $\mu$m or 8.0 $\mu$m, but have NIR detection in the ISPI $K$ band.
In Fig. \ref{Firac}, we have plotted de-reddened [3.6 - 4.5]$_0$ vs [$K$ - 3.6]$_0$ TCD for 
the sources detected in  $K$, 3.6 $\mu$m and 4.5 $\mu$m bands in  all regions studied. 
The following procedure outlined in \citet{2009ApJS..184...18G} has been used
to classify sources as Class I (blue stars) and as  Class II (blue squares) YSOs.
To estimate the reddening, we have used the  NIR TCD (cf. \S 3.1.4). 
Stars having $[J-H]$ color $ \ge 0.6$ mag and lying above the
CTTS locus or its extension are traced back to CTTS locus or its extension to get their intrinsic colors.
The difference between the intrinsic color and the observed one would give the extinction value.
Once we have the extinction value for the individual stars, we generated an
extinction map for the whole BRC region. 
The extinction values in the sky plane were calculated with a resolution of 5 arcsec by 
taking the mean of extinction value of stars in a box having a size of 17 arcsec, and
then were used to deredden the remaining stars. 
In the above procedure, we have assumed the normal extinction law ($R_V$ = 3.1) to back-trace the 
stars to the CTTS locus. In many SFRs, $R_V$ tends to deviate from normal, 
preferably towards the higher values, 
for example: $R_V$ = 3.7 \citep[][Carina region]{2014A&A...567A.109K}, $R_V$= 3.3 \citep[][NGC 1931]{2013ApJ...764..172P},
$R_V$ = 3.5 \citep[][NGC  281]{2012PASJ...64..107S} and $R_V$ = 3.7 \citep[][Be 59]{2008MNRAS.383.1241P}. 
But the difference between $R_V$ = 3.1 and  $R_V$ = 3.7 would make only a small effect in near and mid IR region, since for $\lambda > \lambda_I$, the reddening law can be taken as a universal quantity \citep[][]{1989ApJ...345..245C,1995ApJS..101..335H}.

\subsubsection{Step 3: $H$, $K$ two-band YSO classification scheme:}

\begin{figure*}
\centering\includegraphics[height=8cm,width=14cm,angle=0]{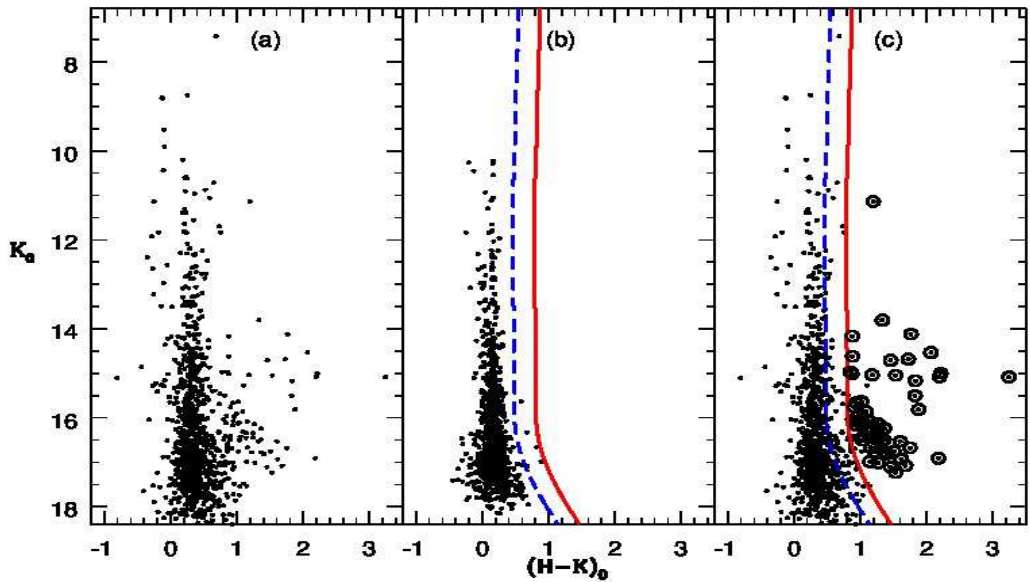}
\caption{\label{Fband} $K{_0}/(H-K){_0}$ CMD for (a) stars in the SFO 55 region, (b) stars in the field
region and (c) stars in the SFO 55 region  with the marked probable IR excess sources (circles).
The blue dashed curve is the outer envelop of the dereddened field stars and
the thick red curve separates the distribution of probable IR excess stars from that of MS stars.
 }
\end{figure*}

We applied Step 3 to all the detections in the ISPI $H$ and $K$  bands. 
The scheme, explained in detail by \citet{2014A&A...567A.109K}, 
compares the dereddened color-magnitude diagrams (CMDs) of equal area
of the studied region  and nearby field region (cf. \S 2.1).
The dereddened magnitudes and colors were obtained by using the 
extinction map as discussed in the previous section.
In Fig.~\ref{Fband}, as an example, we have plotted the de-reddened NIR CMDs, K$_0$ vs (H - K)$_0$, for both SFO 55 and the nearby field region.
The blue dashed curve is the outer envelop of the field stars and the thick red curve 
is the same curve reddened by A$_V$ = 5 mag to make allowance for the scattering due to the 
clumpy nature of the molecular clouds  \citep{2014A&A...567A.109K}.
A comparison of the CMDs reveals many sources of $(H-K)_0 \gtrsim 0.6$ mag in the BRC region, 
suggesting significant IR excesses in the $K$ band.
All the stars having a color `$(H-K)_0 - \sigma_{(H-K)_0}$' larger than the RED cut-off curve 
\citep[shown as a red solid curve in Fig.~\ref{Fband}, see for detail][]{2014A&A...567A.109K}
 might have an excess emission in the $K$ band and thus can be
considered as probable YSOs \citep[see also][]{2004ApJ...616.1042O,2012ApJ...759...48M}. 
While these stars are probably dominated by YSOs, still they could be contaminated by variable stars, dusty
asymptotic giant branch stars, unresolved planetary nebulae and background galaxies
\citep{2008AJ....136.2413R, 2011ApJS..194...14P}.
However, since there are no stars located in the field CMD redward of the 
RED envelope and neither of these IR excess sources
fall on similar positions of previously identified contaminants (cf. \S 3.1.1), we assume
they are most probably YSOs with IR excess.

\subsubsection{Step 4: $J,H,K$ three-band YSO classification scheme:}

\begin{figure*}
\centering\includegraphics[height=4cm,width=4cm,angle=0]{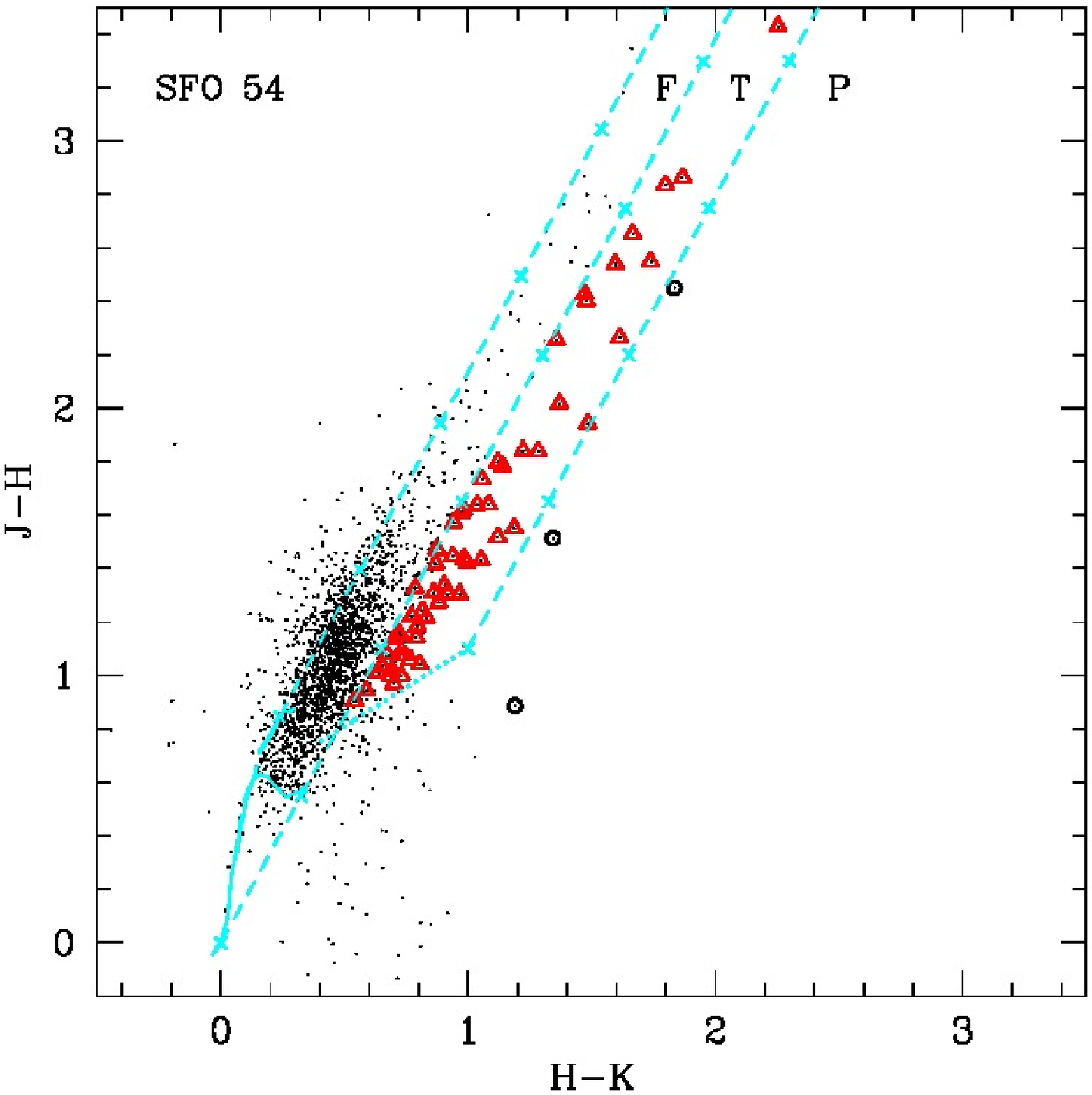}
\centering\includegraphics[height=4cm,width=4cm,angle=0]{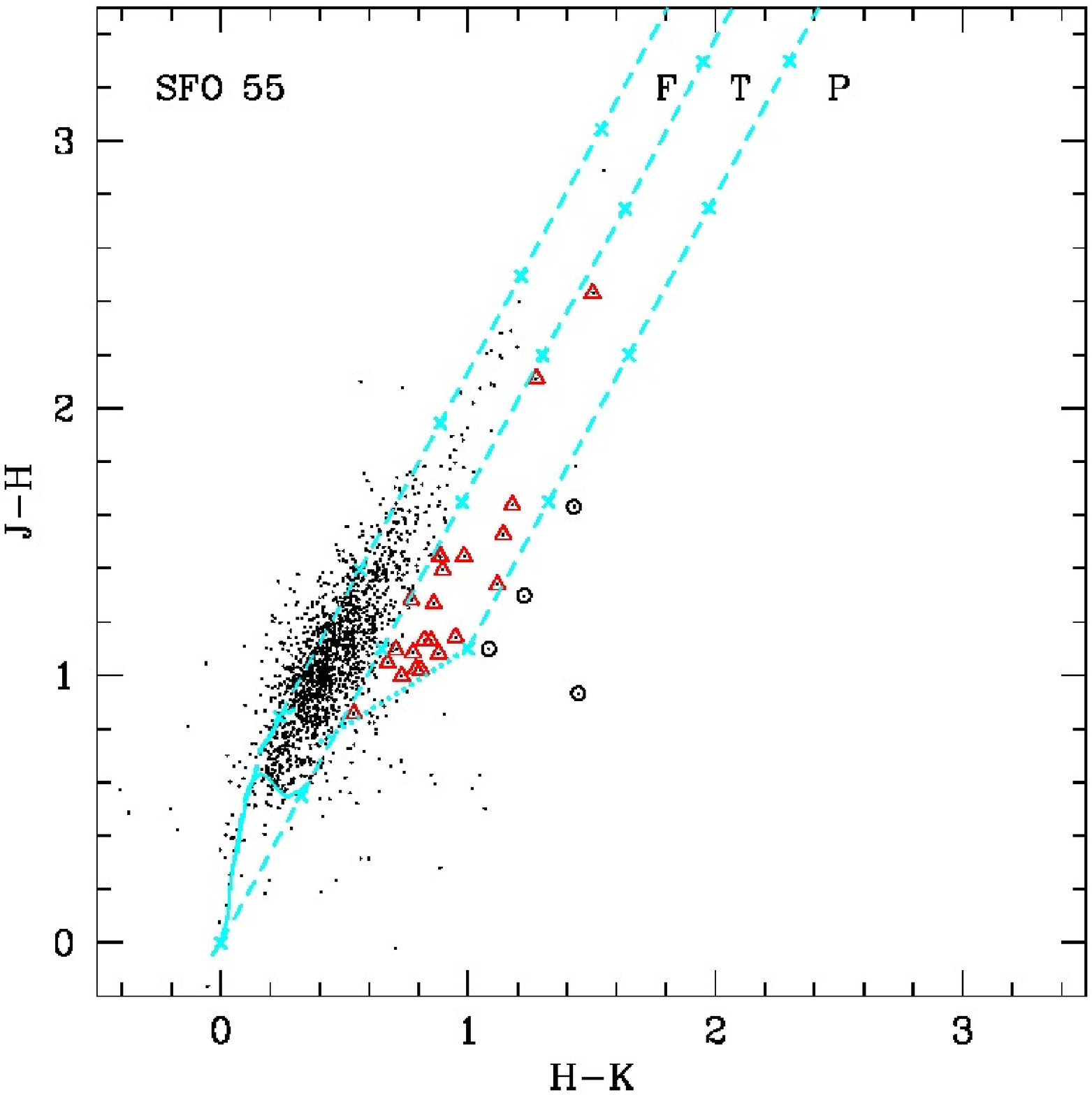}
\centering\includegraphics[height=4cm,width=4cm,angle=0]{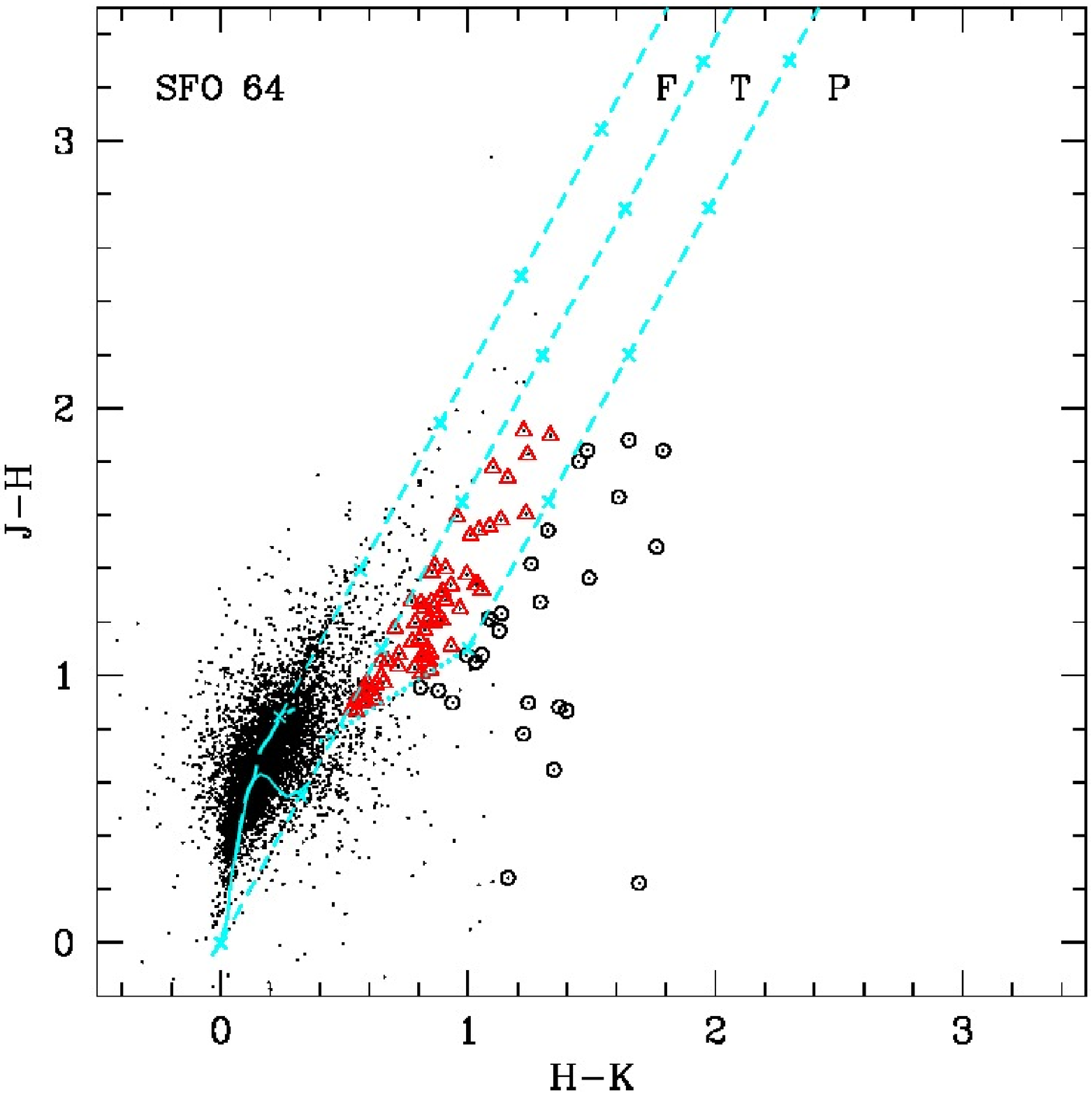}
\centering\includegraphics[height=4cm,width=4cm,angle=0]{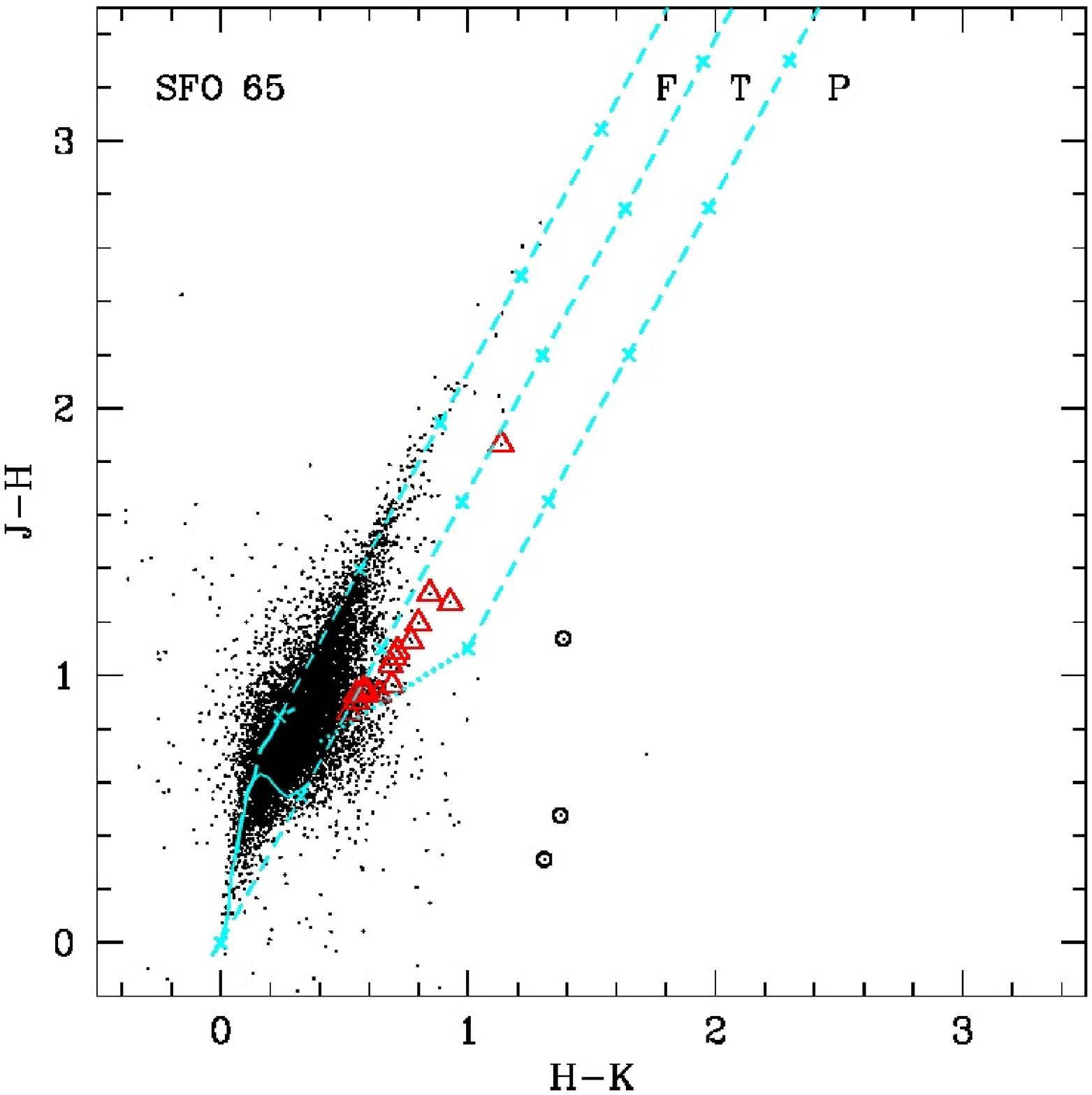}
\centering\includegraphics[height=4cm,width=4cm,angle=0]{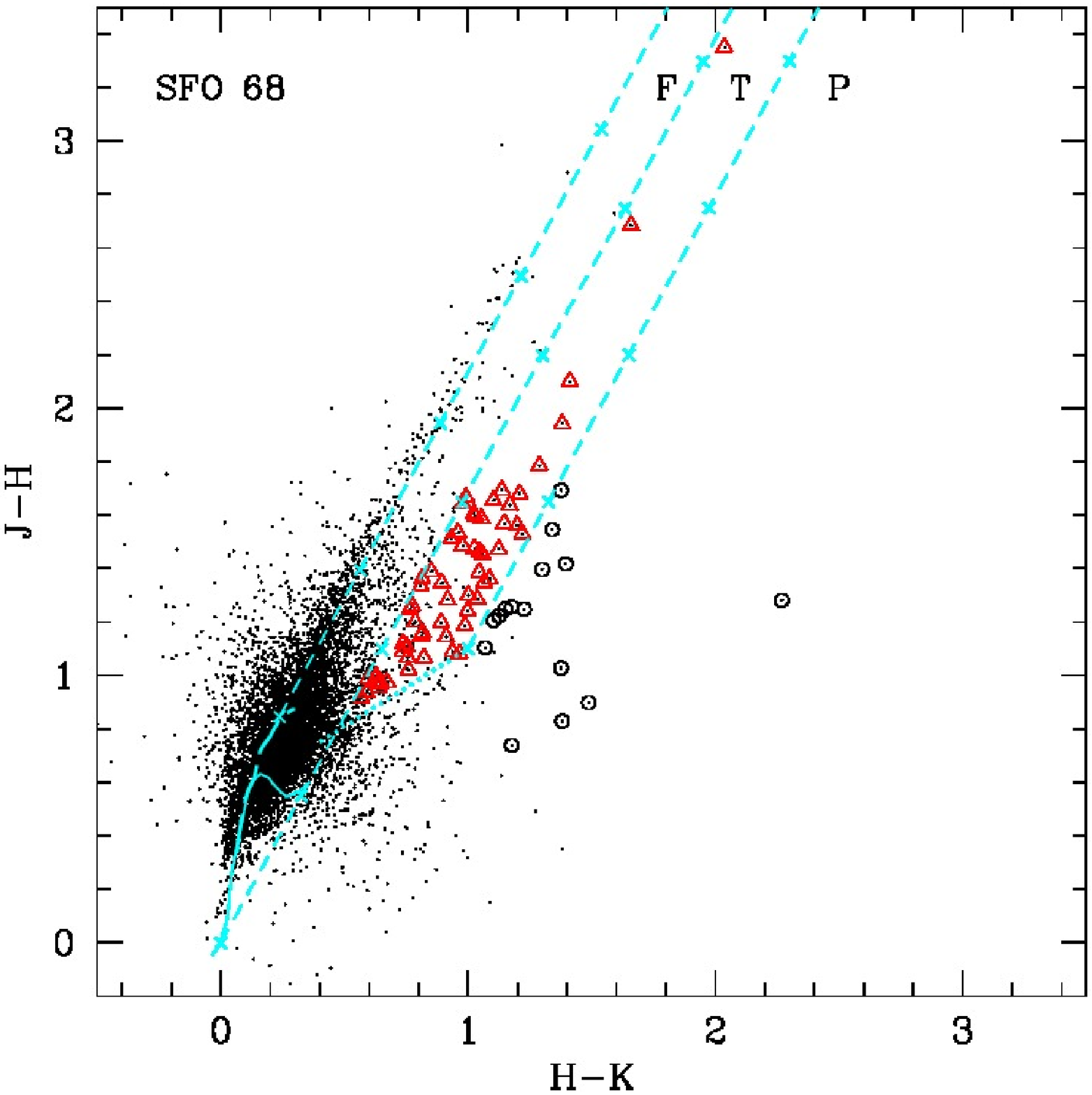}
\centering\includegraphics[height=4cm,width=4cm,angle=0]{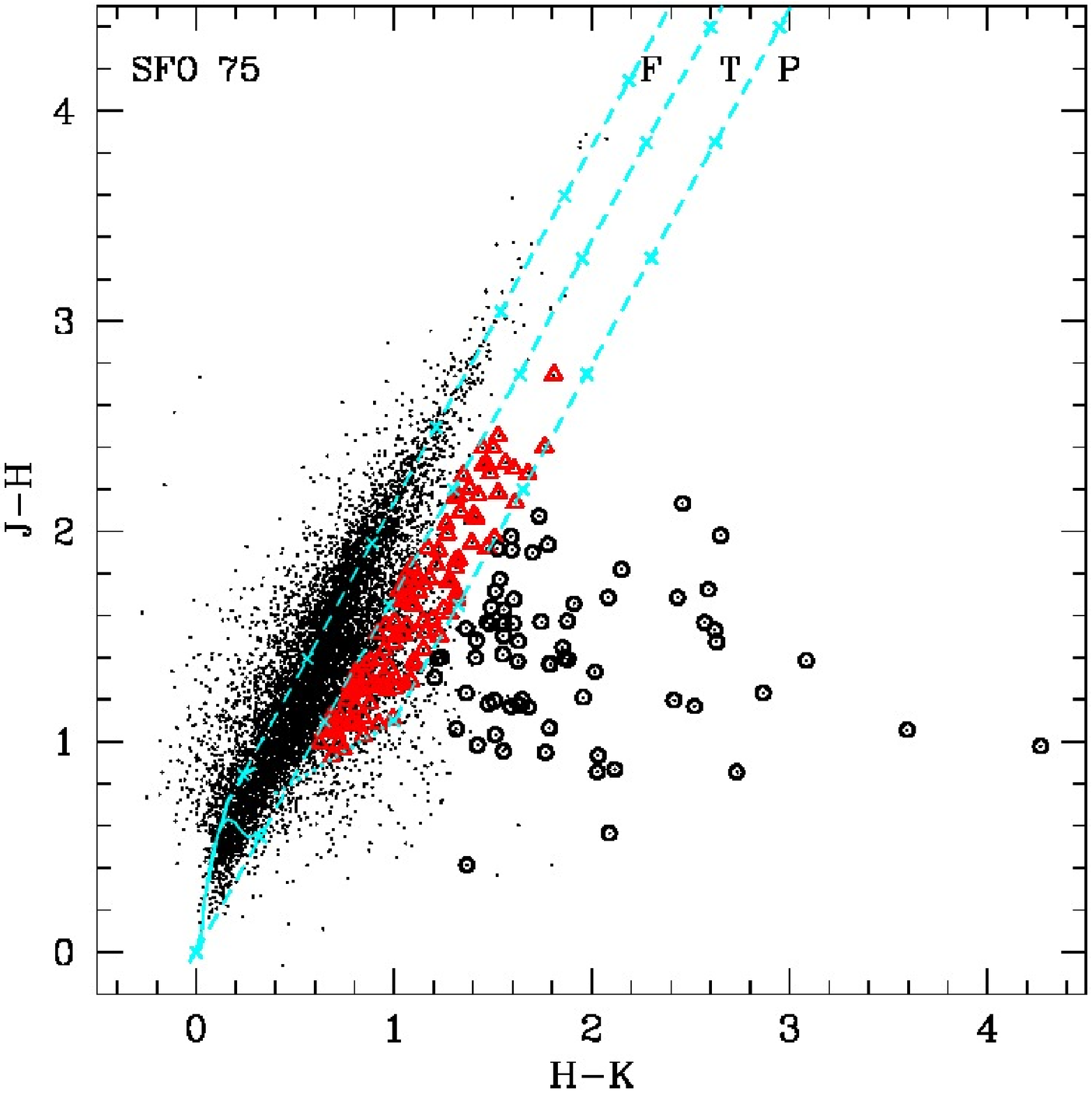}
\centering\includegraphics[height=4cm,width=4cm,angle=0]{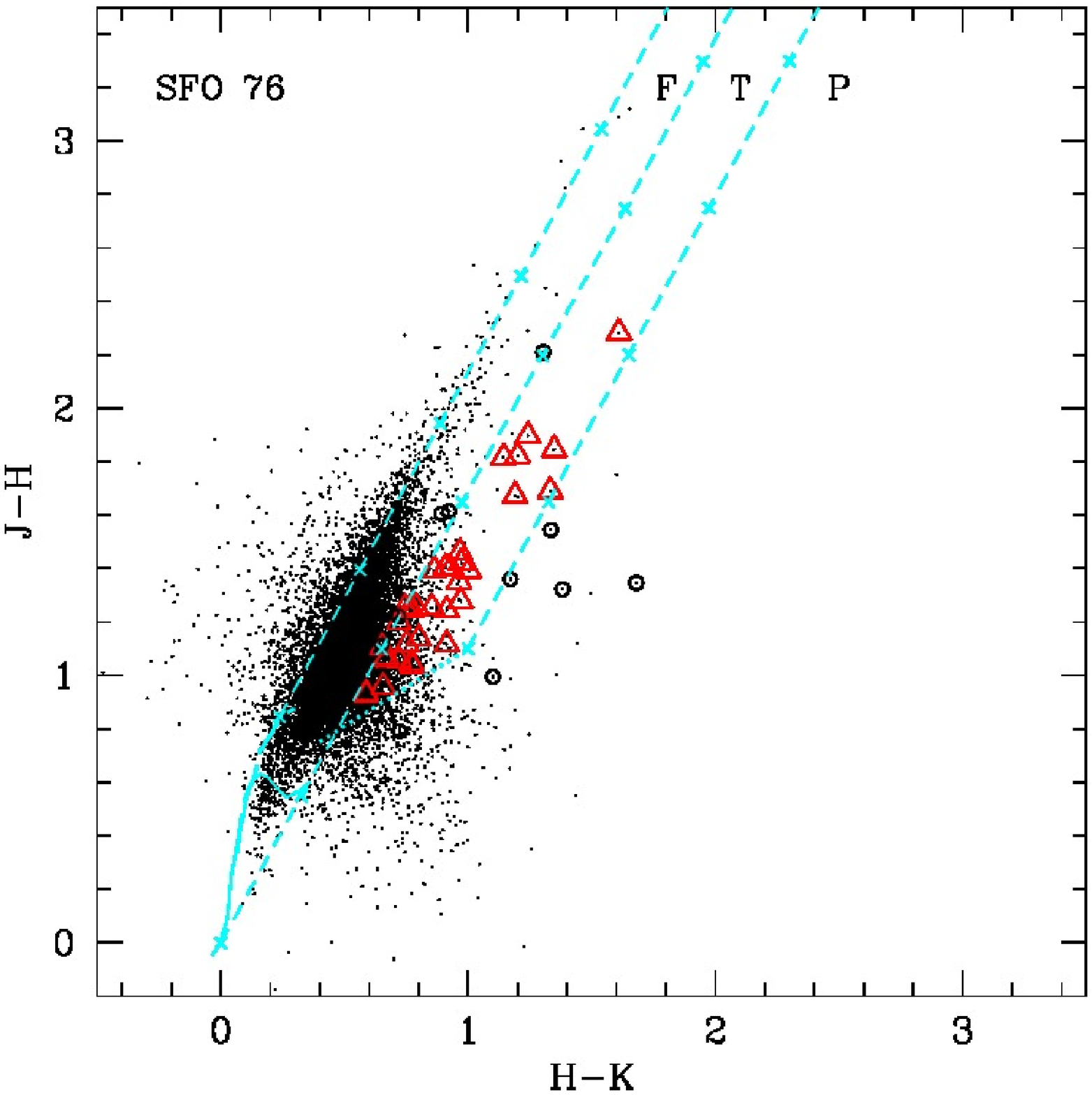}
\centering\includegraphics[height=4cm,width=4cm,angle=0]{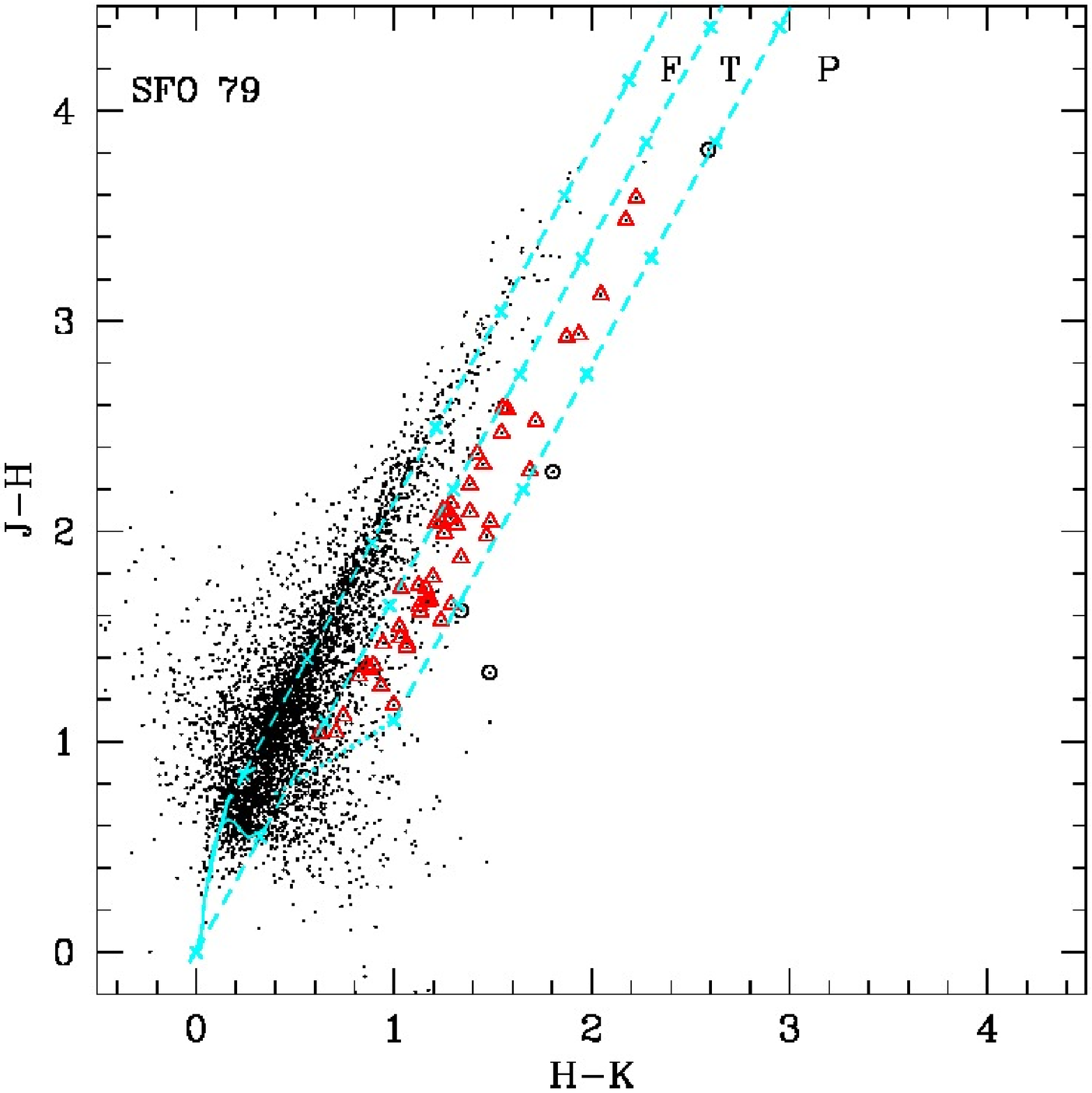}
\caption{\label{Fccd} NIR TCD for the stars in all regions studied.
The continuous and thick dashed curves represent the unreddened MS and giant branch
\citep{1988PASP..100.1134B}, respectively. The dotted line indicates the loci of unreddened CTTSs \citep{1997AJ....114..288M}.
The parallel dashed lines are the reddening vectors drawn from the tip (spectral type M4) of the
giant branch (left reddening line), from the base (spectral type A0) of the MS branch (middle reddening
line) and from the tip of the intrinsic CTTS line (right reddening line).
The crosses on the reddening vectors show an increment of $A_V$ = 5 mag. The sources marked with open triangles and circles
are identified CTTSs and probable IR excess sources, respectively.
 }
\end{figure*}

The sources detected in all three ISPI bands ($JHK$)
have been used to further classify YSOs according to their evolutionary stages by using the conventional NIR TCD.
In Fig.~\ref{Fccd}, we have plotted the NIR TCD for  the stars in all the regions studied.
All the ISPI magnitudes and colors have been converted into the California Institute of
Technology (CIT) system\footnote{http://www.astro.caltech.edu/$\sim$jmc/2mass/v3/transformations/}.
The solid and thick dashed curves represent the un-reddened MS and
giant branch \citep{1988PASP..100.1134B}, respectively. The dotted line indicates the locus
of un-reddened CTTSs \citep{1997AJ....114..288M}. All the curves and lines are also in the
CIT system. The parallel dashed lines are the reddening vectors drawn from the tip
(spectral type M4) of the giant branch (``upper reddening line"), from the base
(spectral type A0) of the MS branch (``middle reddening line") and from the tip of the
intrinsic CTTS line (``lower reddening line"). The extinction ratios
$A_J/A_V = 0.265, A_H/A_V = 0.155$ and $A_K/A_V=0.090$ have been taken from
\citet{1981ApJ...249..481C}. 
We classified the sources according to three regions in this
diagram \citep[cf.][]{2004ApJ...608..797O}.
The `F' sources are located between the upper and middle reddening lines and are considered
to be either field stars (MS stars, giants) or Class III and Class II sources with small
NIR excess. `T' sources are located between the middle and lower reddening lines. These sources
are considered to be mostly CTTSs (or Class II objects) with large NIR excess. There may be an
overlap of Herbig Ae/Be stars in the `T' region \citep{1992ApJ...397..613H}. `P' sources are
those located in the region redward of the lower reddening line and are most likely Class I
objects \citep[protostellar-like objects:][]{2004ApJ...608..797O}. It is worthwhile to mention
also that \citet{2006ApJS..167..256R} have shown that there is a significant overlap between
protostars and CTTSs. 
All sources have been designated as CTTSs if they satisfy the
criteria that they fall in the `T' region of the NIR TCD (Fig.~\ref{Fccd}) and lie redward of
the blue dotted cut-off line of the de-reddened CMD (cf. Fig.~\ref{Fband}). They are shown as open
triangles in Fig.~\ref{Fccd}. We have also plotted the previously identified 
probable IR excess sources (open circles) having $J$ band  detection.
Most of these sources are located in the `P'  region in
Fig.~\ref{Fccd}, which means that they are most likely Class I objects.

\subsection{The YSO sample and its completeness}

We have thus identified and classified 44 to 433 YSOs in our BRCs based on their excess emission in IR.
In total 1347 YSOs have been identified, out of which 790 are Class I protostars. 
We have made a catalog of the YSOs for each BRC identified in the present study. 
In Table \ref{Tyso}, we have given a sample of these YSOs
along with their positions, magnitudes at various bands and the flags for the scheme used to classify them.
A complete table is available in electronic form only. 

\begin{figure*}
\centering\includegraphics[height=6cm,width=6cm,angle=0]{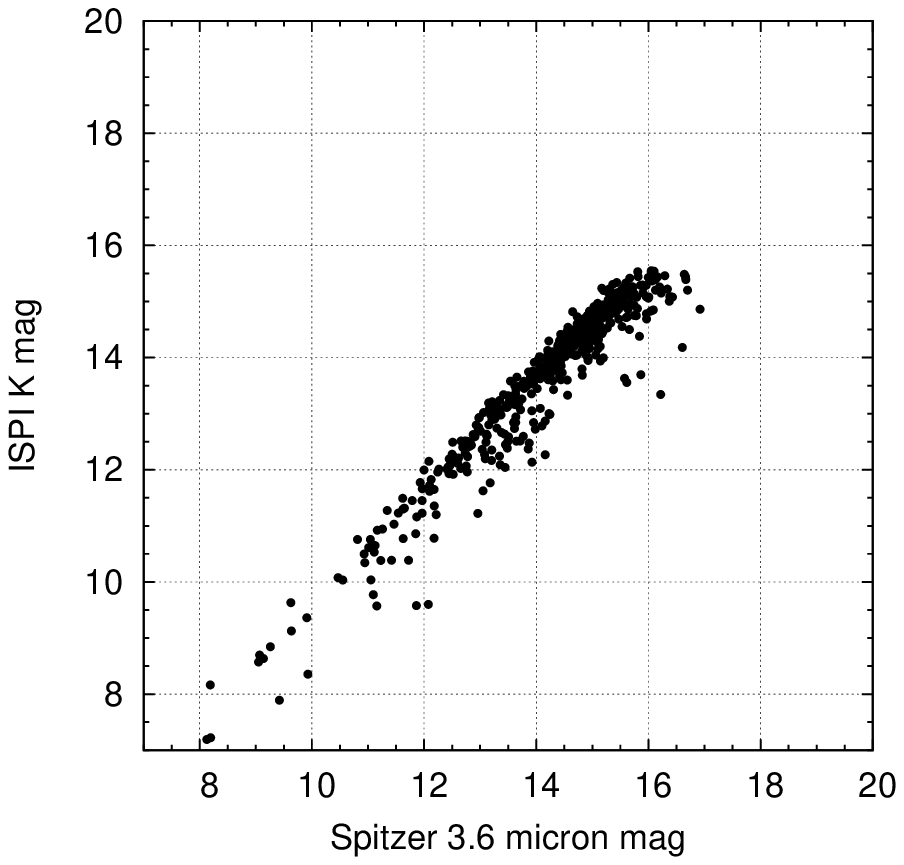}
\centering\includegraphics[height=6cm,width=8cm,angle=0]{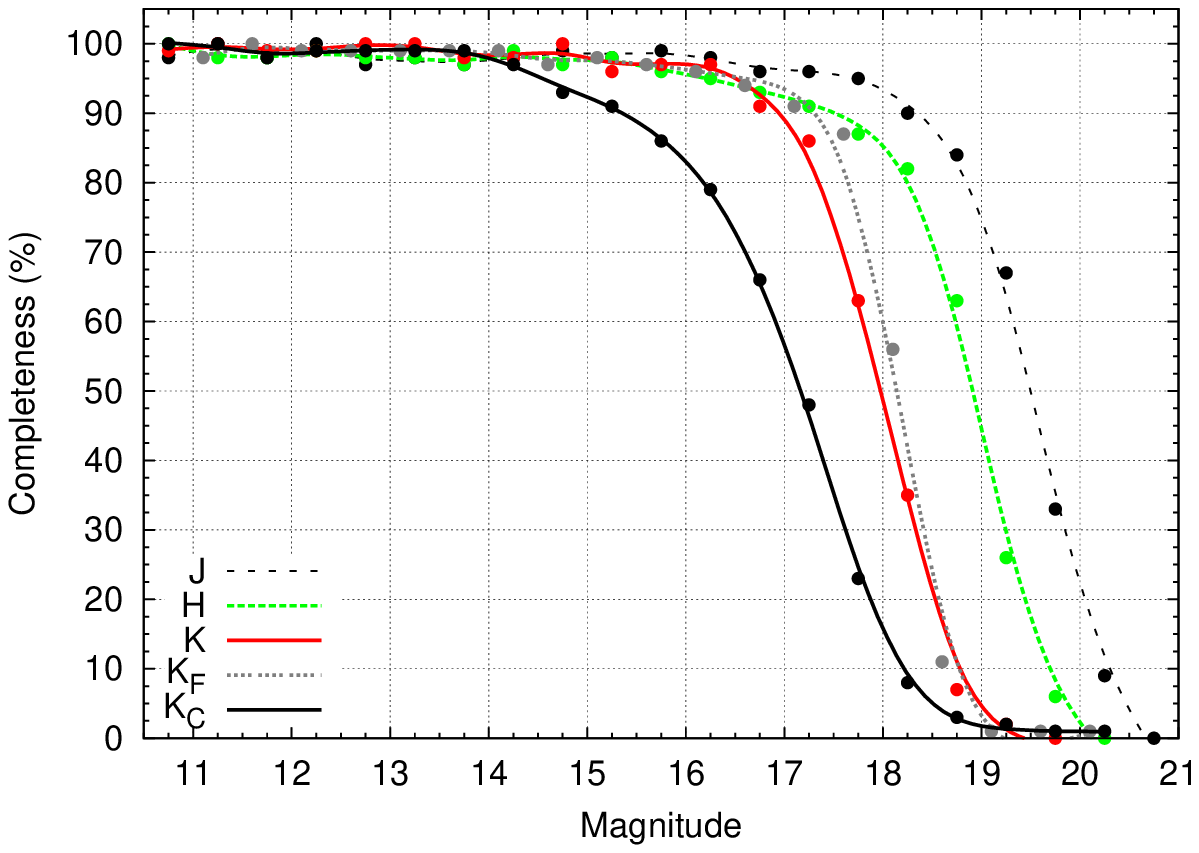}
\centering\includegraphics[height=8cm,width=8cm,angle=0]{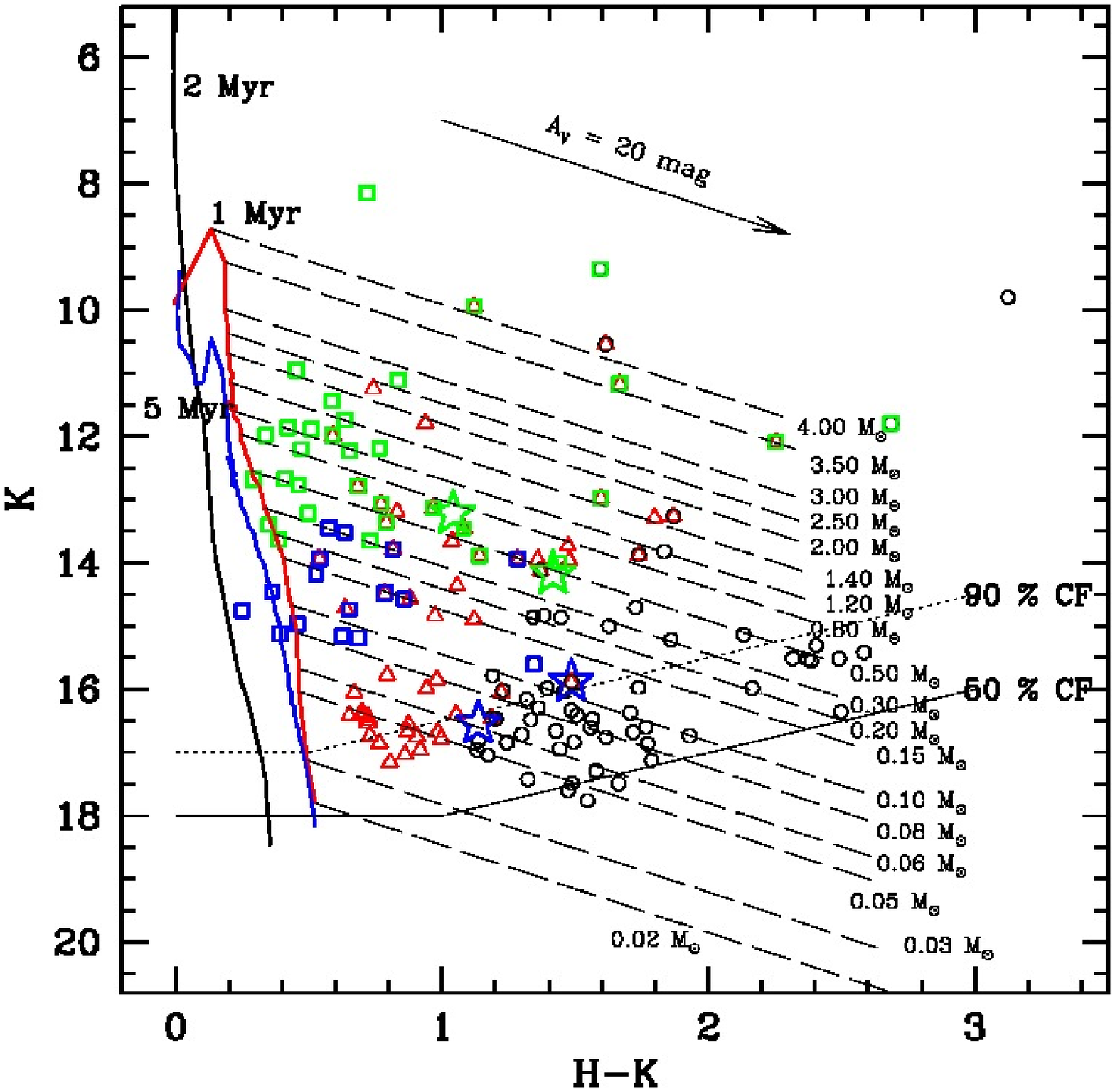}
\caption{\label{Fcft} 
(Top Left): Comparison between the ISPI $K$ band and $Spitzer$ 3.6 $\mu$m photometry. 
(Top Right): Completeness factor of the photometry in different ISPI bands in the SFO 54 region.
 Black dashed, dotted green and solid red curves are the 
smoothened bezier curves for the data points for completeness in the $J$, $H$ and $K$ bands, respectively. 
$K_F$ (dotted grey curve) and $K_C$ (solid black curve) represent similar curves 
in $K$ band for the field and SFO 64 regions, respectively. 
(Lower Panel):  $K$ vs ($H-K$) CMD for the YSOs detected in the SFO 54 region along with the
theoretical MS isochrone of  2 Myr (Z = 0.02, solid black curve) by \citet{2008AA...482..883M}, 
and the PMS isochrones of ages 1 and 5 Myr (solid red and solid blue curves) by
\citet{2000AA...358..593S} (for masses $>$ 1.2 M$_\odot$) and \citet{1998A&A...337..403B} (for masses $<$ 1.2 M$_\odot$), 
all corrected for the distance and the foreground reddening.
The symbols are same as in Figs. \ref{Firac}, \ref{Fband} and \ref{Fccd}.
Slanting parallel dashed lines show the reddening vectors for PMS stars of different masses.
The dotted and solid broken lines represent the 90\% and 50\% completeness limits for the data. See \S 3.2 for a detail.
}
\end{figure*}

To understand the level of star formation at present or in the past in our BRCs,
it is important to know the completeness limits in terms of masses for the sample of YSOs 
identified in each region.
The photometric data may be incomplete due to various reasons, e.g., background nebulosity, 
crowding of stars, the detection limit etc.
For the current sample, the YSOs are identified either, if they are detected in at-least two $Spitzer$ IRAC channels
(3.6 $\mu$m and 4.5  $\mu$m, Steps 1 and 2), or if they are detected in two ISPI $H$ and $K$ bands (Steps 3 and 4).
In  Fig. \ref{Fcft} (top left panel), we have plotted IRAC 3.6 $\mu$m mag vs. ISPI $K$ mag for the sources detected in the SFO 54 region.
From the figure, it is clear that the faintest source in 3.6 $\mu$m band ($\sim$16.5 mag) corresponds to those of $K\sim$15.5 mag. The $K$ band has yielded
much fainter detections ($\sim$18.5 mag), therefore the completeness  of our photometric data taken from ISPI
will dictate the  completeness of the current sample of YSOs.

To determine the completeness factor (CF) for ISPI data, we used the ADDSTAR routine of DAOPHOT II.
This method has been used by various authors 
\citep[see][and references therein]{2007MNRAS.380.1141S,2008AJ....135.1934S}.
Briefly, the method consists of randomly adding artificial stars of known magnitudes and
positions into the original frame. The frames are reduced by using the same procedure used for the original frame.
The ratio of the number of stars recovered to those added in each magnitude interval gives the CF as a function of magnitude.
The luminosity distribution of artificial stars is chosen in
such a way that more stars are inserted into the fainter magnitude bins.
In all about 15\% of the total stars are added so that the crowding characteristics of the original frame do not change
significantly \citep[see][]{1991A&A...250..324S}.
As an example, the CF for the SFO 54 region (distance = 0.95 kpc) is given in the top right panel of Fig. \ref{Fcft} for different ISPI bands.
In the same panel, we have also plotted the $K$ band CFs for a nearby field region (shown by grey dotted curve `$K_F$') as well as for the SFO 64 region (distance = 2.7 kpc, 
shown by black solid curve `$K_C$'), representing a field free from nebulosity and that of higher crowding, respectively.
Higher crowding, in the case of SFO 64 can be attributed to its intrinsic property or its farther distance.
The field region has the highest completeness because of the lack of nebular backgrounds. 
Out of SFO 64 and SFO 54, both having a similar level of nebulosity, SFO 64 shows a
lower CF  because of its larger distance and/or higher crowding.

In the lower panel of Fig. \ref{Fcft}, we have plotted the ($H-K$) vs $K$ CMD for the SFO 54 region along with the 
theoretical MS isochrone of 2 Myr (Z = 0.02) by \citet{2008AA...482..883M}  and the pre-main sequence (PMS) isochrones of age 1 and 5 Myr by
 (\citet{2000AA...358..593S} (for mass $>$ 1.2 M$_\odot$) and \citet{1998A&A...337..403B} (for mass $<$ 1.2 M$_\odot$), 
all corrected for the distance \citep[cf.][]{2009A&A...497..789U} and 
the foreground expected reddening \citep[$A_{K_{foreground}}=$ 0.15$\times$D, where D is the distance in kpc,][]{2005ApJ...619..931I}. 
The slanting parallel dashed lines represent the reddening vectors for PMS stars of different masses.
The dotted and solid broken lines represent the 90\% and 50\% completeness limits for the data as inferred 
from the CF calculated earlier. We have taken into account the effect of color ($H-K$) on the completeness limit of this CMD.
We can easily see that the 90\% and 50\% completeness limits correspond to $\sim$0.04 M$_\odot$ 
and  $\sim$0.03 M$_\odot$ YSOs, respectively.
However, the photometric errors at this level should be higher ($\pm$0.1 mag to $\pm$0.2 mag), therefore, 
the corresponding errors in the derived mass should be of the order of $\sim$ $\pm$0.02 M$_\odot$ to $\pm$0.03 M$_\odot$.
The  90\% and 50\% completeness limits
for the YSOs embedded deeply in the molecular cloud (corresponding to the peak $A_K$ value taken from Table \ref {Tp2}) 
are considered to be $\sim$0.15$\pm$0.03 M$_\odot$ and  $\sim$0.08$\pm0.04$ M$_\odot$, respectively.
In the last column of Table \ref {Tlog}, we have given the  90\% completeness limit for YSOs  corrected for the 
foreground and peak (including the foreground) reddening for all the studied regions.

The PMS isochrones were used to derive the CF for the sample of YSOs in terms of mass. 
The errors in the mass of YSOs  quoted above comes mainly from the large photometric errors at faint levels. 
In the $H-K$ vs. $K$ diagram (Fig. \ref{Fcft} lower panel), the PMS isochrones are almost vertical; 
therefore any change of age in the lower mass regime would shift the isochrones in the vertical direction (magnitude axis), 
and typically it would cause to 0.02 M$_\odot$ difference for the age difference of 4 Myr (cf. Fig. \ref{Fcft}. Lower Panel). 
From Fig. \ref{Fccd} and Fig. \ref{Fcft} and its siblings for the rest of our BRCs we 
find that all of them are regions of recent star formation harboring YSOs of various 
evolutionary stages from Class 1 (protostars) to at-least Class II (CTTSs), 
leading to the lower error values in  mass determination of YSOs due to their comparatively smaller age spreads.

\subsection{Distribution of molecular cloud around the BRCs}

\begin{figure*}
\centering
\includegraphics[height=7cm,width=8cm]{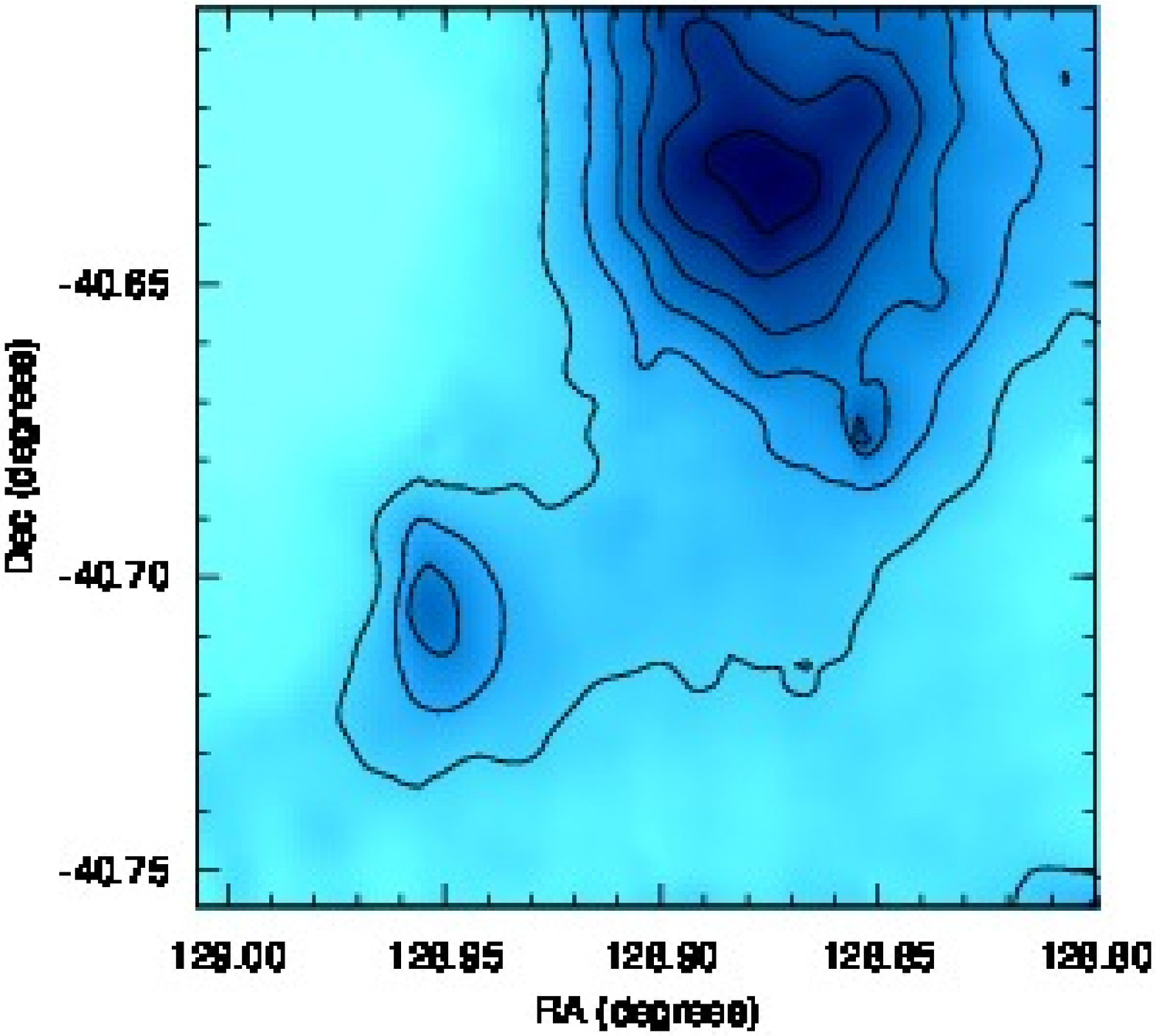}
\includegraphics[height=7cm,width=8cm]{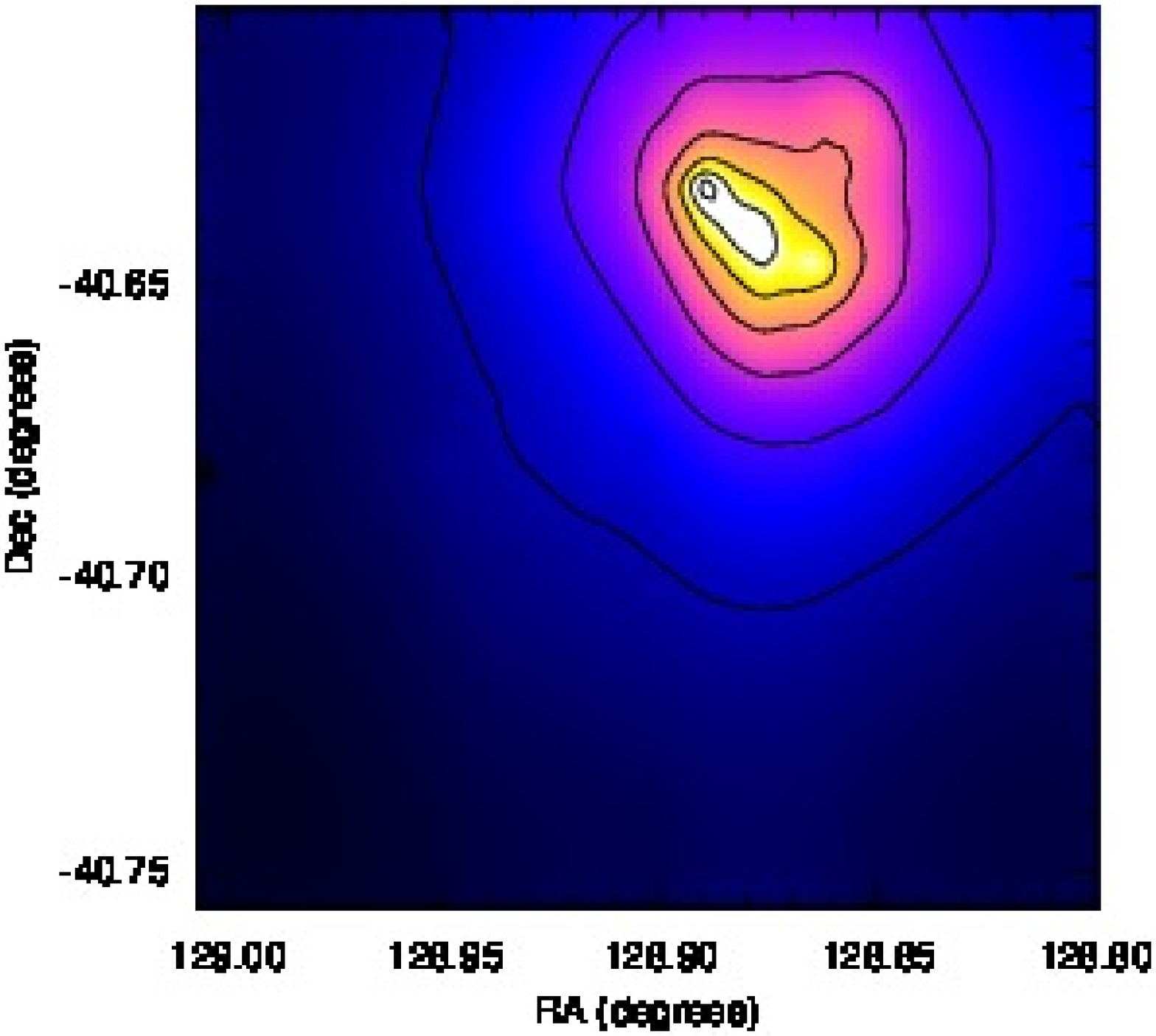}
\caption{\label{Fiso} (Left Panel): Extinction map smoothed to a resolution of 5 arcsec for the SFO 54 region.
The contours are drawn with a step size of $A_K$ = 0.1 mag starting from the 
lowest contour approximately 
equal to the mean $A_K$ value for the selected active region (cf. Table \ref {Tp2}).
(Right Panel): Surface isodensity contours of YSOs detected in the same region with the same resolution.
The contours are shown with a step size of 2 stars/arcmin$^2$ with the lowest contour approximately 
equal to the mean number density in arcmin square for the selected active region (cf. Table \ref {Tp2}).
Both the maps have FOV of $\sim10\times10$ arcmin$^2$.
}
\end{figure*}

To study the relationship between the distribution of YSOs and the molecular clumps in the regions, 
we have derived $A_K$ extinction maps using the $(H - K)$ colors of the MS stars \citep[cf.][]{2011ApJ...739...84G}.
This map has also been used to quantify the amount of extinction within each sub-regions 
of the studied BRCs and to characterize the structures of the molecular clouds  within each sub-regions
\citep{2013MNRAS.432.3445J, 2011ApJ...739...84G, 2009ApJS..184...18G}.
The sources showing excess emission in IR can lead to overestimation of extinction 
values in the derived maps. The excess emission can be of the order of $\sim$0.25 mag in $H-K$ 
color \citep[][]{2012PASJ...64..107S, 2008MNRAS.383.1241P,2011MNRAS.415.1202C} 
which will correspond to the overestimation of $A_K$ value by $\sim$0.4 mag. 
This will increase cloud mass estimate which we are going to derive later 
(\S 4.2.7). Therefore, to improve the quality of the extinction maps, 
the candidate YSOs and probable contaminating sources (cf. \S 3.1) must be excluded from the stars used.
In order to determine the mean value of $A_K$ we used the nearest neighbor (NN) 
method as described in detail in
\citet{2005ApJ...632..397G} and \citet{2009ApJS..184...18G} 
to determine the mean value of $A_K$.  Briefly, at each position in a uniform grid of 5 arcsec, 
we calculated the  mean value of $(H - K)$ colors of five nearest stars.
The sources deviating above
3$\sigma$ were excluded to calculate the final mean color of each point. 
To convert $(H -K)$ color excesses into $A_K$ we used the relation  $A_K$ = 1.82 $\times$ ($(H - K)_{obs} - (H - K)_{int}$). This is derived from the reddening law by \citet{2007ApJ...663.1069F}.
We have assumed $(H - K)_{int}$ = 0.2 mag as an average intrinsic color for all stars in young
clusters \citep[see.][]{2008ApJ...675..491A, 2009ApJS..184...18G}. 
Of-course, the intrinsic color depends on the spectral type, i.e., mass of the MS stars,
so, this assumption introduces errors in reddening estimation. 
But this effect will be small as can be seen in Fig. \ref{Fcft} (bottom panel). 
The MS for stars for all mass/spectral range is almost vertical. The standard deviation 
in their $H-K$ color is of the order of 0.1 mag, which corresponds to the error of $\sim$0.1 $A_K$ 
in the extinction map. 
To eliminate the foreground contribution in the extinction measurement, we
used only those stars with $A_K >$ 0.15$\times$D, where D is the distance in kpc \citep{2005ApJ...619..931I}
to generate the extinction map.  The extinction maps smoothened to a resolution of 5 arcsec 
and reaching down upto $A_K\sim$2.8 mag were generated for all the regions studied.
However, the derived $A_K$ values are to be considered as a lower limits, 
because the sources with higher extinction may not be detected in our study.
As an example, we have shown in Fig.~\ref{Fiso} (Left Panel) the derived extinction map for the SFO 54 region.

\subsection{Spatial distribution of YSOs in the region}

By analyzing the stellar density distribution
morphology in relation to the molecular cloud structure, observational
analyses can address the link between star
formation, gas expulsion, and the dynamics of the clusters as well as
how these processes guide the evolution of young clusters \citep{2005ApJ...632..397G}.
To study the density distribution of YSOs in the region we have  generated their surface density maps
using the NN method as described by  \citet{2005ApJ...632..397G}.
We have taken the radial distance necessary to encompass the 5th nearest YSOs
and computed the local surface density (cf. Fig. \ref{Fiso}, Right Panel) in a grid size of 5 arcsec.
To facilitate comparisons between the stellar density and the gas column density, we
adopted the grids identical to the grid size of the extinction map for each region.

\begin{figure*}
\centering\includegraphics[height=4.2cm,width=4.0cm]{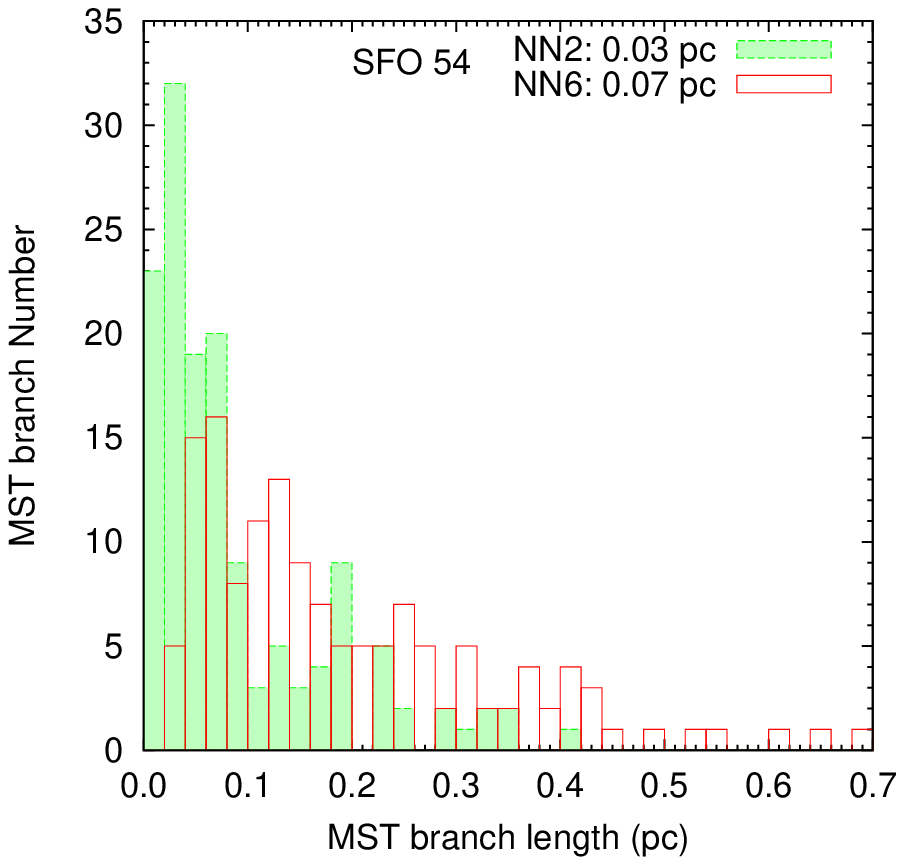}
\centering\includegraphics[height=4.2cm,width=4.0cm]{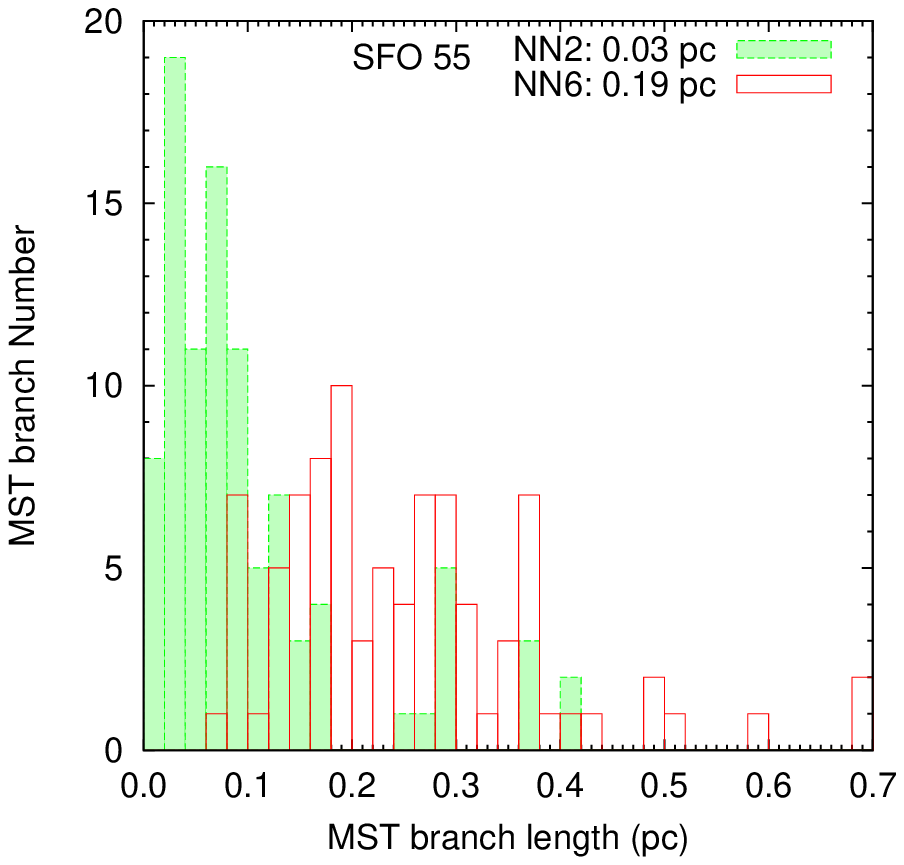}
\centering\includegraphics[height=4.2cm,width=4.0cm]{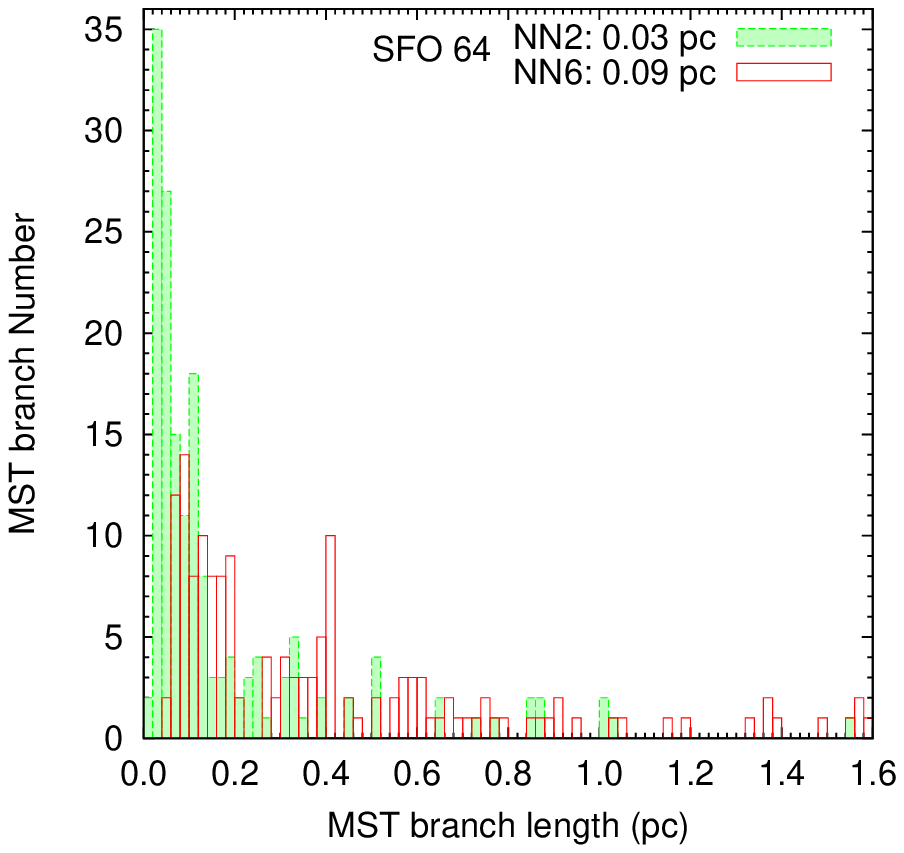}
\centering\includegraphics[height=4.2cm,width=4.0cm]{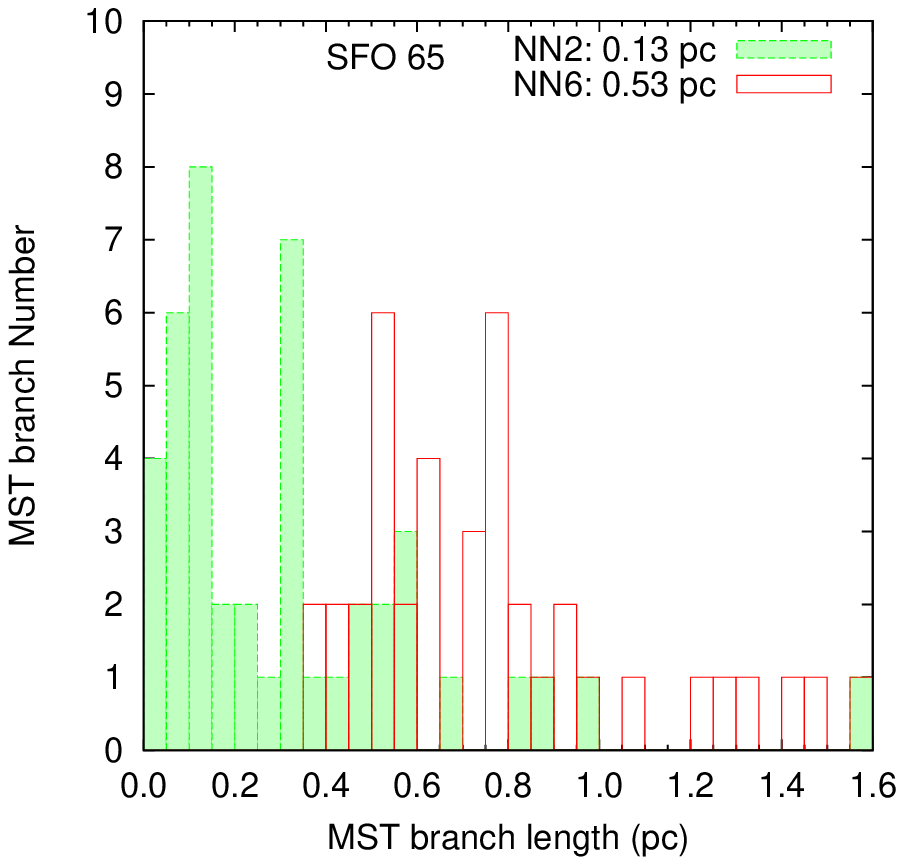}
\centering\includegraphics[height=4.2cm,width=4.0cm]{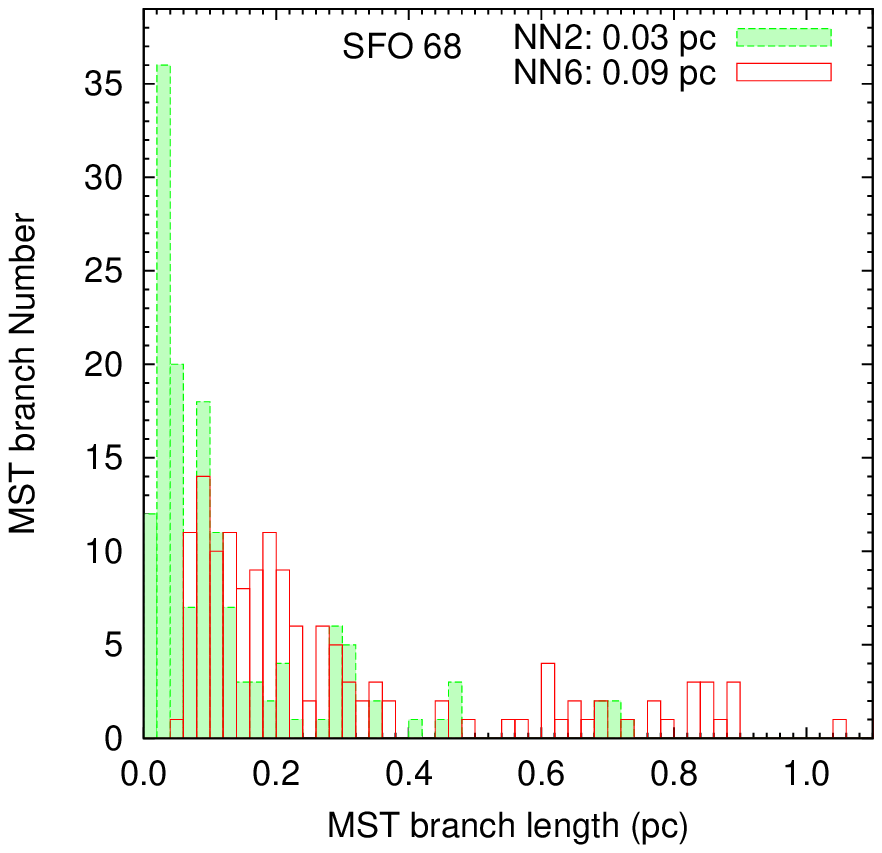}
\centering\includegraphics[height=4.2cm,width=4.0cm]{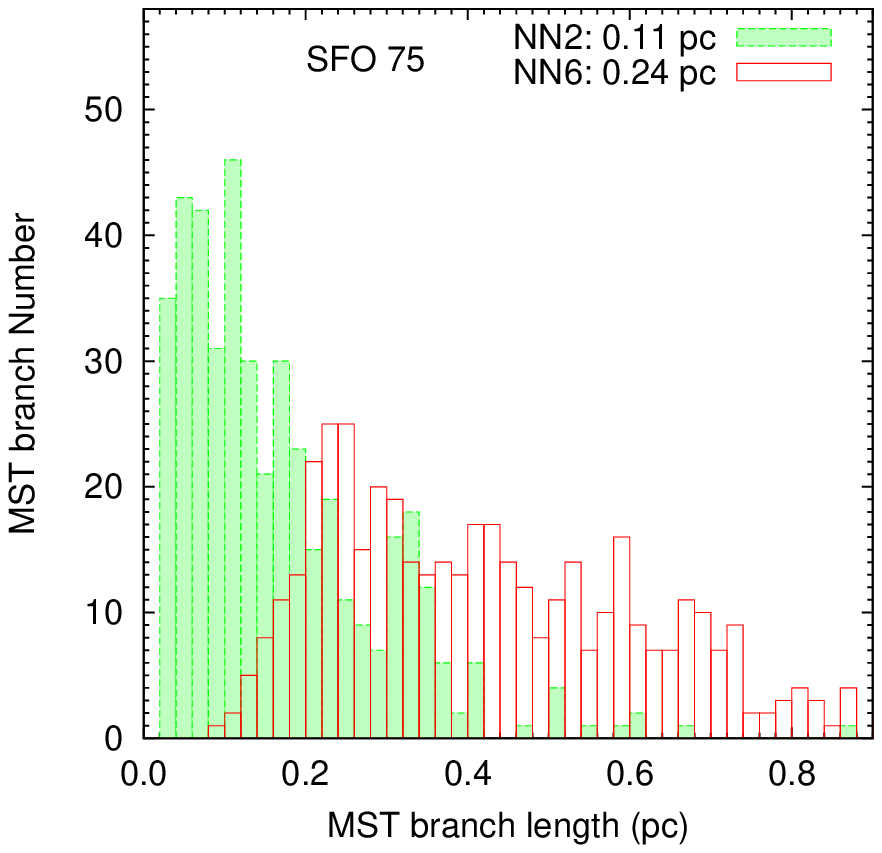}
\centering\includegraphics[height=4.2cm,width=4.0cm]{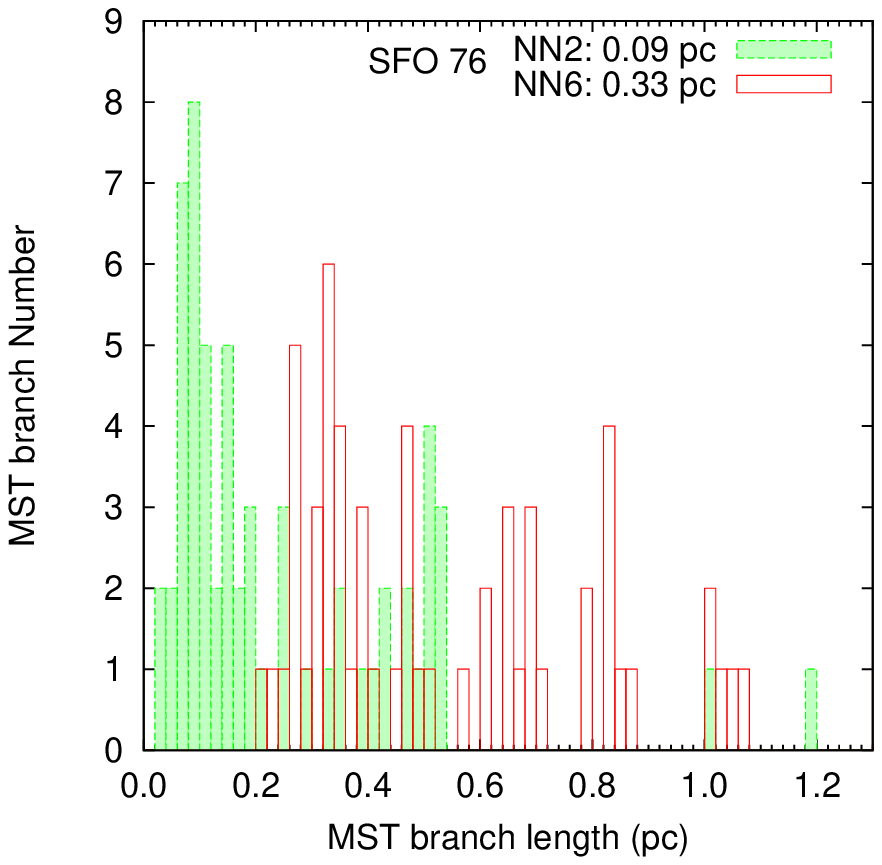}
\centering\includegraphics[height=4.2cm,width=4.0cm]{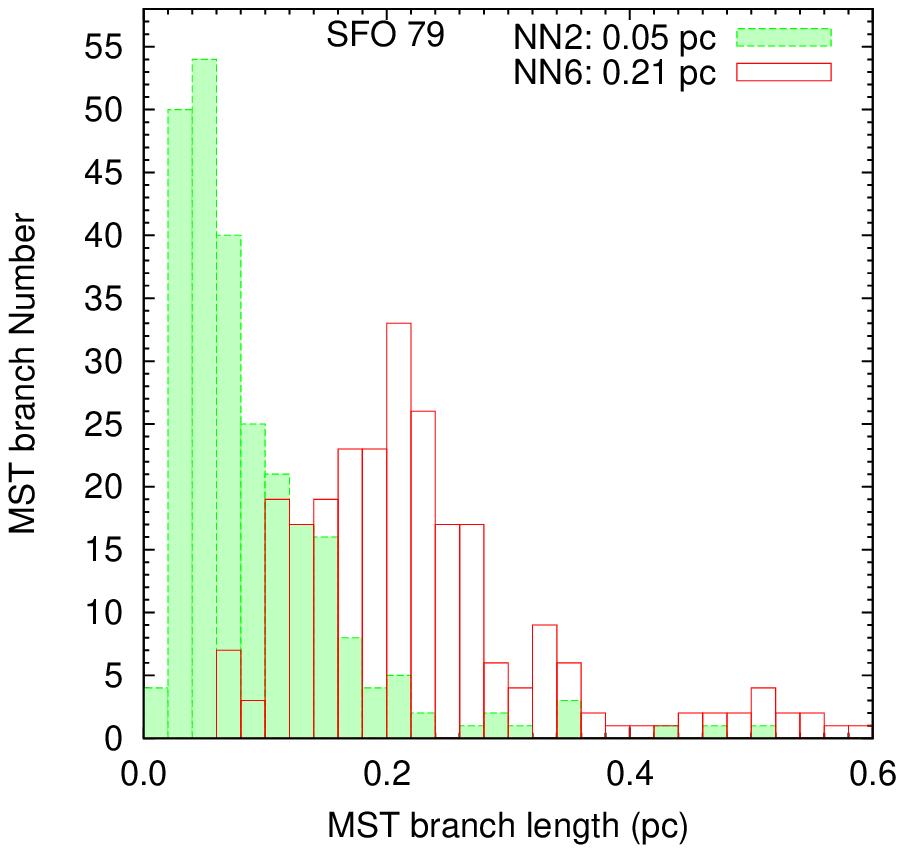}
\caption{\label{Fnn2} Histograms of the  nearest neighbor (NN) lengths for the YSOs in 
the studied regions with a bin size of 0.02 pc.
The red and green histograms represent NN6 ( the projected distance from each YSO to its fifth nearest YSO neighbor)
and NN2 (the projected distance from each YSO to its nearest YSO neighbor), respectively (see the text for detail).
 }
\end{figure*}

The spatial distribution of YSOs in
a region can also be analyzed by deriving the typical spacing between them and comparing this
spacing to the Jeans fragmentation scale for a self-gravitating medium with thermal pressure
\citep{1993AJ....105.1927G}. 
We measure the projected distance from each YSO to its nearest YSO neighbor (NN2), as well as
to its fifth nearest YSO neighbor (NN6; the radial distance from
each source such that a circular area of that radius centered on
the source contains the nearest five neighbors, i.e. a total of six YSOs),
and plotted their histograms with a bin size of 0.02 pc in Fig. \ref{Fnn2} for all the studied BRCs.
All the histograms show a major peak in their distribution along with a couple of smaller peaks indicating groupings/sub-groupings in the regions.

\subsection{Extraction of YSO's cores embedded in the molecular cloud}

\begin{figure*}
\hbox{
\hspace{1.2cm}
\includegraphics[height=7.2cm,width=7.2cm]{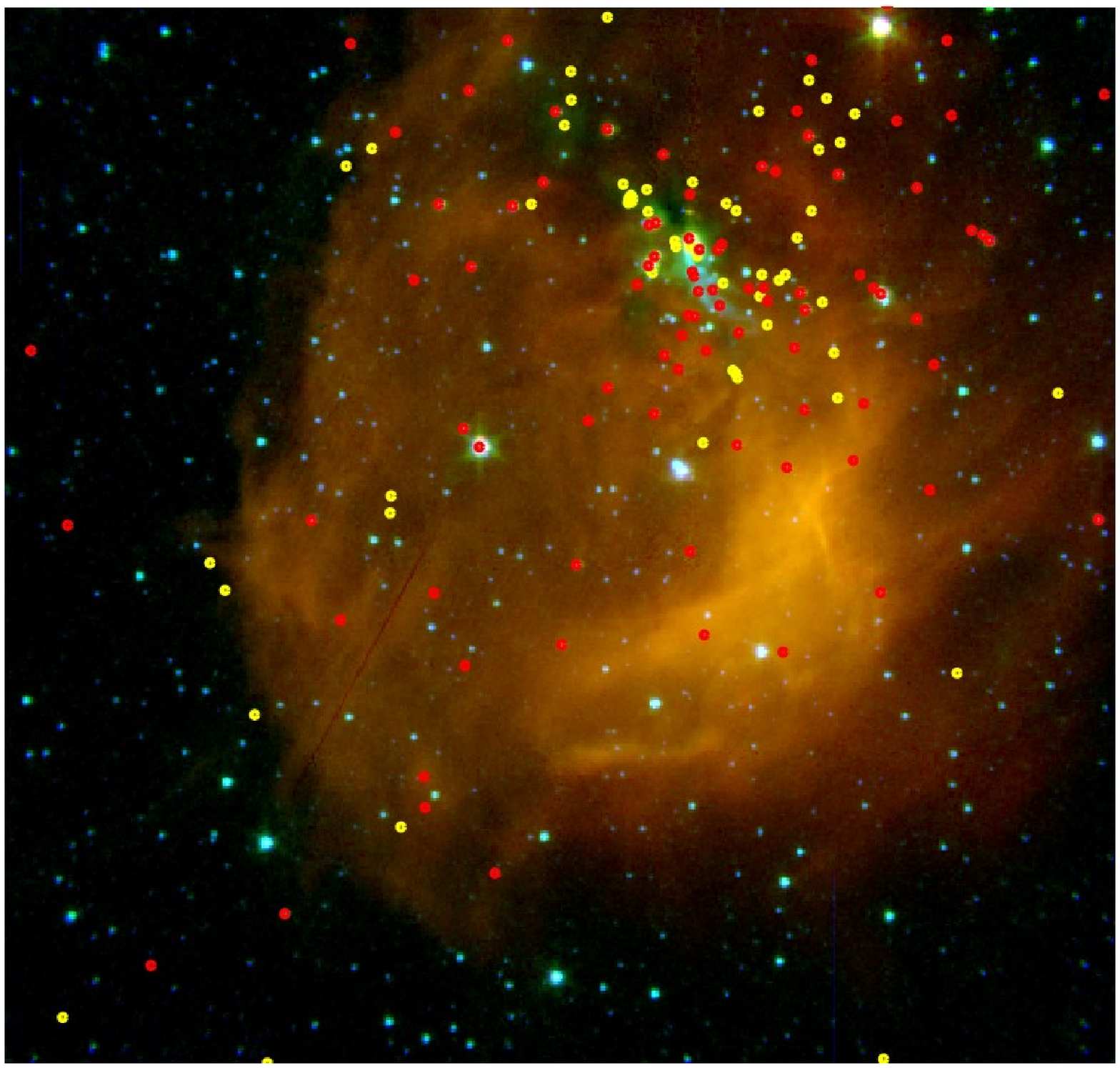}
\hspace{1.7cm}
\includegraphics[height=7.2cm,width=7.2cm]{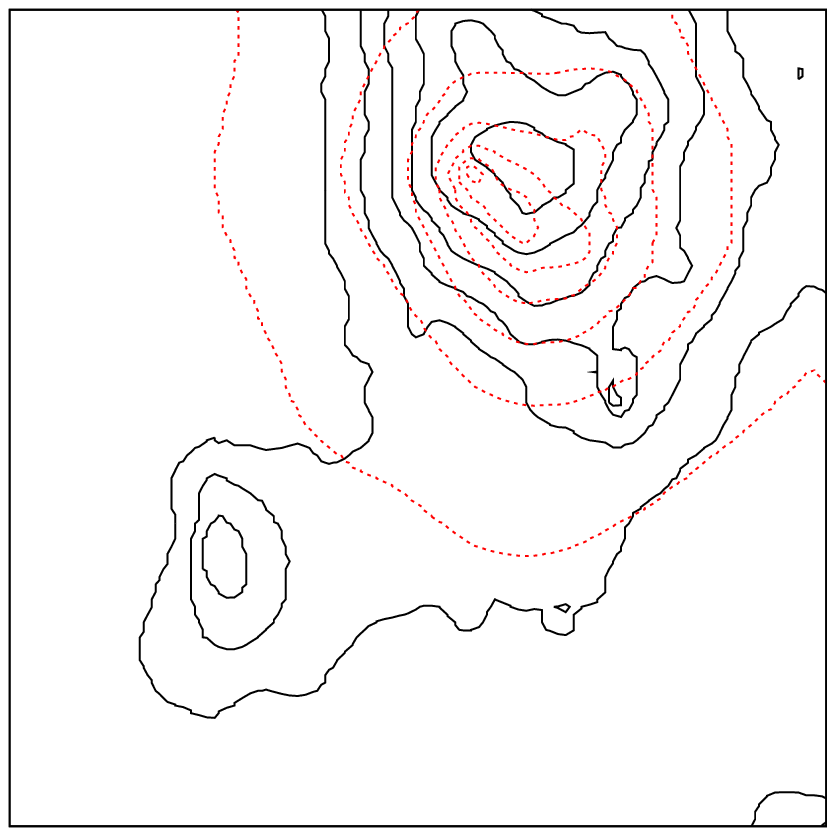}
}
\hbox{
\includegraphics[height=8.0cm,width=9.0cm]{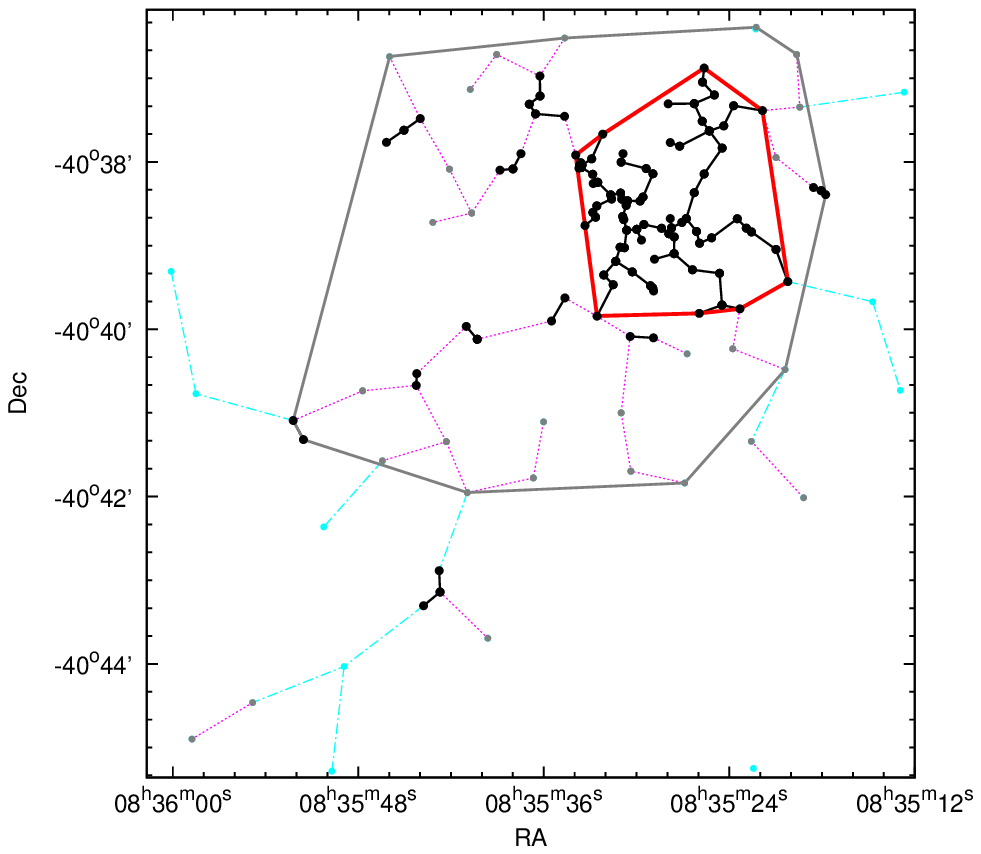}
\includegraphics[height=8.0cm,width=9.0cm]{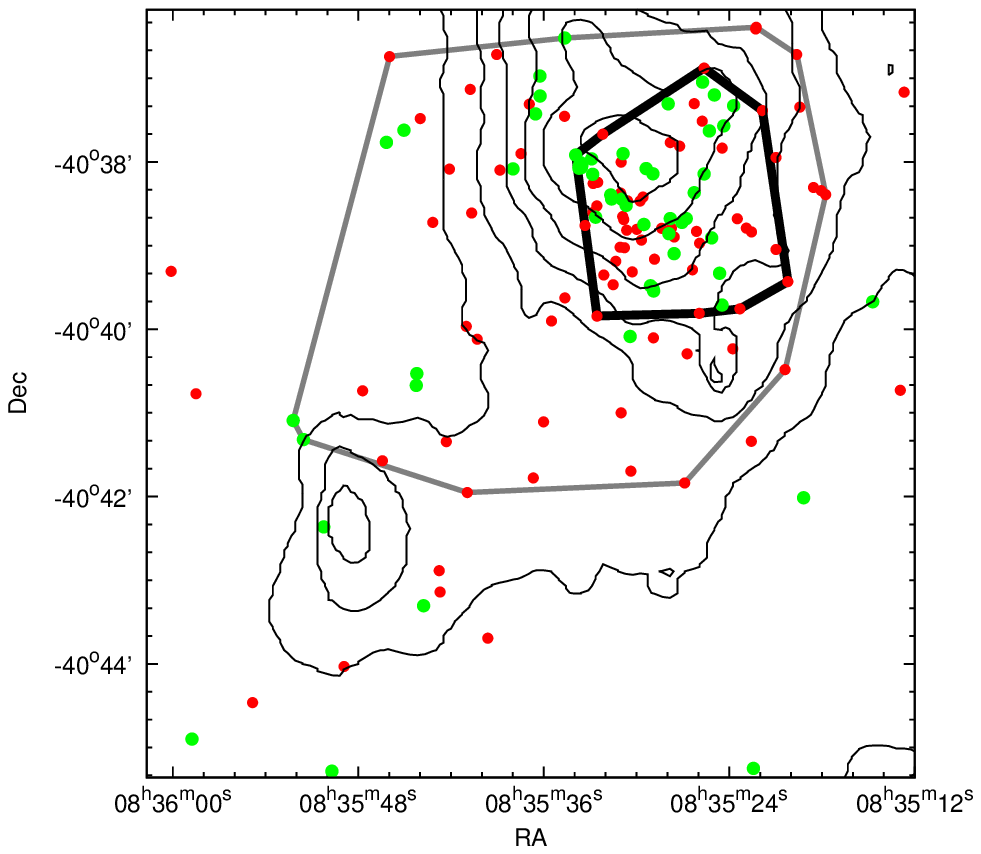}
}
\caption{\label{Fall54} \small (Top Left) Color-composite image of the SFO 54 region obtained 
by combining the $K$ (blue); 3.6 $\mu$m (green)
and 8.0 $\mu$m (red) images for an area of $\sim10\times 10$ arcmin$^2$. The identified YSOs 
(Class I: yellow dots, Class II: red dots) are also plotted.
(Top Right): Isodensity contours for the YSO distribution (red dotted contours) and the 
reddening map (black solid contours) for the same region. 
 The contour levels are the same as in Fig. \ref{Fiso}.
(Bottom Left): Minimal spanning tree (MST) for the identified YSOs in the same 
region along with the convex hull.
The black dots connected with solid lines and grey dots connected with purple dotted lines are the branches smaller than the critical
length for the cores and the active region, respectively. 
The identified core and the  active region are encircled with red and grey
solid lines, respectively.
(Bottom Right): Spatial correlation between the molecular material
inferred from the extinction map (thin black contours) and the 
distribution of YSOs along with the identified cores and active regions 
(thick black and thin grey lines, respectively).  }
\end{figure*}

\begin{figure*}
\hbox{
\hspace{1.2cm}
\includegraphics[height=7.2cm,width=7.2cm]{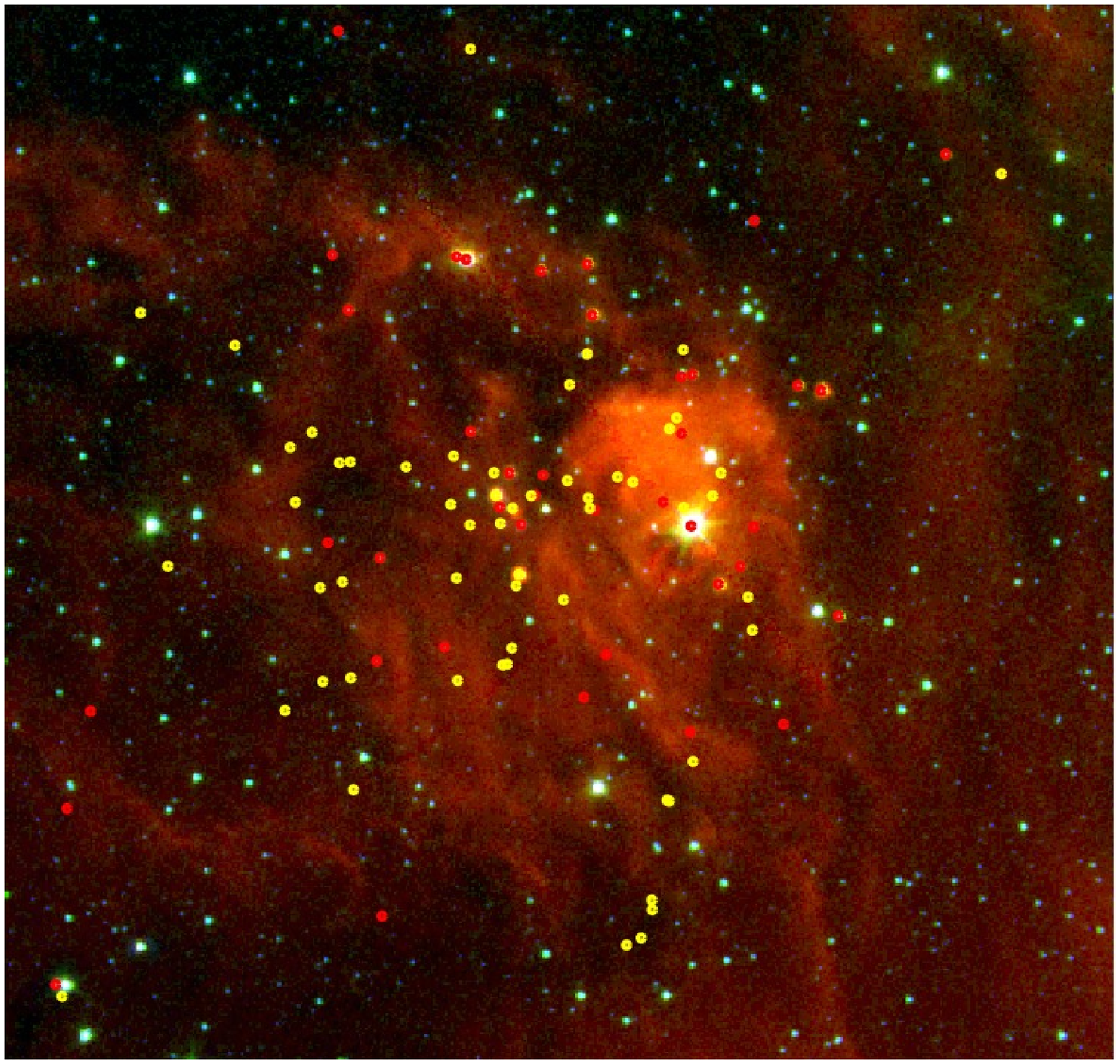}
\hspace{1.7cm}
\includegraphics[height=7.2cm,width=7.2cm]{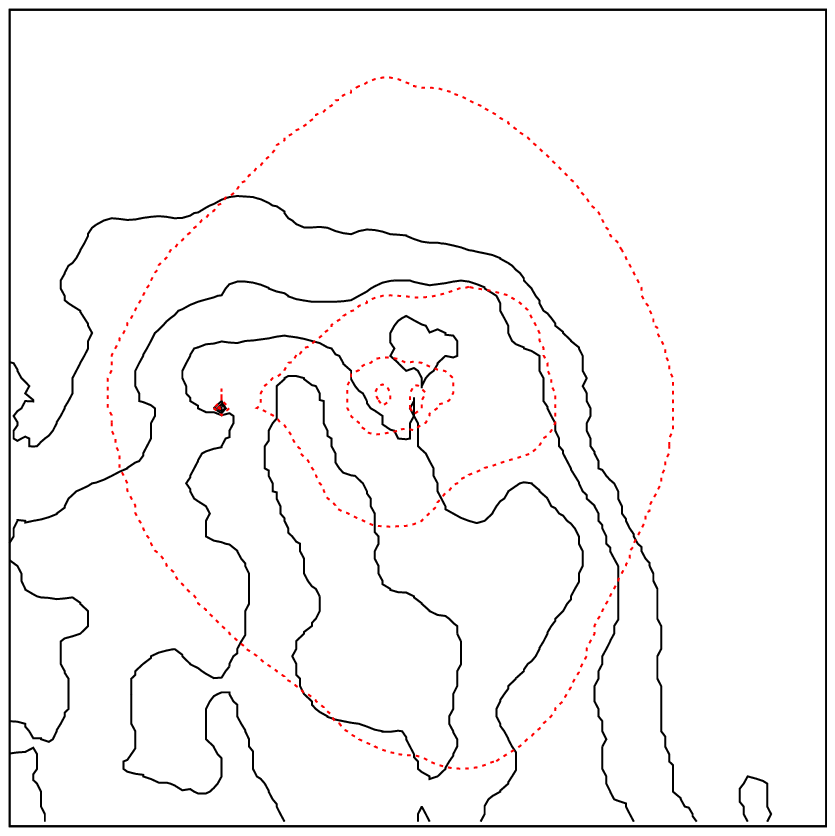}
}
\hbox{
\includegraphics[height=8.0cm,width=9.0cm]{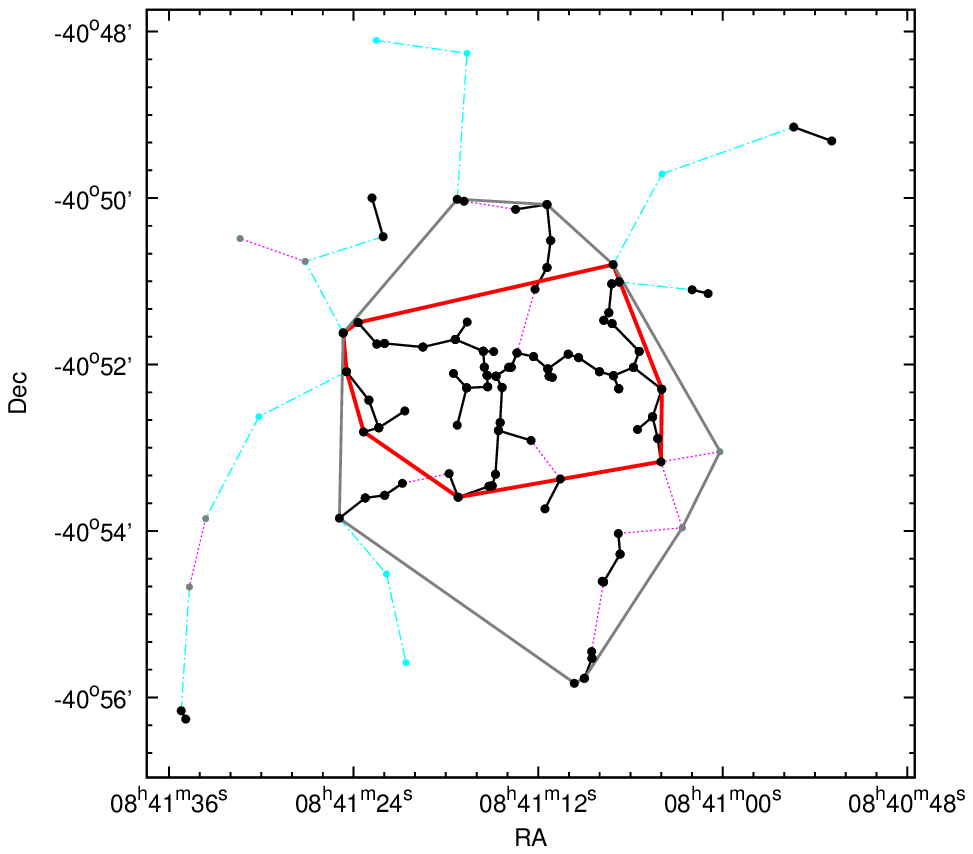}
\includegraphics[height=8.0cm,width=9.0cm]{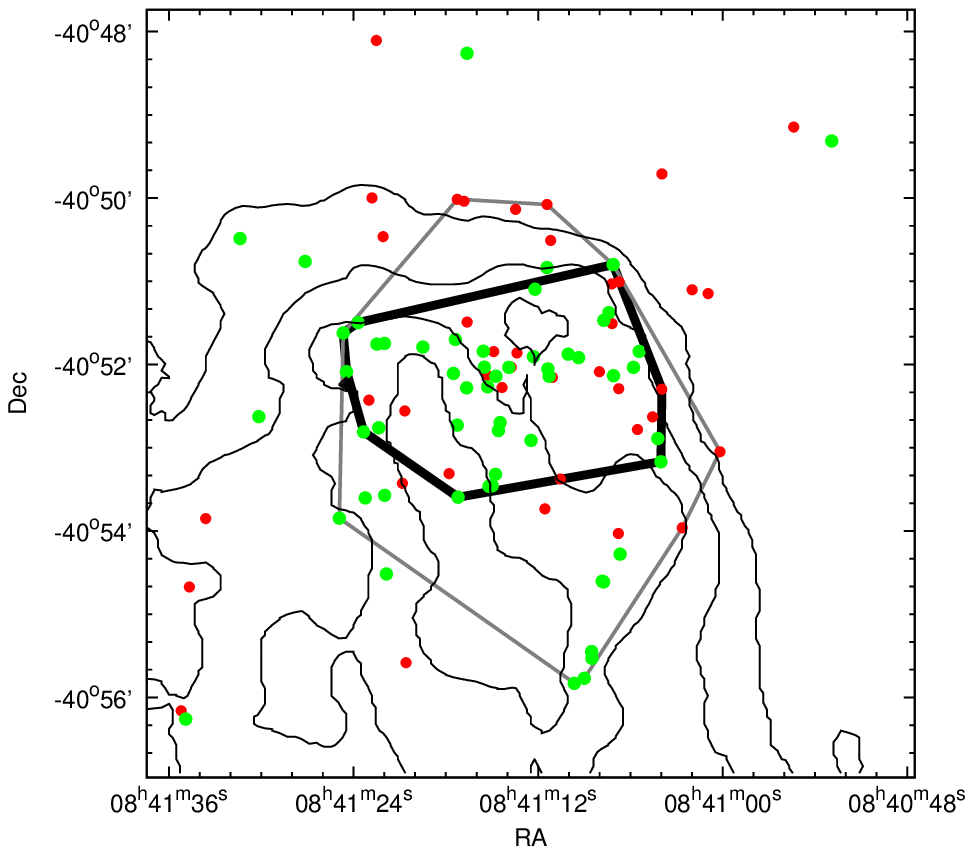}
}
\caption{\label{Fall55} Same as Fig. \ref {Fall54}, but for SFO 55.}
\end{figure*}

\begin{figure*}
\hbox{
\hspace{1.2cm}
\includegraphics[height=7.2cm,width=7.2cm]{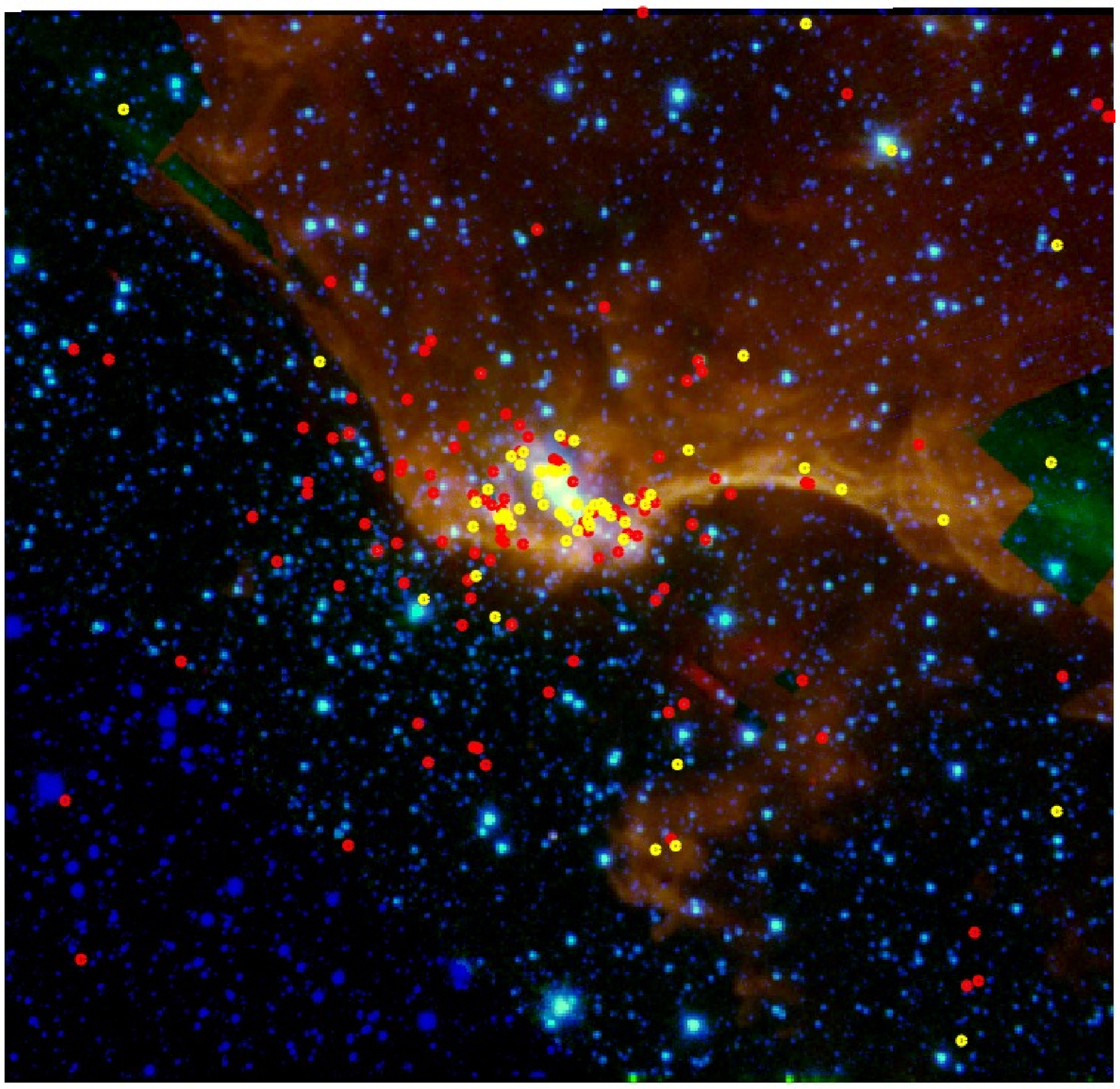}
\hspace{1.7cm}
\includegraphics[height=7.2cm,width=7.2cm]{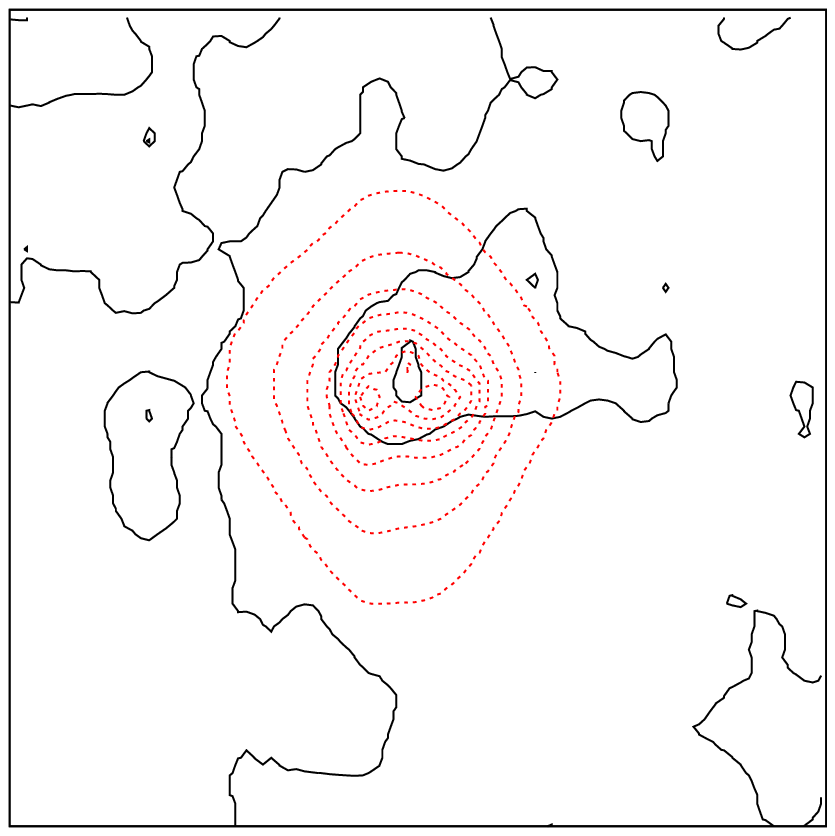}
}
\hbox{
\includegraphics[height=7.5cm,width=8.5cm]{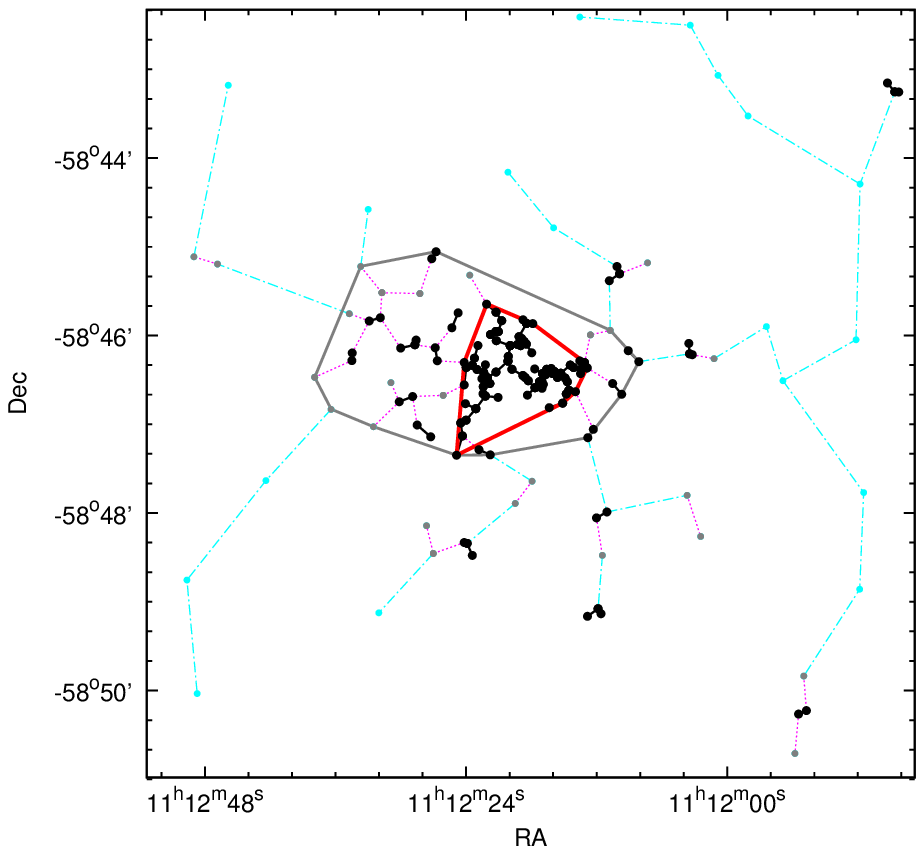}
\hspace{0.5cm}
\includegraphics[height=7.5cm,width=8.5cm]{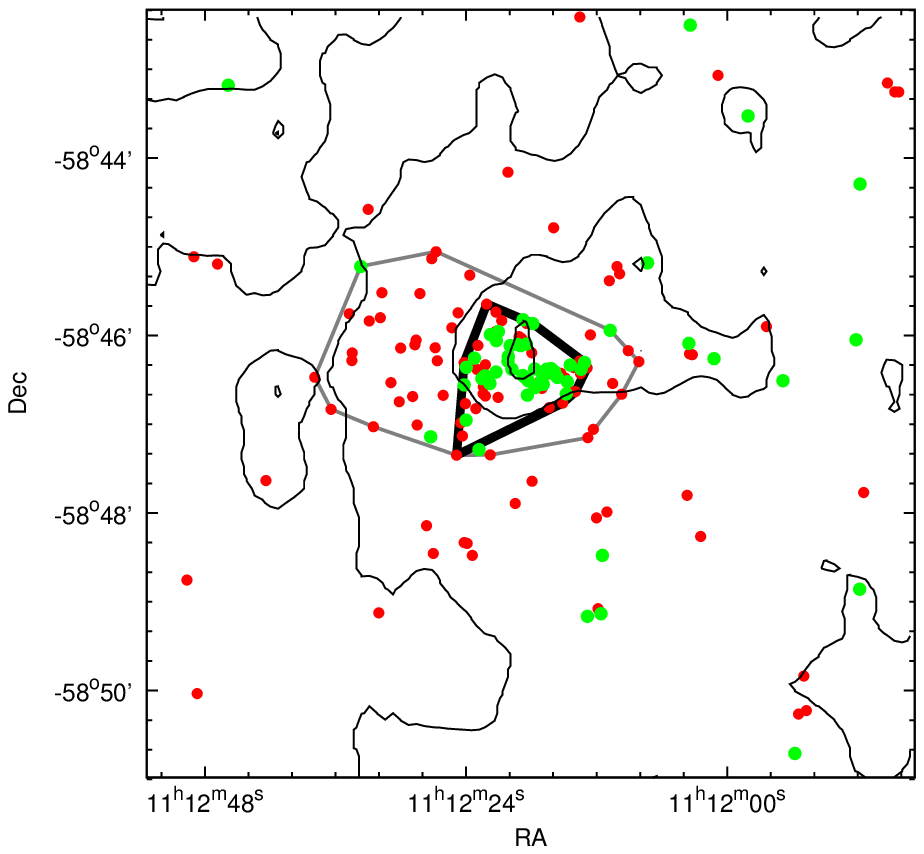}
}
\caption{\label{Fall64} Same as Fig. \ref {Fall54}, but for SFO 64.  }
\end{figure*}

\begin{figure*}
\hbox{
\hspace{1.2cm}
\includegraphics[height=7.2cm,width=7.2cm]{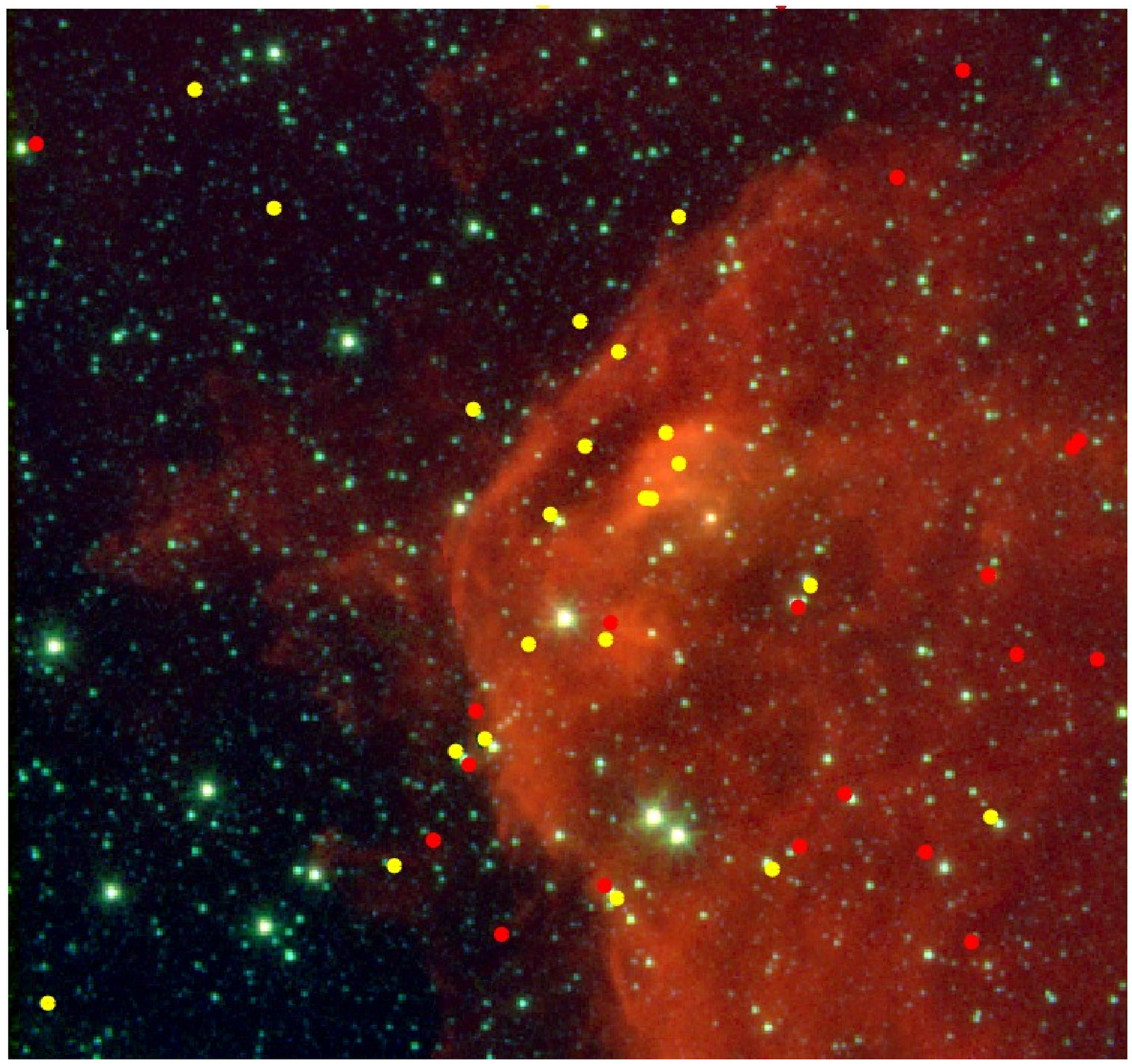}
\hspace{1.7cm}
\includegraphics[height=7.2cm,width=7.2cm]{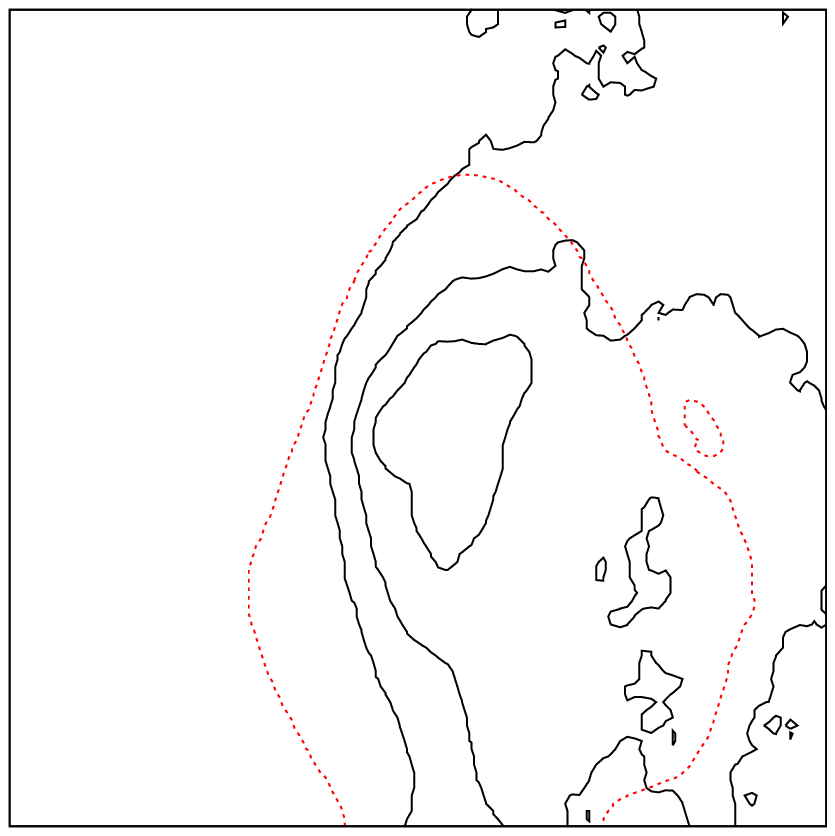}
}
\hbox{
\includegraphics[height=7.5cm,width=8.5cm]{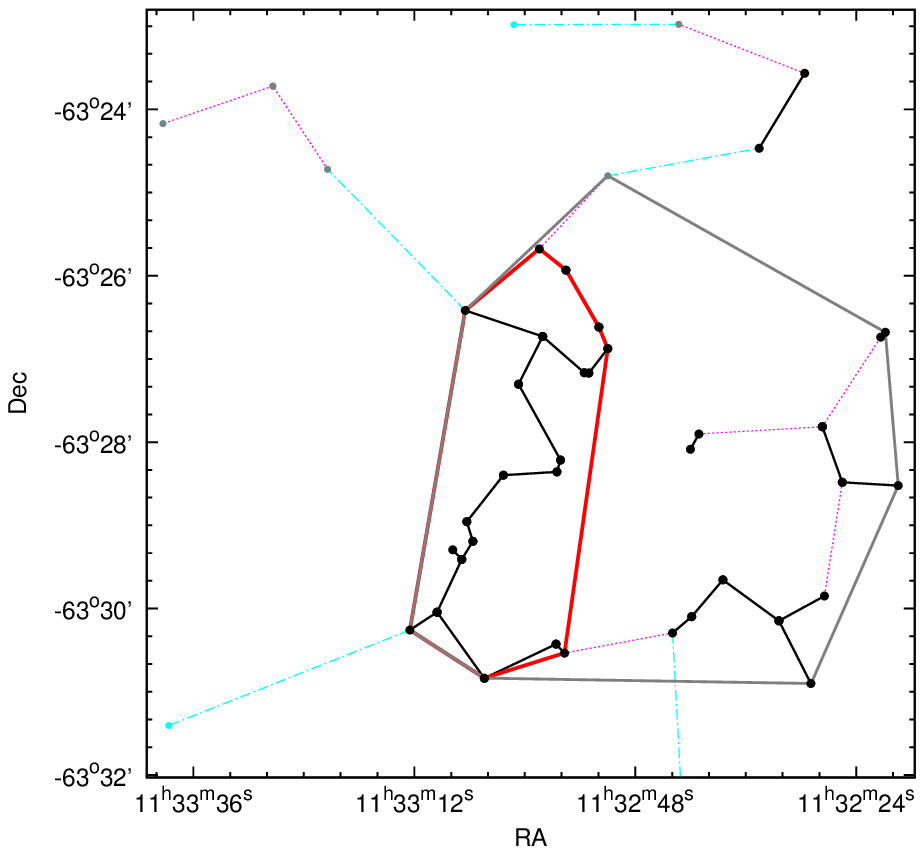}
\hspace{0.4cm}
\includegraphics[height=7.5cm,width=8.5cm]{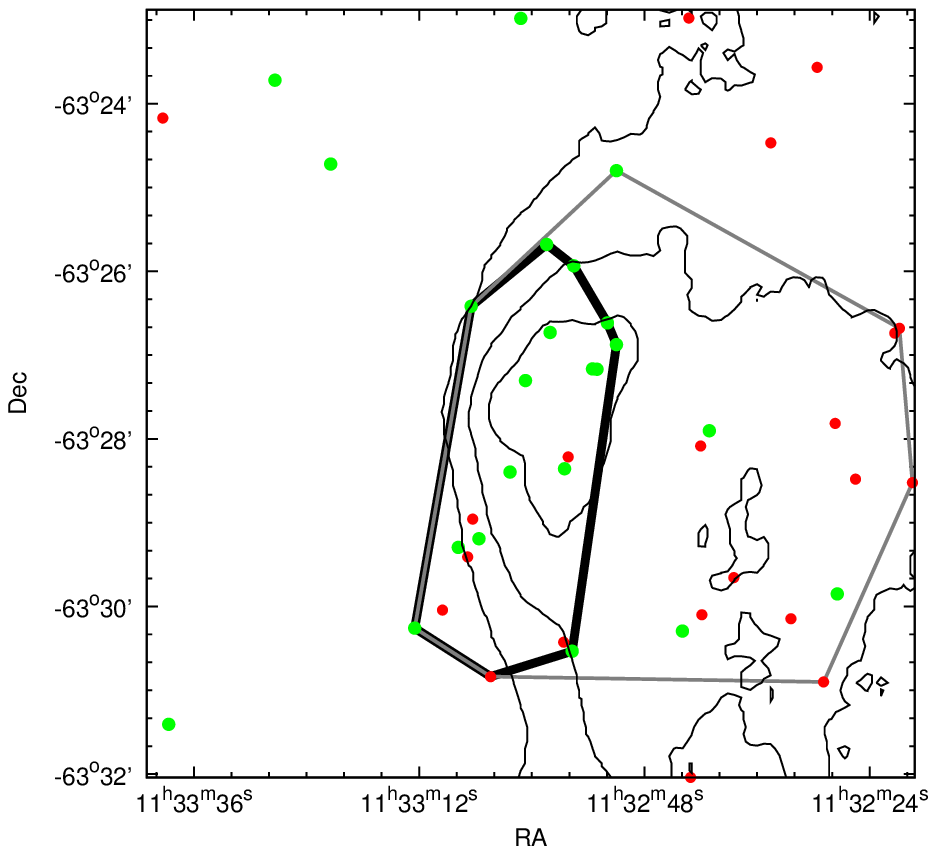}
}
\caption{\label{Fall65} Same as Fig. \ref {Fall54}, but for  SFO 65.  }
\end{figure*}

\begin{figure*}
\hbox{
\hspace{1.2cm}
\includegraphics[height=7.2cm,width=7.2cm]{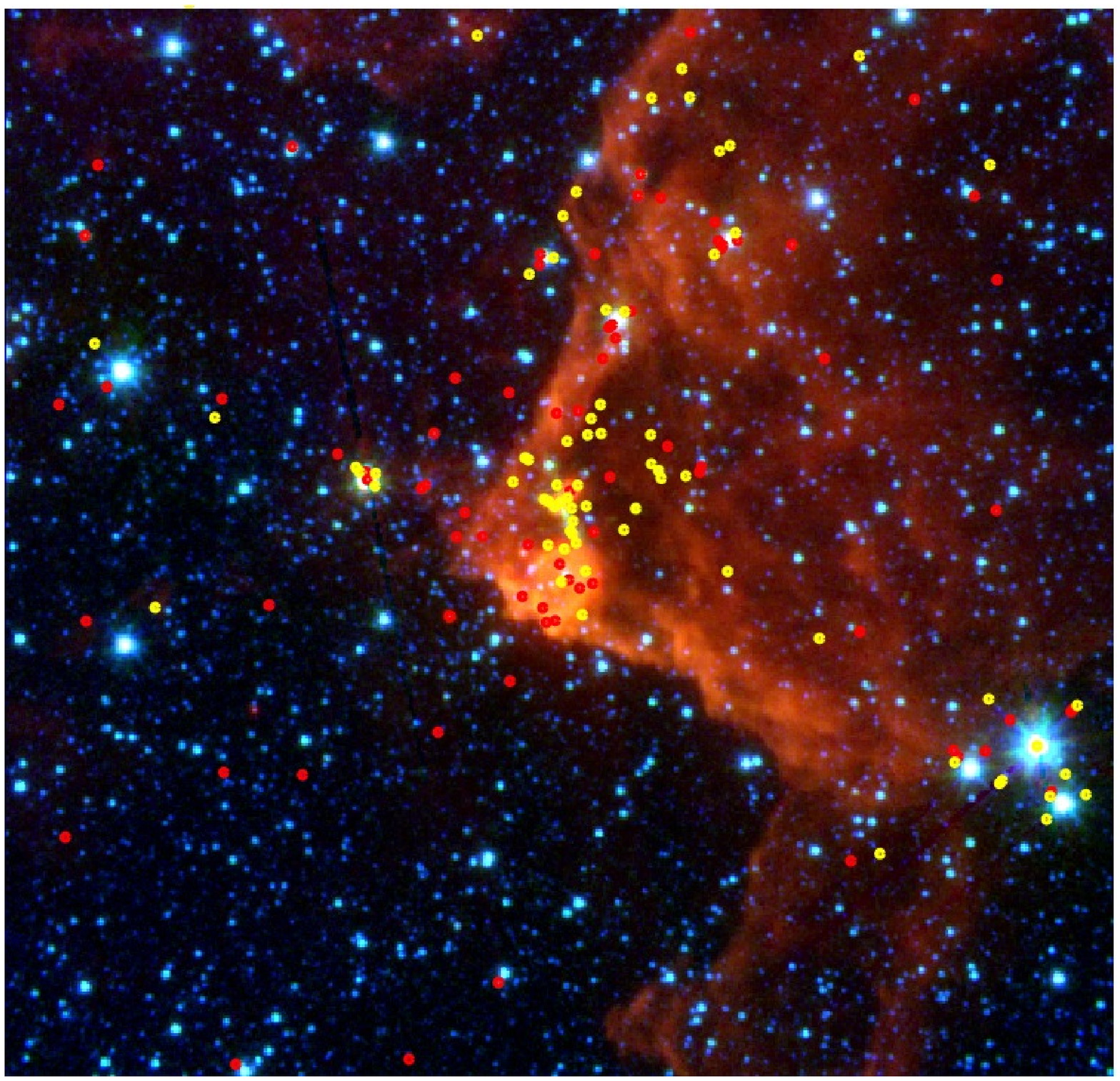}
\hspace{1.7cm}
\includegraphics[height=7.2cm,width=7.2cm]{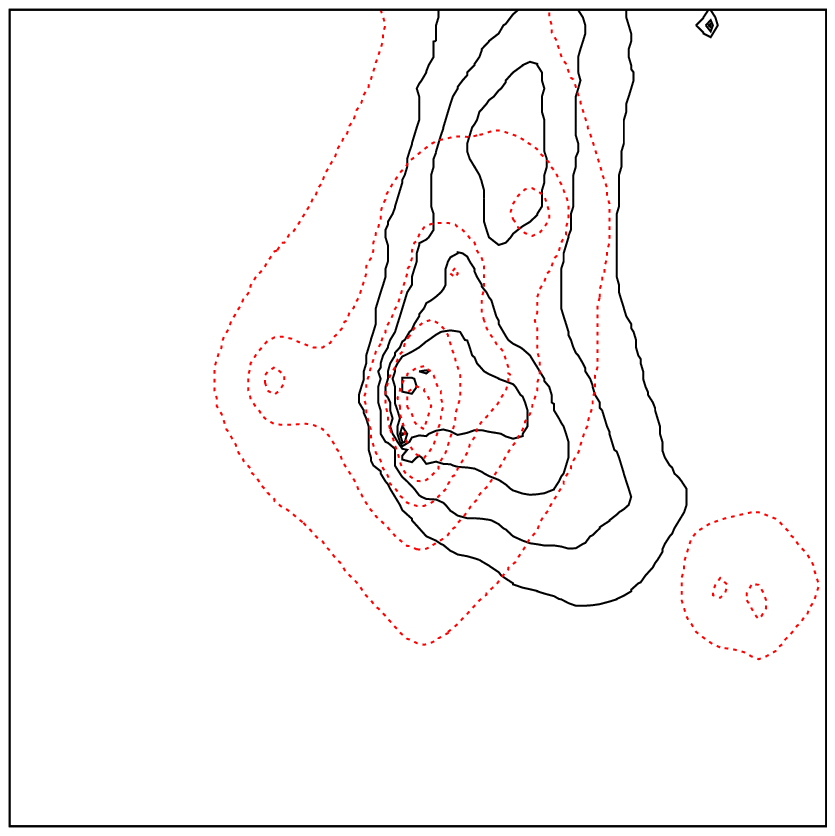}
}
\hbox{
\includegraphics[height=7.5cm,width=8.5cm]{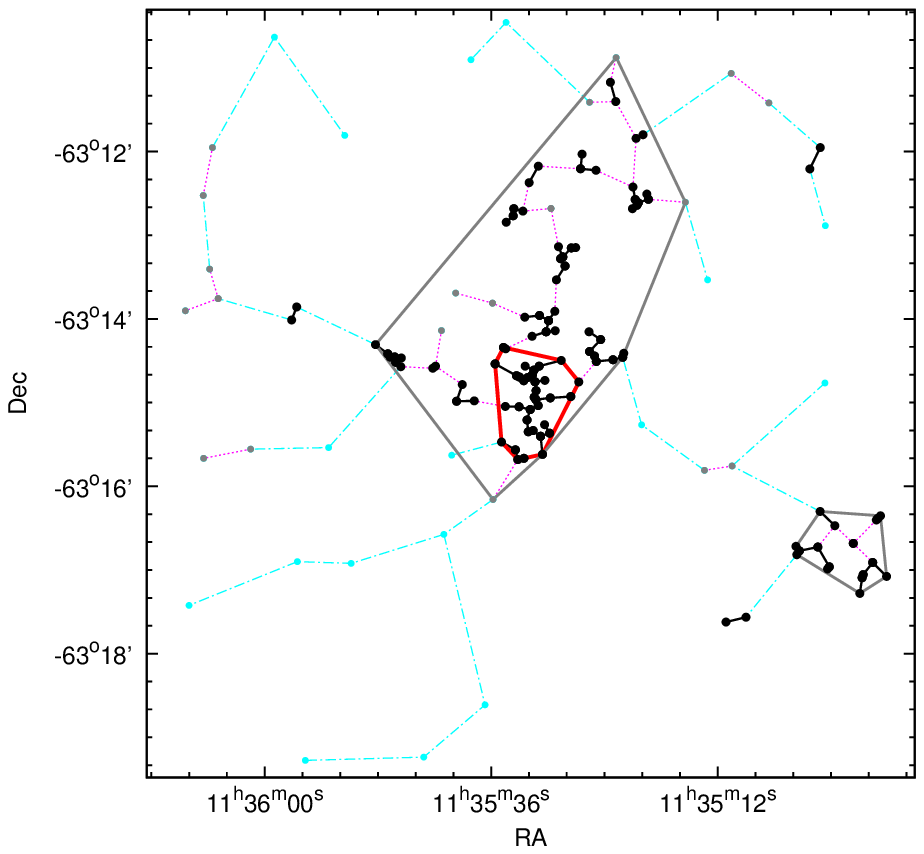}
\hspace{0.4cm}
\includegraphics[height=7.5cm,width=8.5cm]{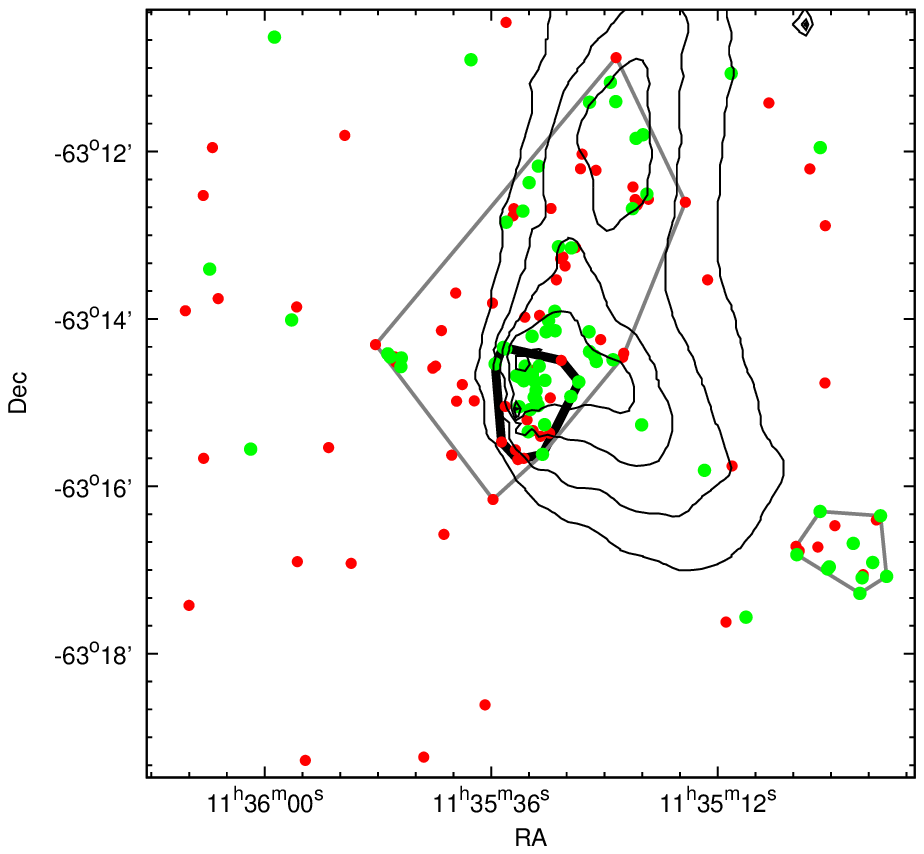}
}
\caption{\label{Fall68} Same as Fig. \ref {Fall54}, but for  SFO 68. }
\end{figure*}

\begin{figure*}
\hbox{
\hspace{1.2cm}
\includegraphics[height=7.2cm,width=7.2cm]{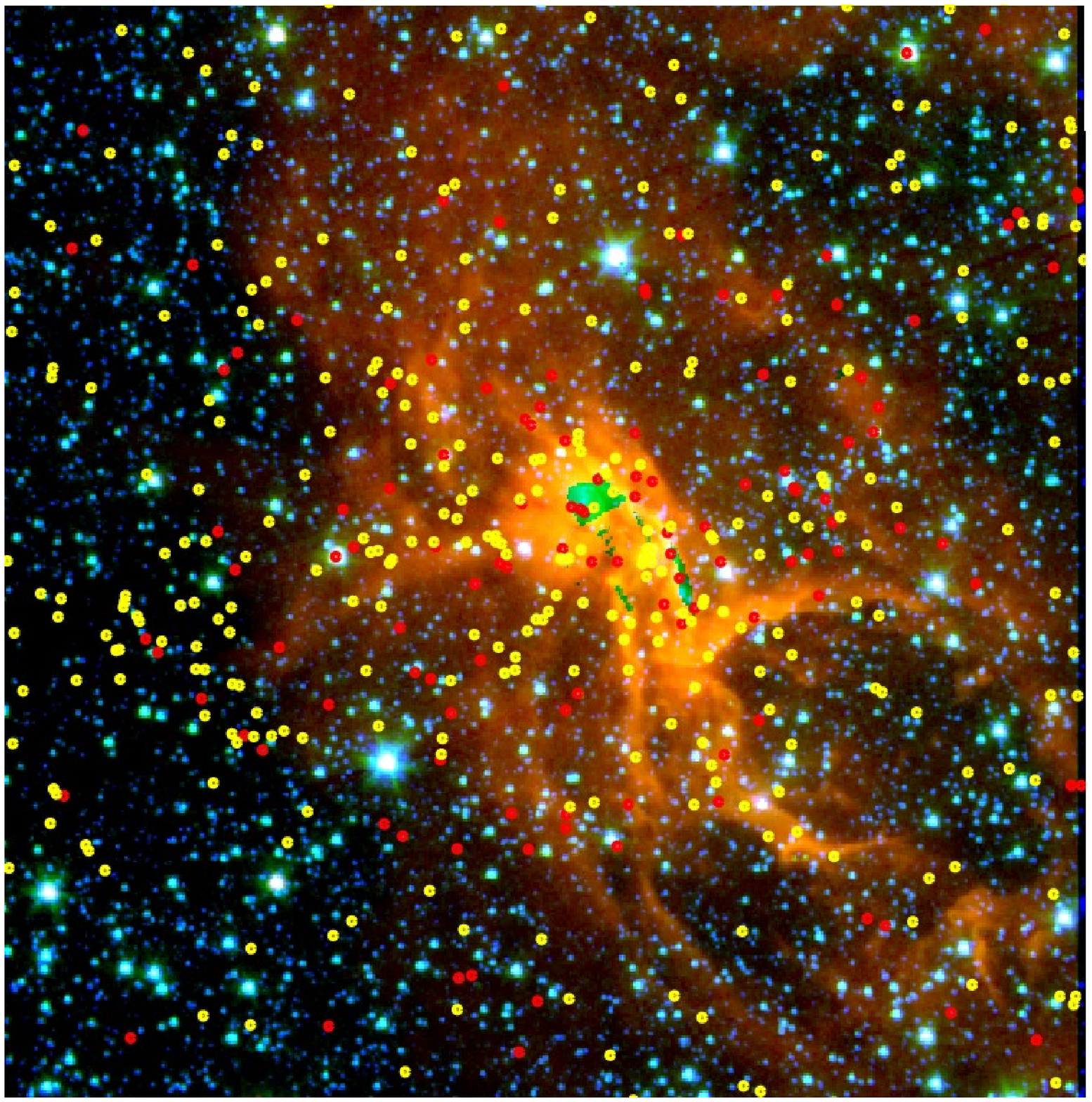}
\hspace{1.7cm}
\includegraphics[height=7.2cm,width=7.2cm]{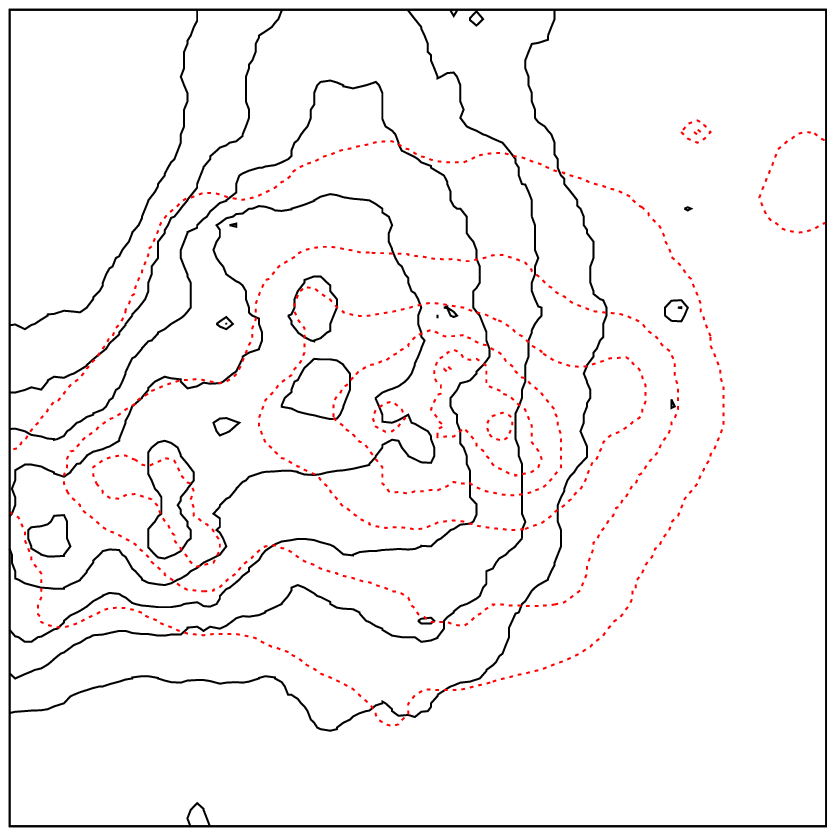}
}
\hbox{
\includegraphics[height=7.8cm,width=8.8cm]{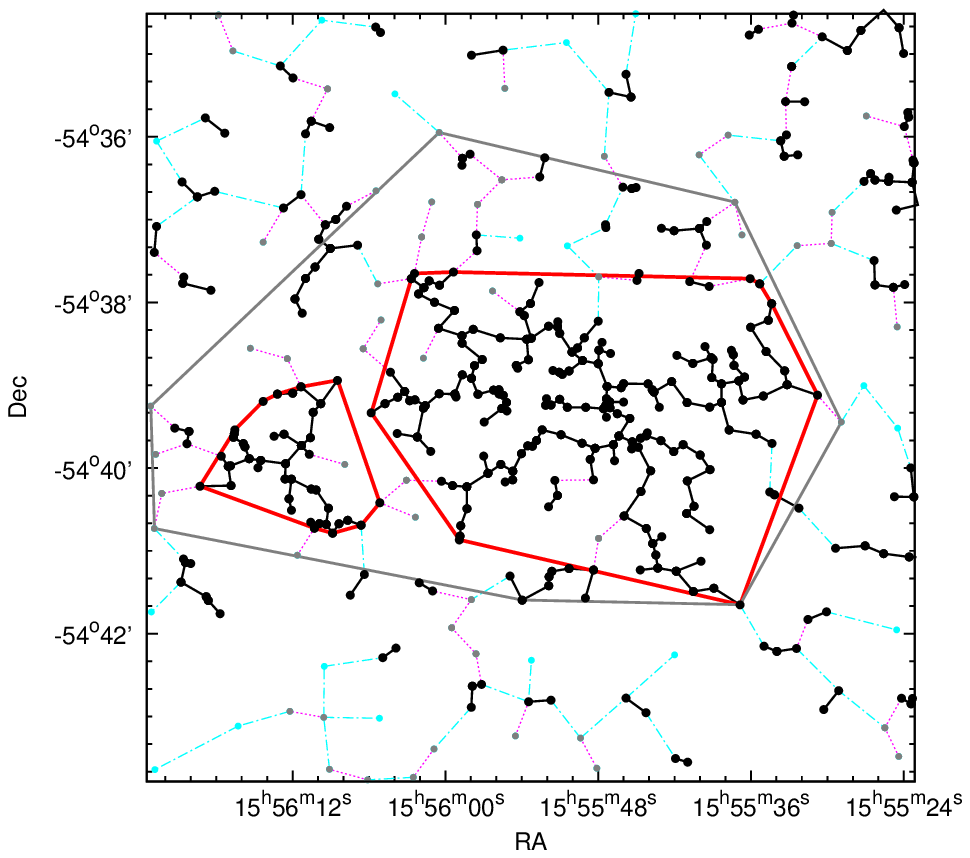}
\hspace{0.4cm}
\includegraphics[height=7.8cm,width=8.8cm]{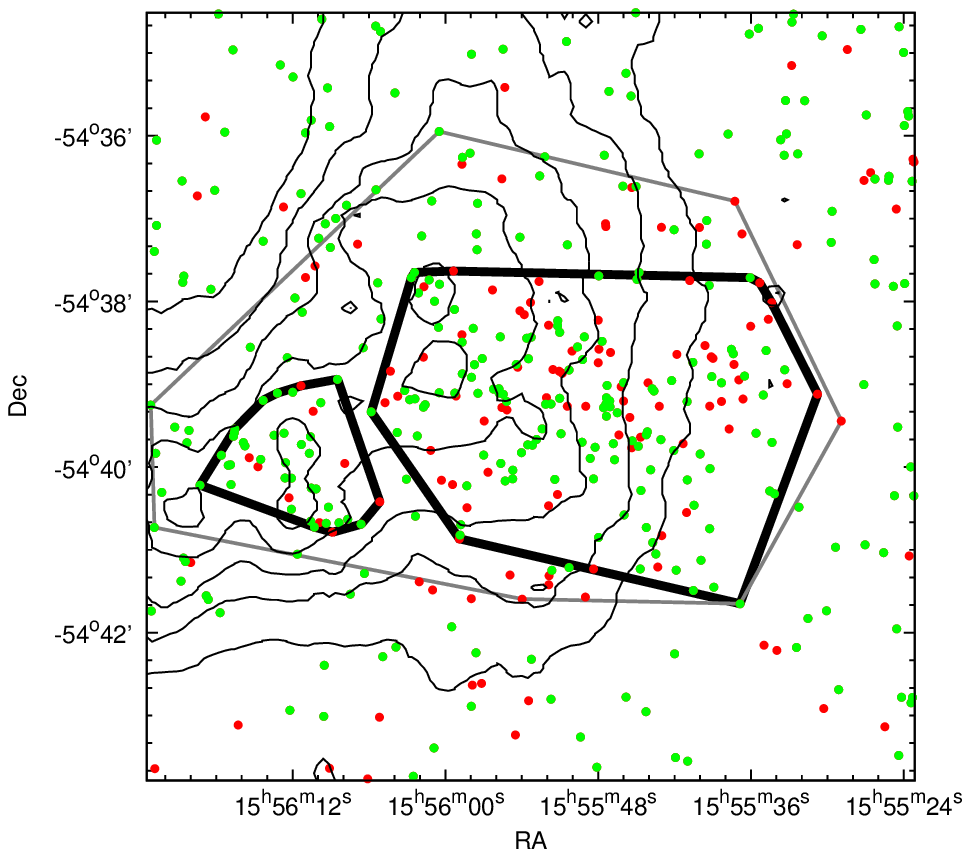}
}
\caption{\label{Fall75} Same as Fig. \ref {Fall54}, but for SFO 75.}
\end{figure*}

\begin{figure*}
\hbox{
\hspace{1.2cm}
\includegraphics[height=7.2cm,width=7.2cm]{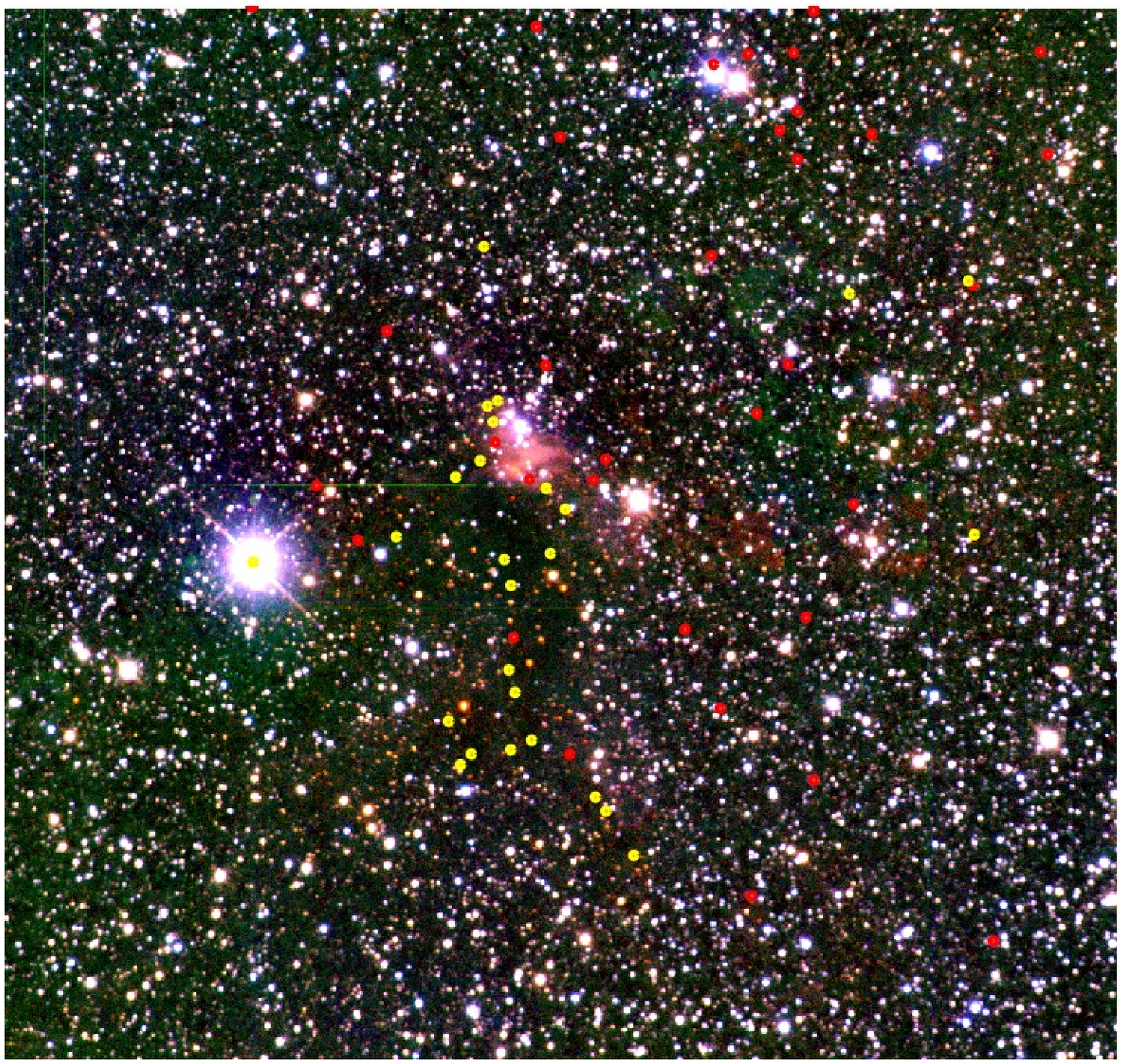}
\hspace{1.7cm}
\includegraphics[height=7.2cm,width=7.2cm]{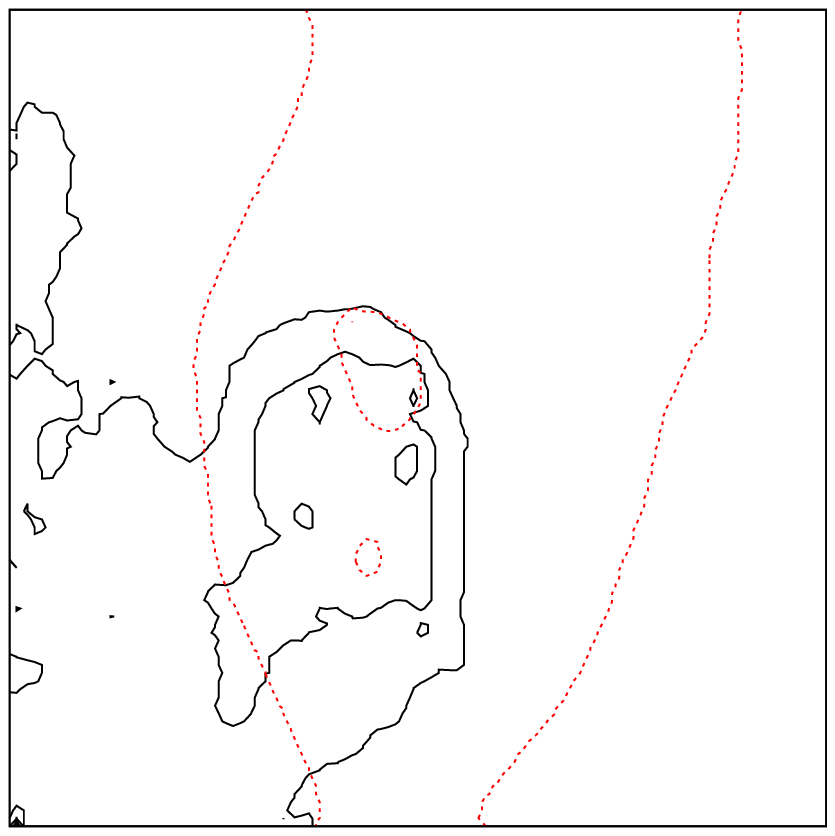}
}
\hbox{
\includegraphics[height=7.5cm,width=8.6cm]{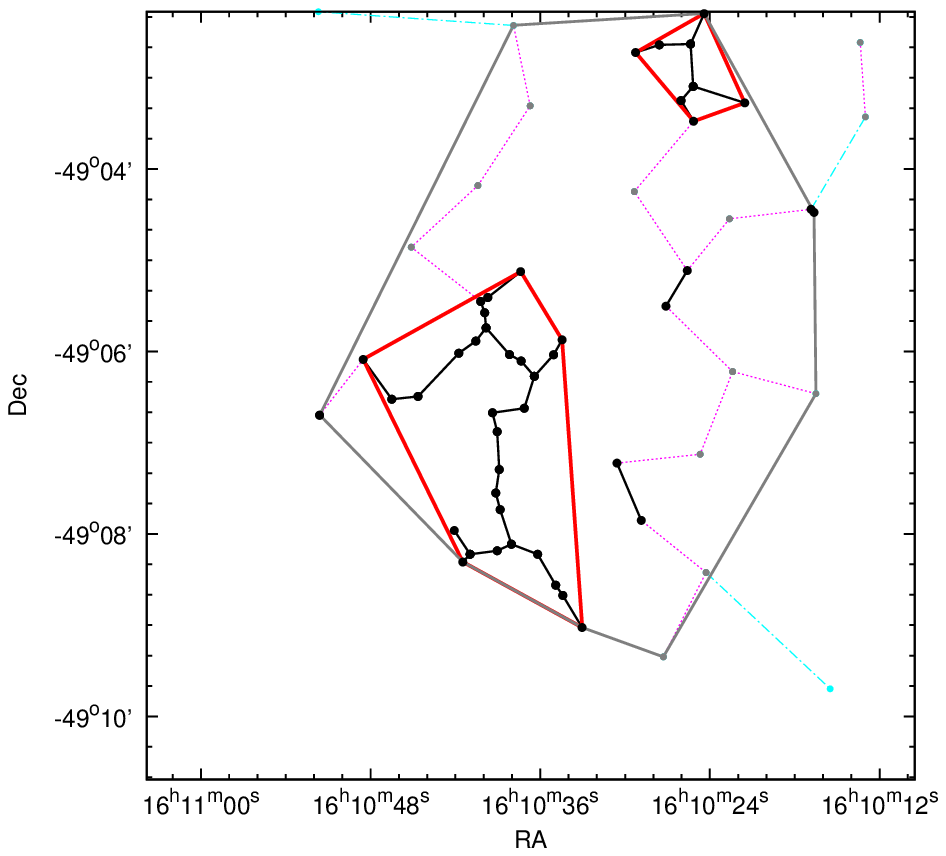}
\hspace{0.4cm}
\includegraphics[height=7.5cm,width=8.6cm]{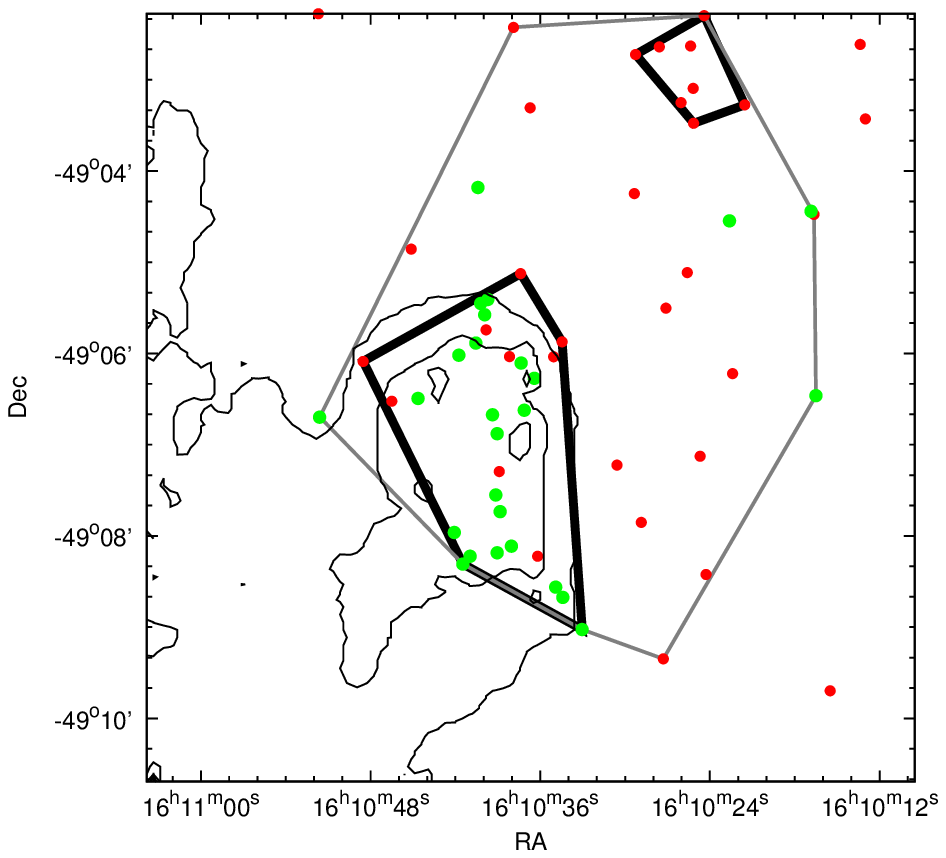}
}
\caption{\label{Fall76} Same as Fig. \ref {Fall54}, but for SFO 76.}
\end{figure*}

\begin{figure*}
\hbox{
\hspace{1.2cm}
\includegraphics[height=7.2cm,width=7.2cm]{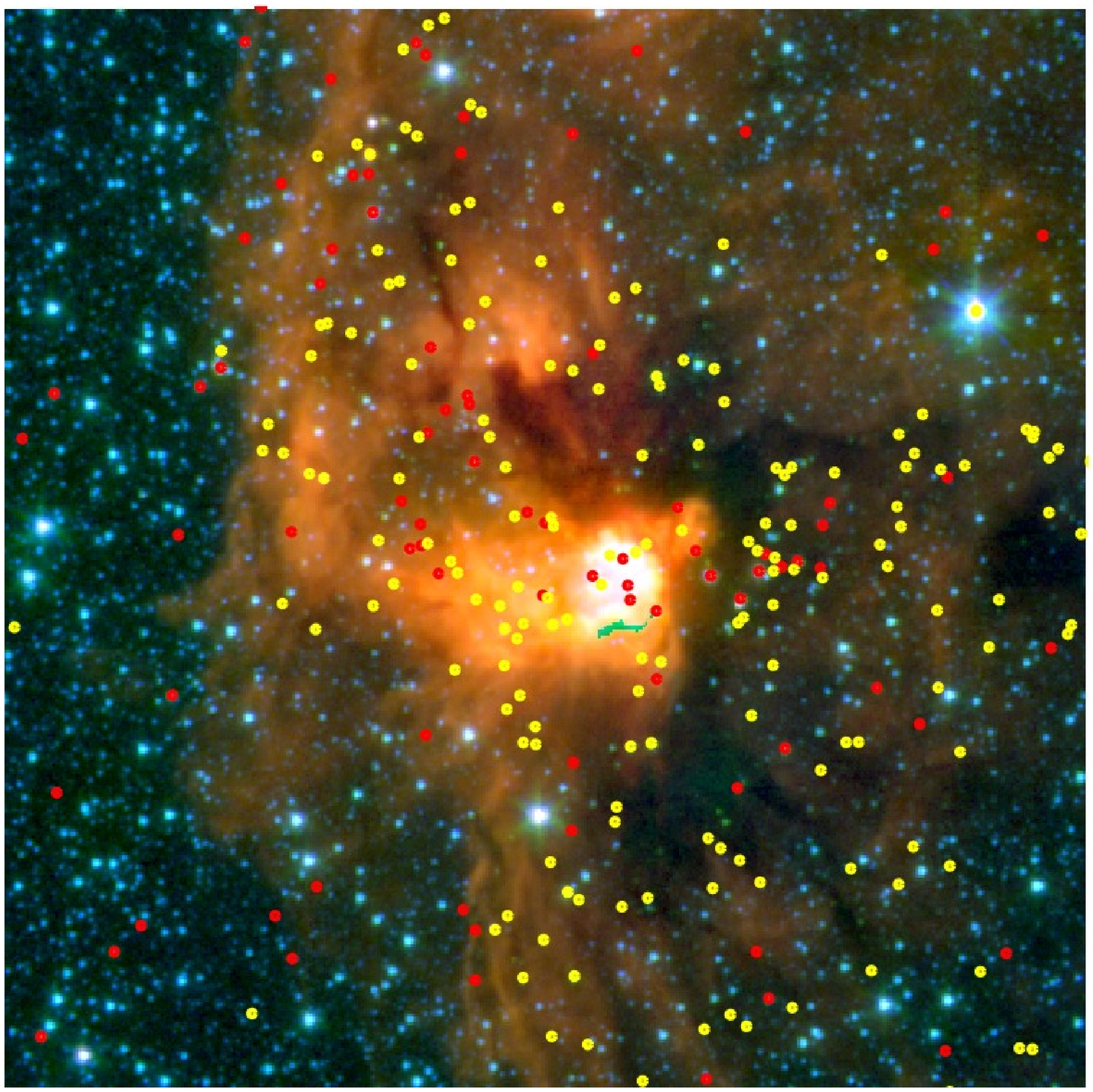}
\hspace{1.2cm}
\includegraphics[height=7.2cm,width=7.2cm]{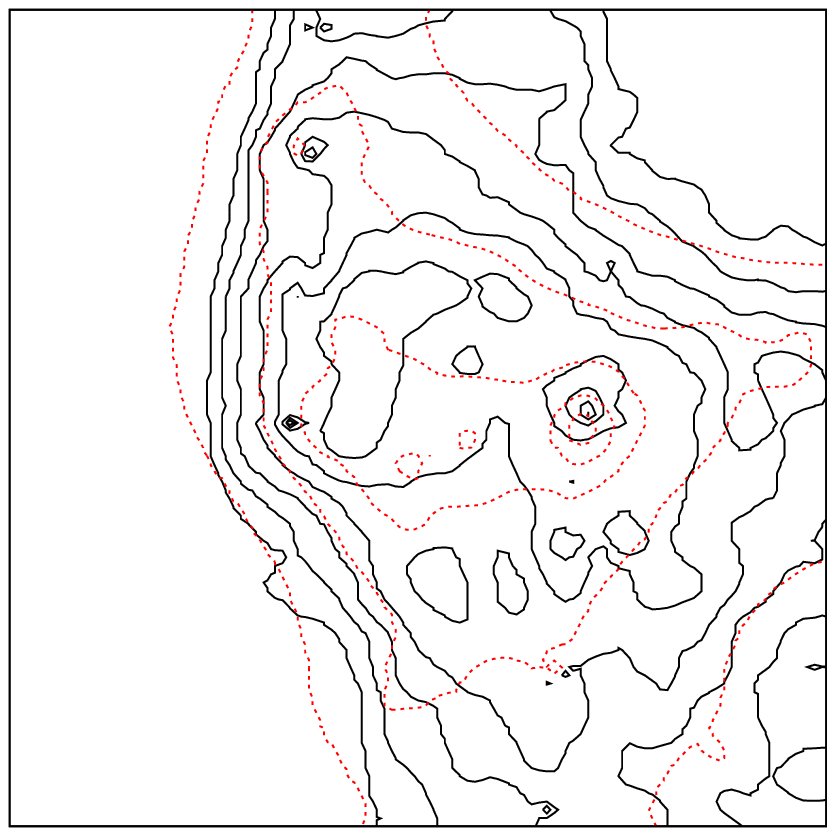}
}
\hbox{
\includegraphics[height=8.0cm,width=8.5cm]{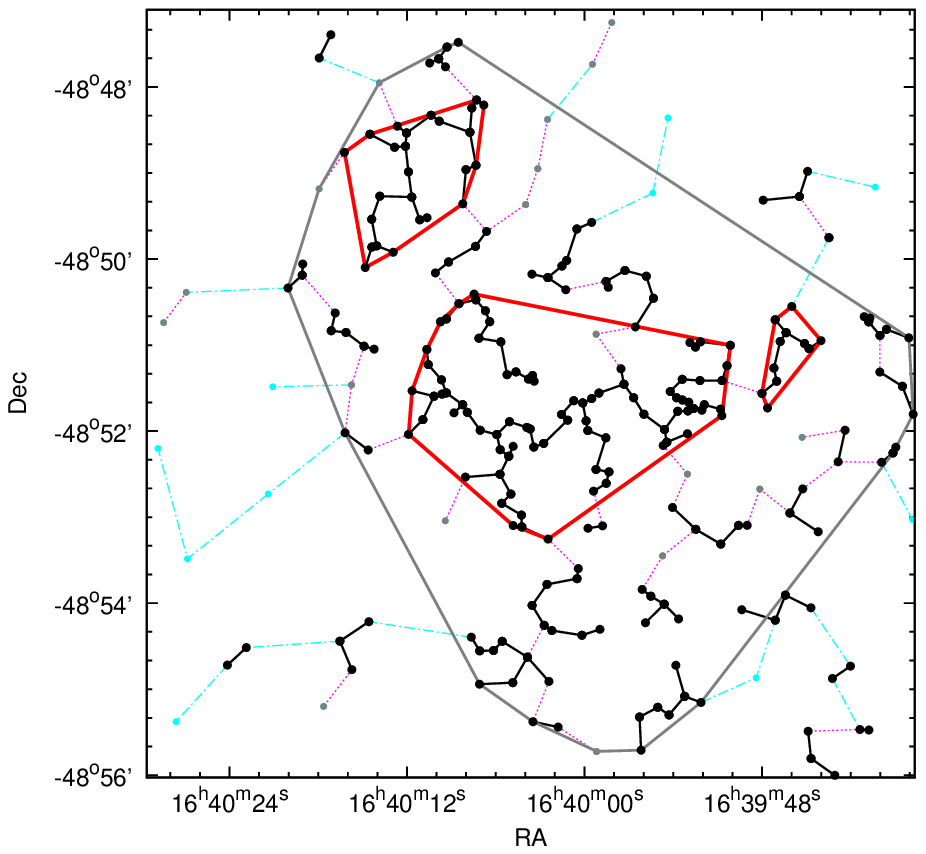}
\includegraphics[height=8.0cm,width=8.5cm]{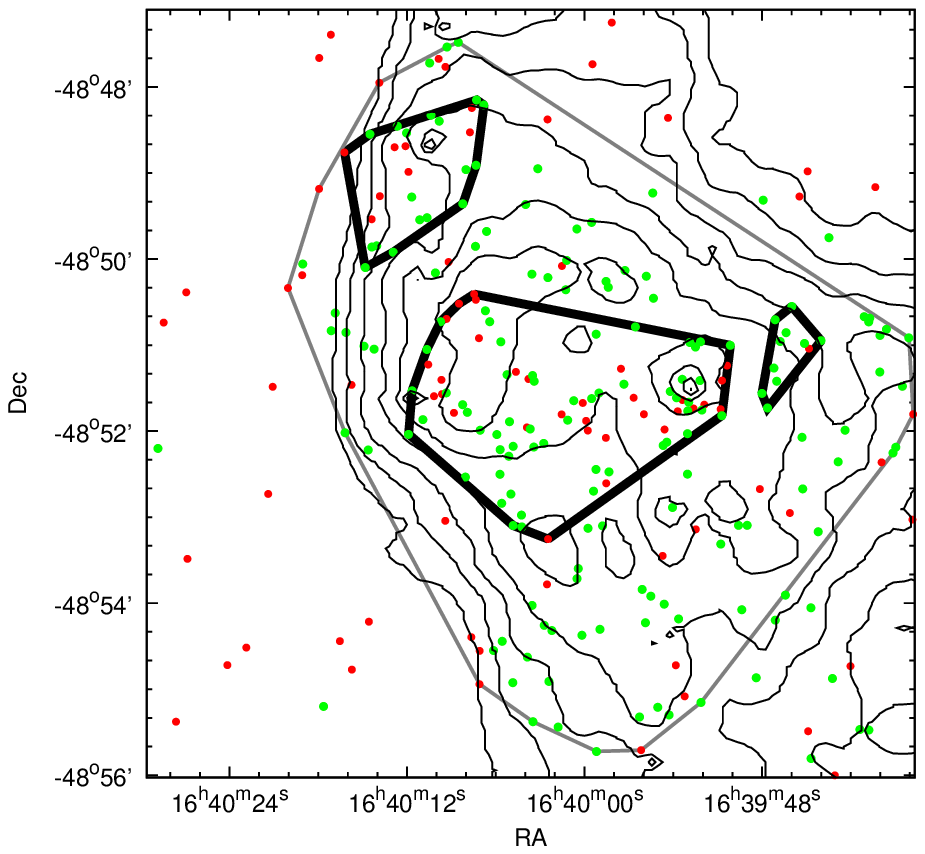}
}
\caption{\label{Fall79} Same as Fig. \ref {Fall54}, but for SFO 79.}
\end{figure*}

All the studied BRCs contain a number of sub-groups/cores of YSOs (cf. Figs. \ref{Fall54} - \ref{Fall79}, top left panels) presumably
due to fragmentation of the molecular cloud.
Physical parameters of these cores, which might have formed in a single star-forming event, 
play a very important role in the study of star formation.
\citet{2009ApJS..184...18G} have applied an empirical method based on the minimal sampling tree (MST)
technique to a sample of 36 young stellar clusters to isolate groupings (cores) from the
more diffuse distribution of YSOs in nebulous regions.
This method  effectively isolates the 
sub-structures without any type of smoothening
\citep[e.g.,][]{2004MNRAS.348..589C, 2006A&A...449..151S, 2007MNRAS.379.1302B, 2009MNRAS.392..868B, 2009ApJS..184...18G}. 
The sub-groups detected in this way have no bias regarding  the shapes
of the distribution and preserve the underlying
geometry of the distribution \citep{2009ApJS..184...18G}.
In  Figs. \ref{Fall54} - \ref{Fall79} (bottom left panels), we have plotted the derived MSTs for the YSOs  
in the BRC regions studied. The different color dots and lines are the positions of the
YSOs and the MST branches, respectively. A close inspection of these figures reveals that all the
regions exhibit in-homogeneous structures and there is one major and several other concentrations of YSOs
distributed throughout the regions. 

\begin{figure*}
\centering\includegraphics[height=3.4cm,width=4.0cm]{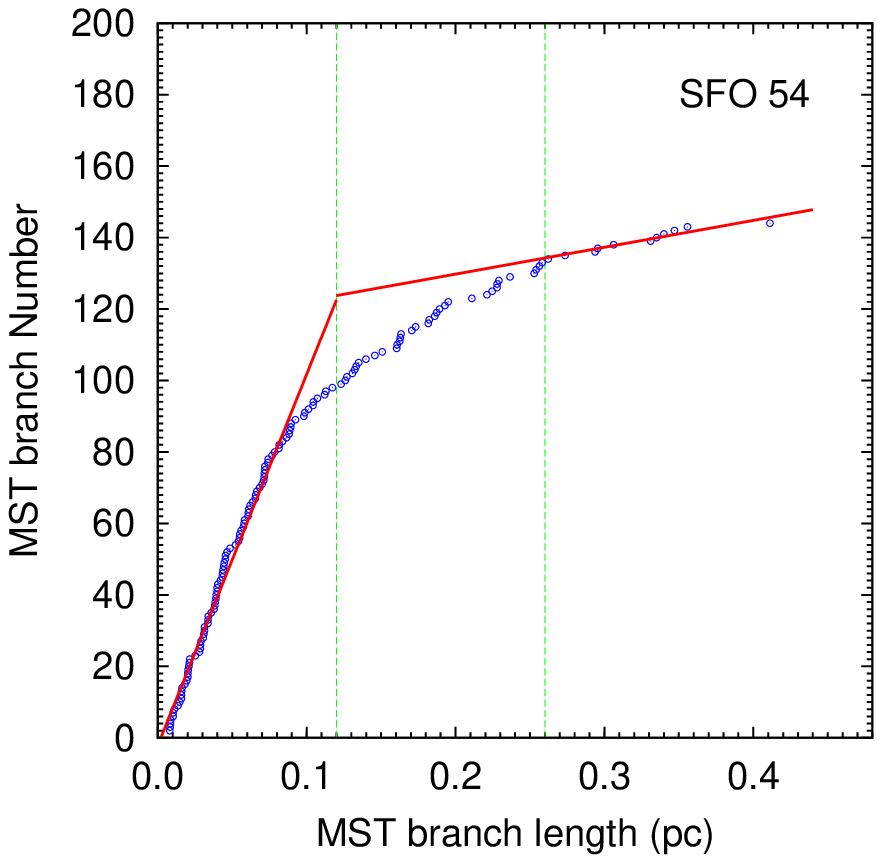}
\centering\includegraphics[height=3.4cm,width=4.0cm]{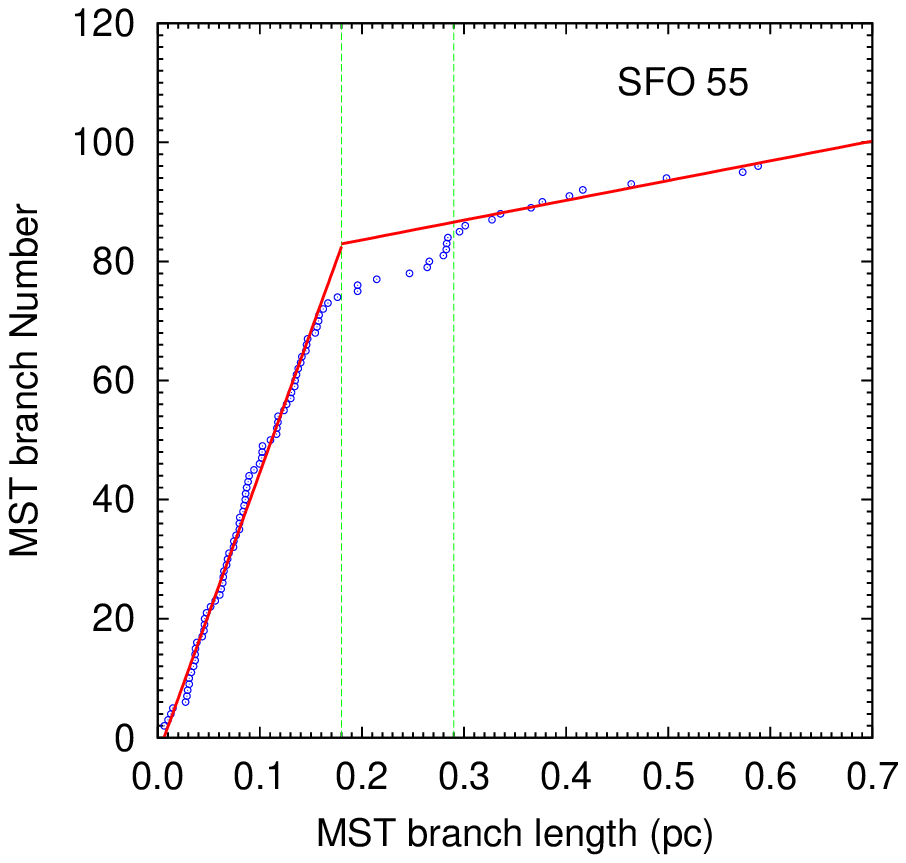}
\centering\includegraphics[height=3.4cm,width=4.0cm]{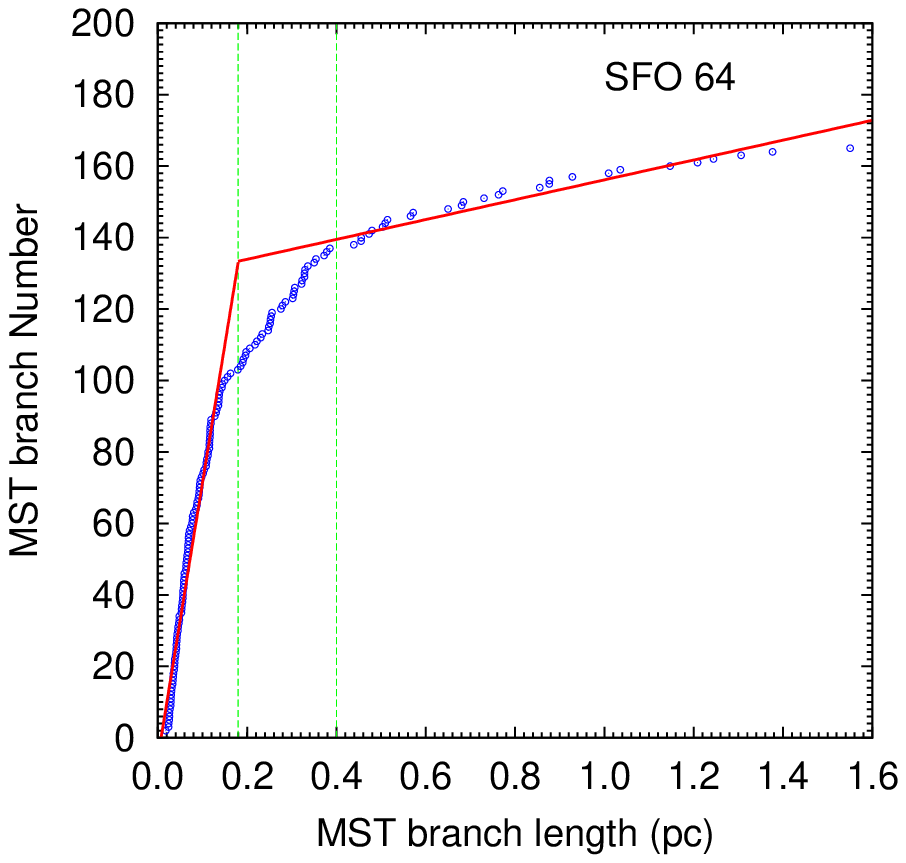}
\centering\includegraphics[height=3.4cm,width=4.0cm]{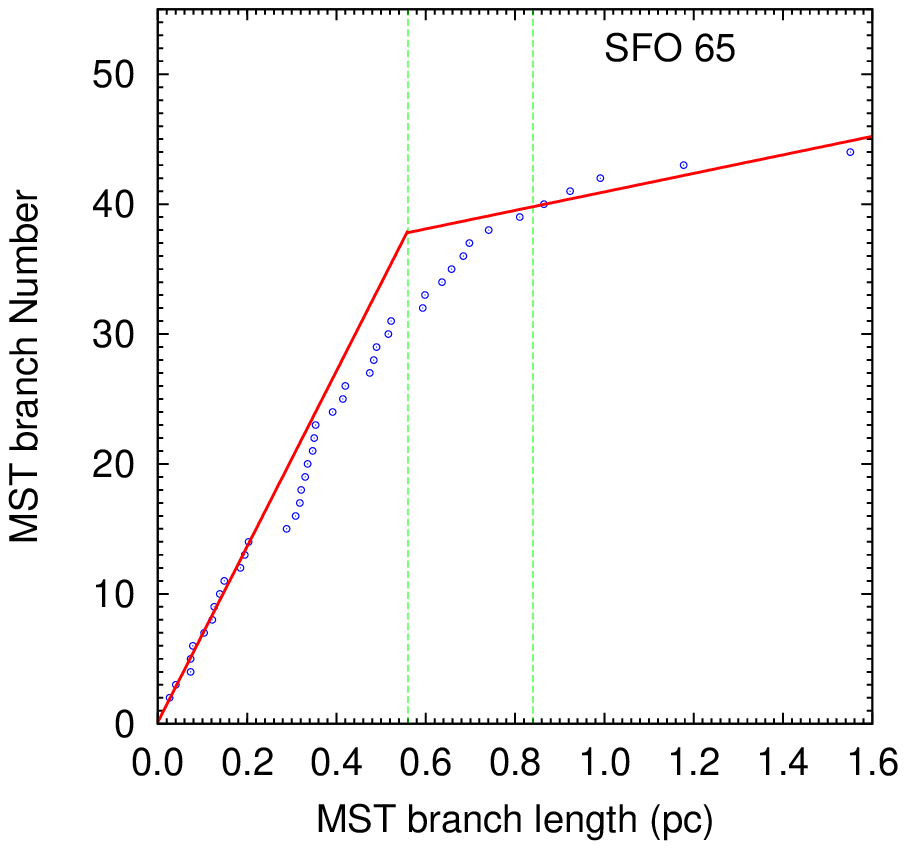}

\centering\includegraphics[height=3.4cm,width=4.0cm]{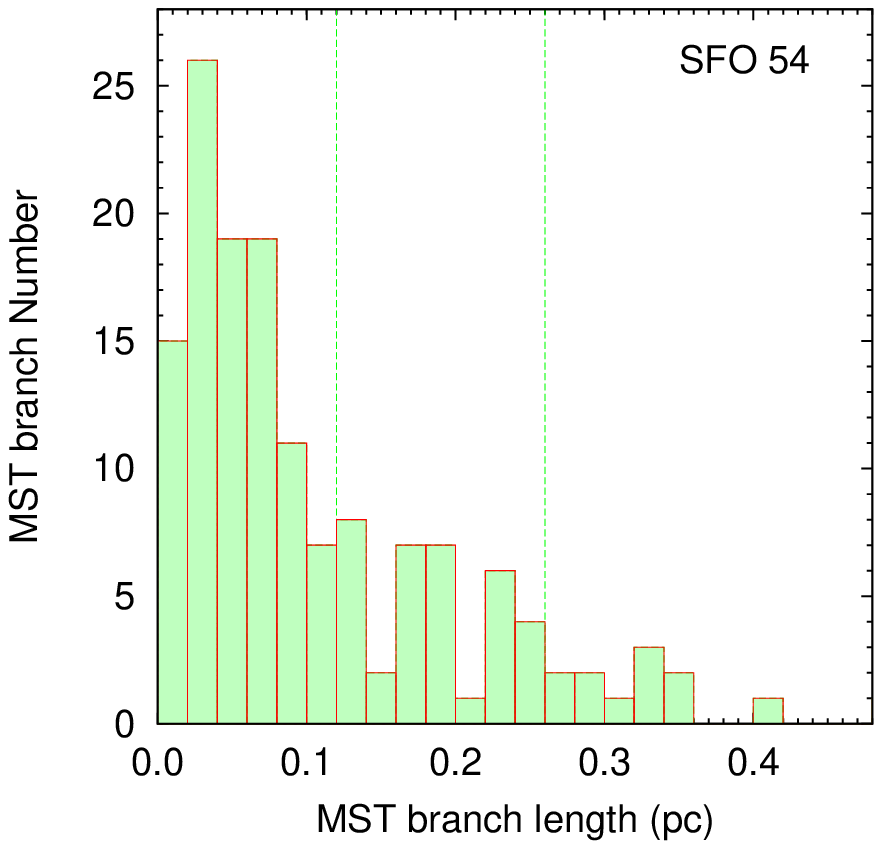}
\centering\includegraphics[height=3.4cm,width=4.0cm]{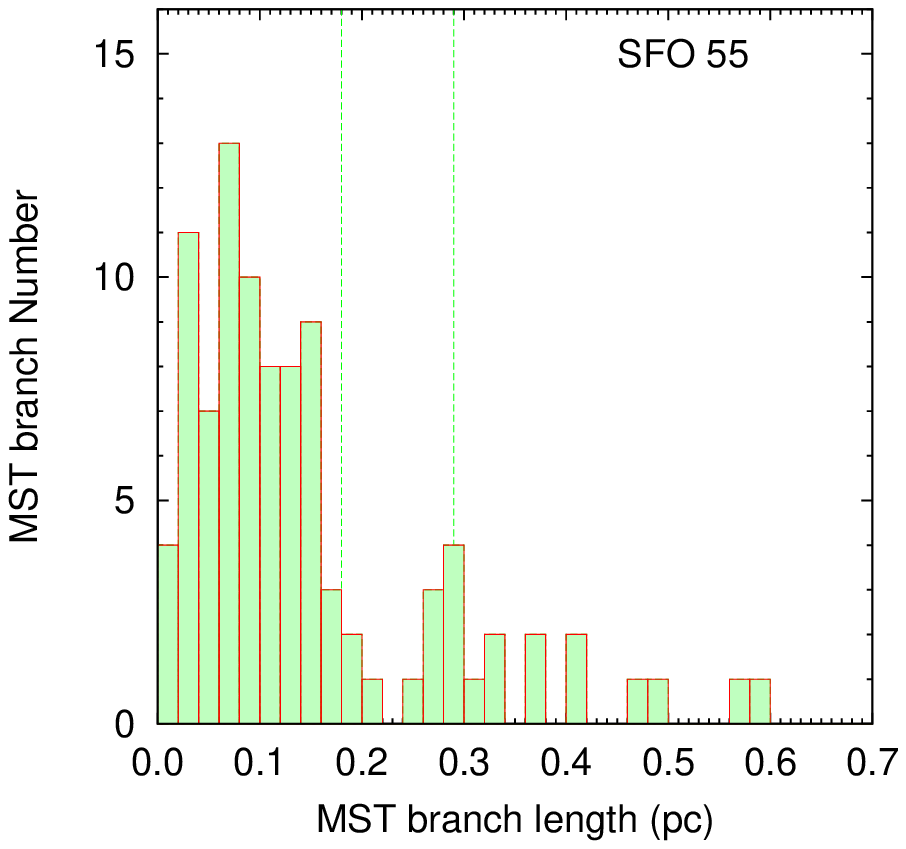}
\centering\includegraphics[height=3.4cm,width=4.0cm]{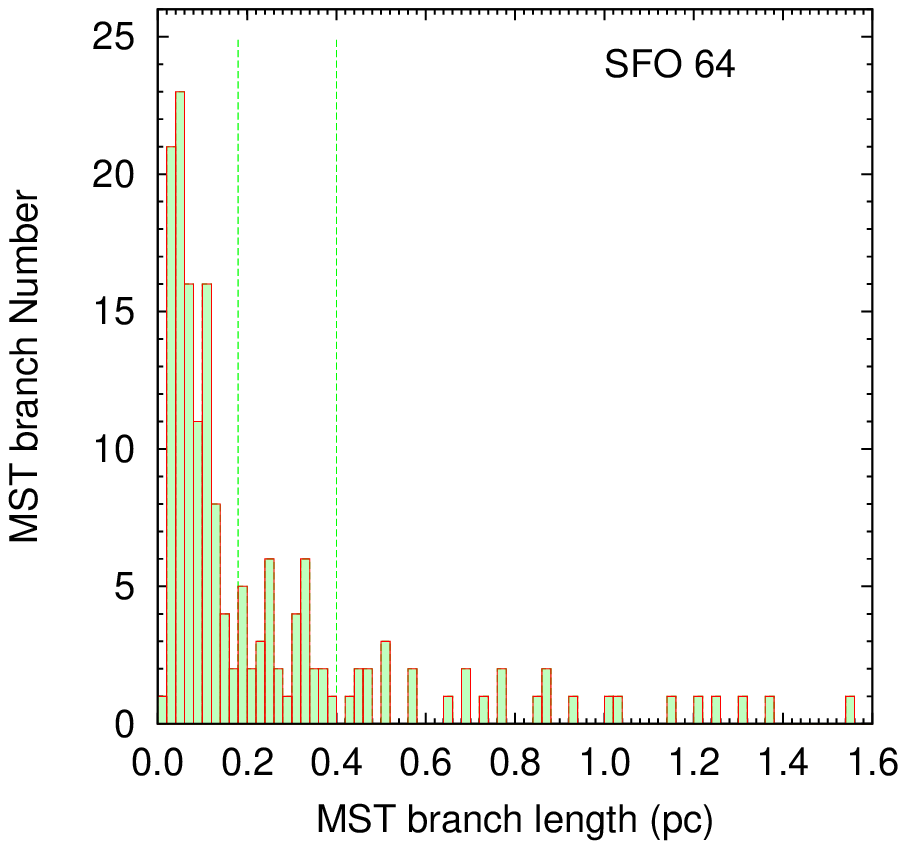}
\centering\includegraphics[height=3.4cm,width=4.0cm]{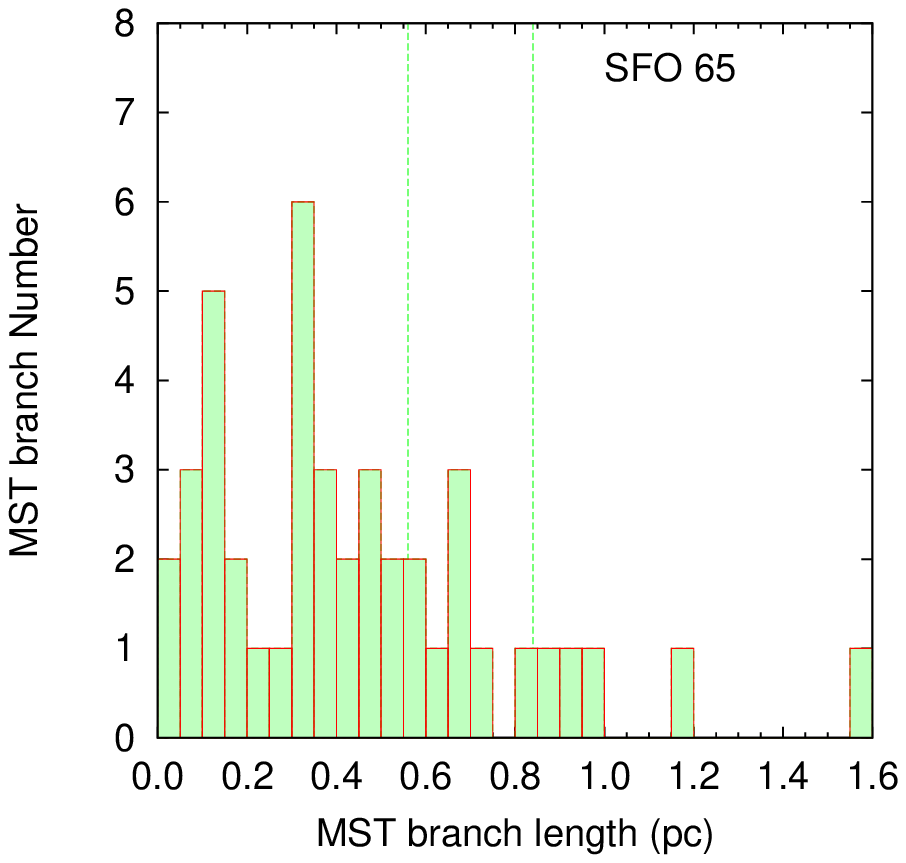}

\centering\includegraphics[height=3.4cm,width=4.0cm]{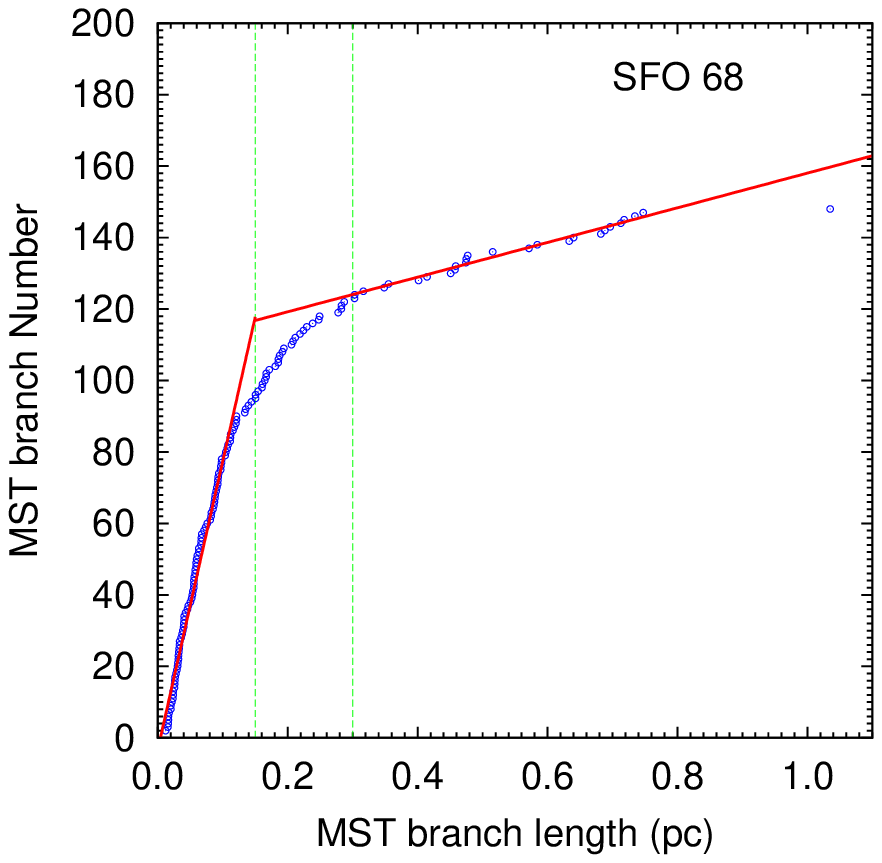}
\centering\includegraphics[height=3.4cm,width=4.0cm]{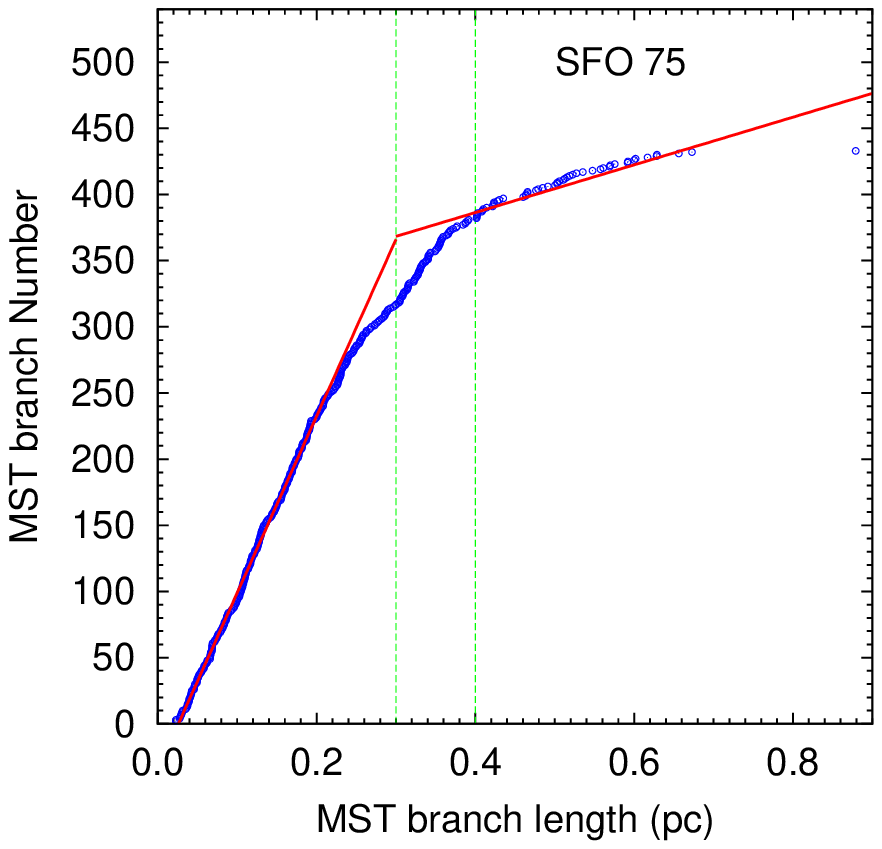}
\centering\includegraphics[height=3.4cm,width=4.0cm]{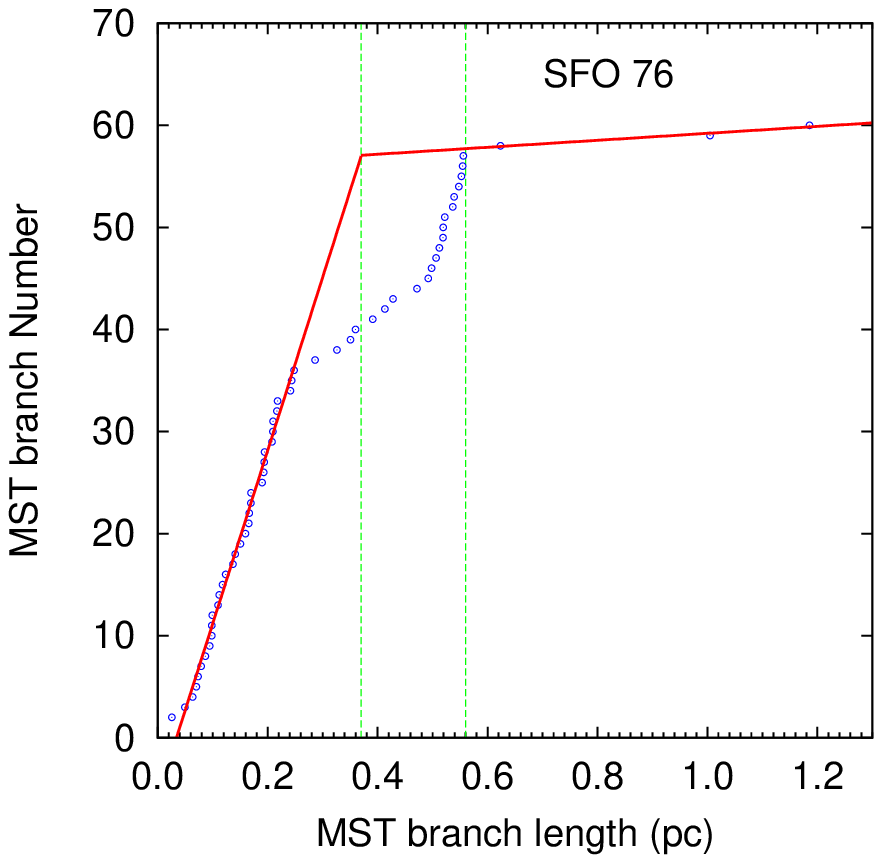}
\centering\includegraphics[height=3.4cm,width=4.0cm]{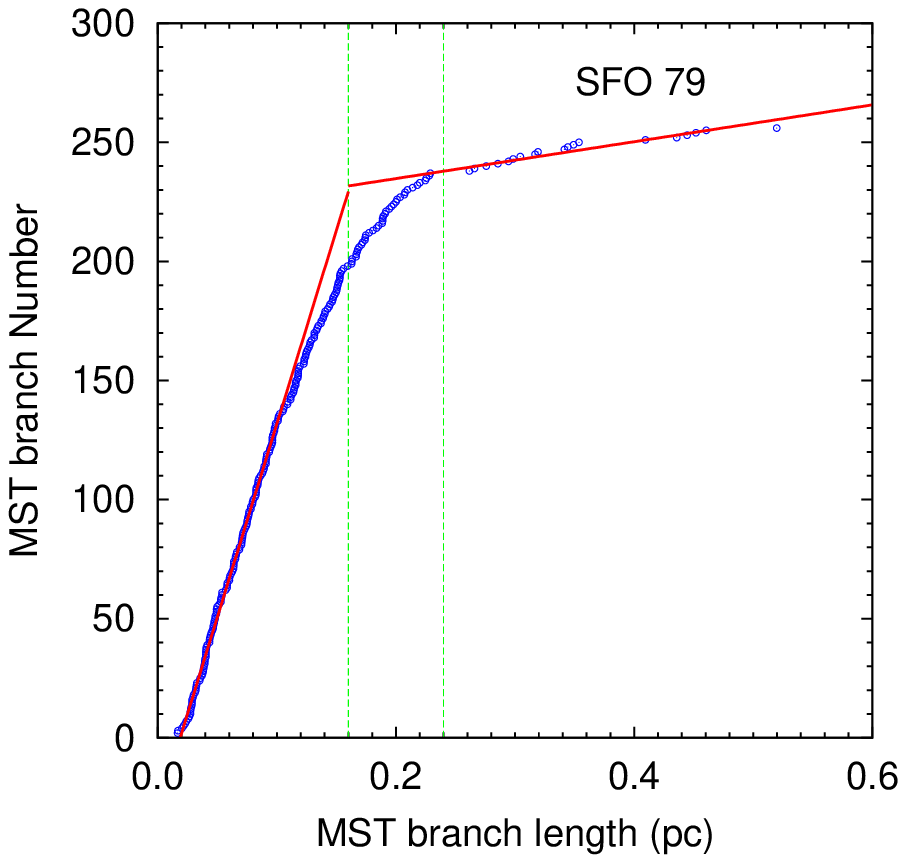}

\centering\includegraphics[height=3.4cm,width=4.0cm]{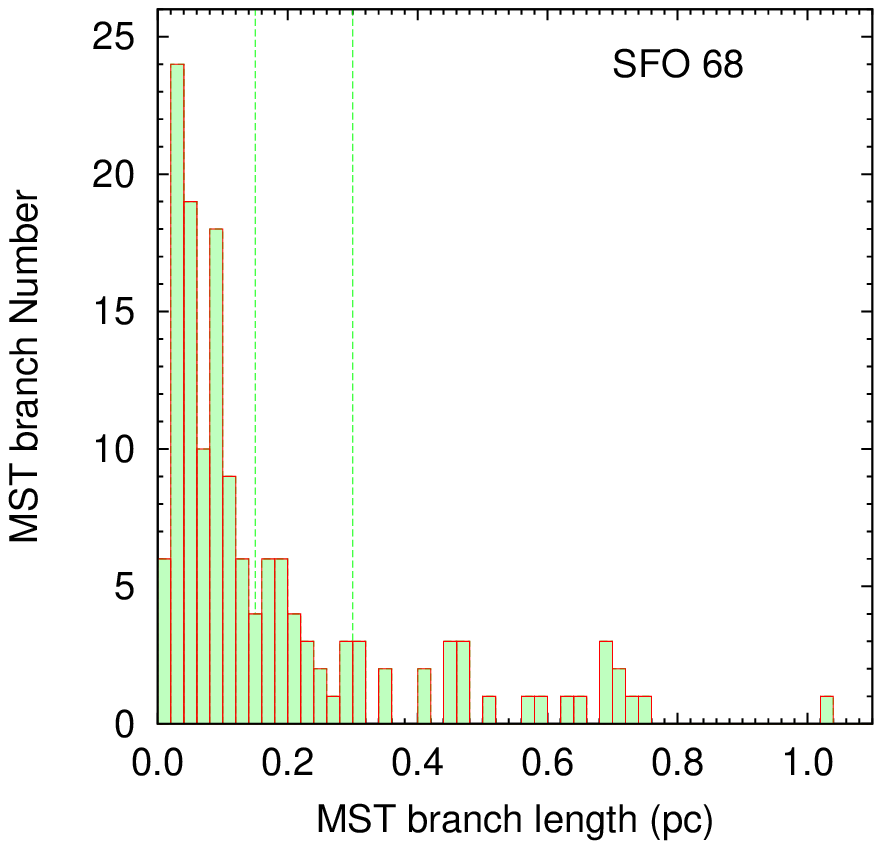}
\centering\includegraphics[height=3.4cm,width=4.0cm]{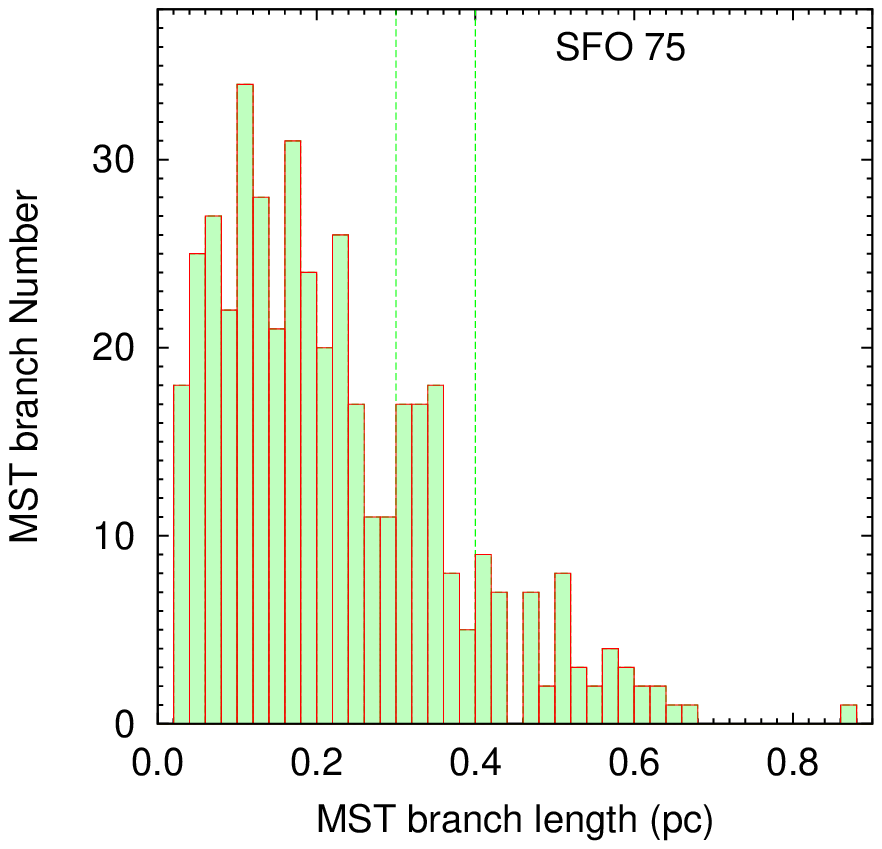}
\centering\includegraphics[height=3.4cm,width=4.0cm]{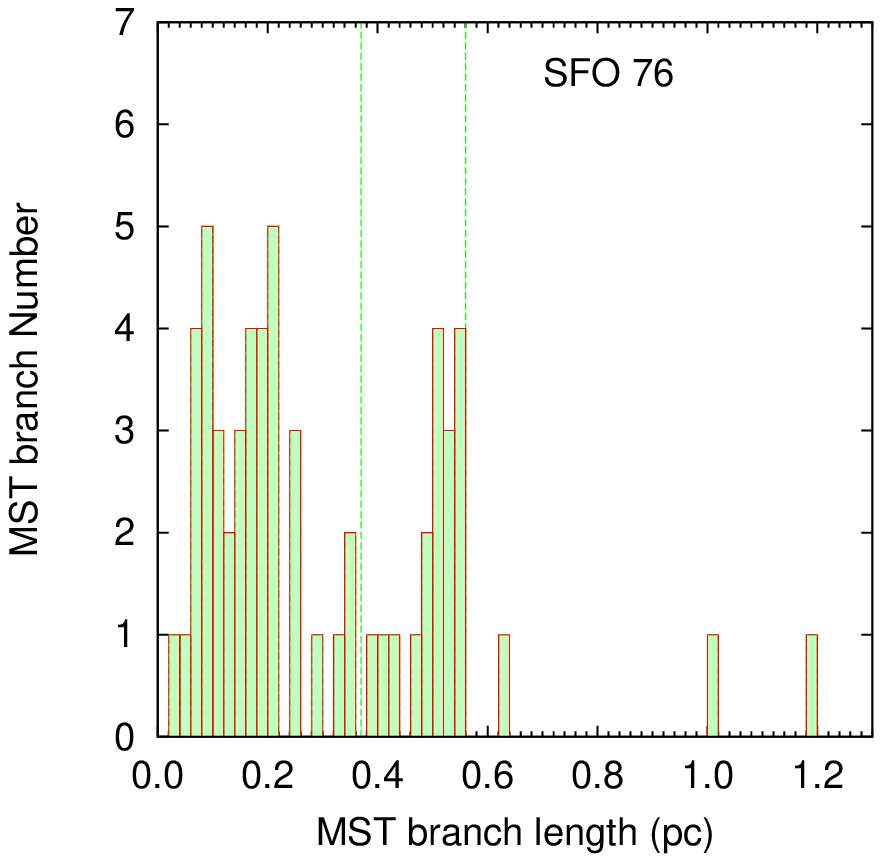}
\centering\includegraphics[height=3.4cm,width=4.0cm]{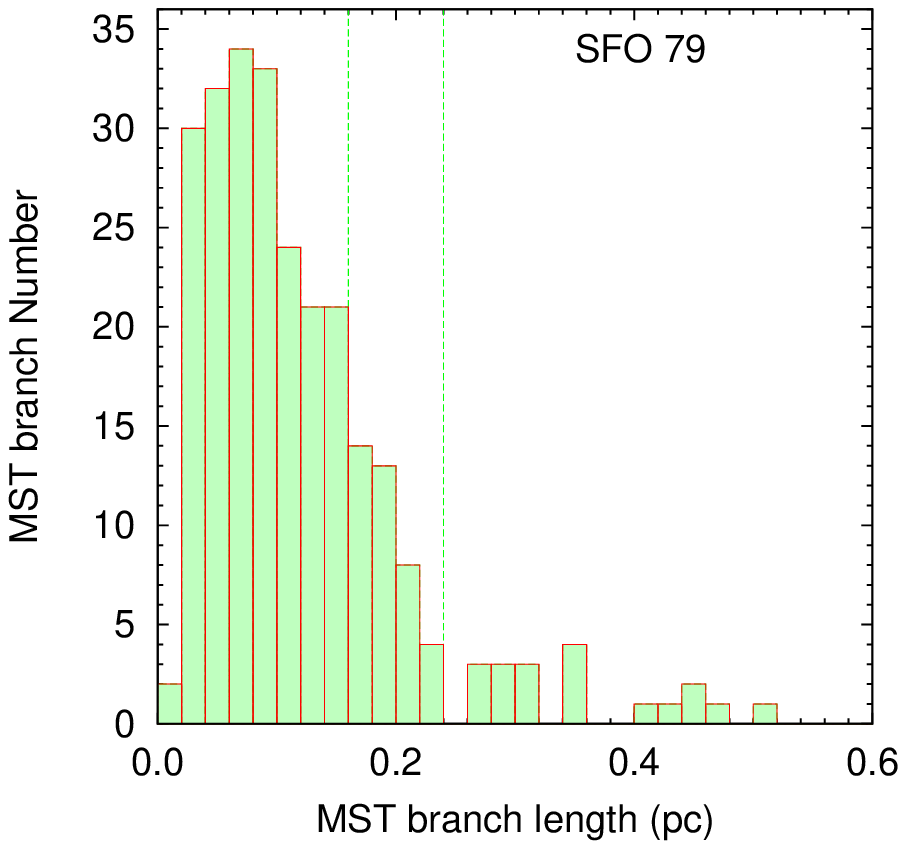}
\caption{\label{Fmst}
Cumulative distribution functions (CDFs) and histograms of MST branch lengths used for critical length analyses of the YSOs.
The CDF plots have  sorted length values on the horizontal axis and a rising integer counting index on the vertical axis. 
The red solid line is a two-line fit to the CDF distribution. 
The inner and outer vertical green lines stand for critical lengths obtained
for the core and the active region, respectively.
 }
\end{figure*}

In order to isolate the sub-structures, we have to adopt a surface density threshold expressed by a critical branch length.
In Fig.~\ref{Fmst}, we have plotted histograms between MST branch lengths and MST branch numbers for the YSOs. 
From this plot, it is clear that they have a peak at small spacings 
and a relatively long tail towards large spacings. These peaked distance distributions
typically suggest a significant sub-region (or sub-regions) above a relatively uniform, elevated surface density. 
By adopting an MST length threshold, we can isolate those sources which are closer
than this threshold, yielding populations of sources that make up local surface density
enhancements. To obtain a proper threshold distance, we have used the similar approach as already demonstrated
by \citet{2009ApJS..184...18G}.
In Fig.~\ref{Fmst}, we have also plotted the cumulative distribution function (CDF) for the branch length of YSOs for MST. 
In this distribution, we can see three line segments: 
a steep-sloped segment at short spacings, a transition segment that approximates curved character 
of the intermediate length spacings, and a shallow-sloped segment at long spacings \citep[cf.][]{2009ApJS..184...18G}. 
Typically, most of the sources are found in the steep segment, 
where the spacings are small, e.g., in a core of the stellar distribution.
Therefore, to isolate the core region in the BRCs, we have fitted by two true lines in shallow and steep 
segment of the CDF and extended them to connect together. 
We adopted the intersection point between these two lines as the MST critical branch length, 
as shown in Fig. \ref{Fmst} \citep[see also,][]{2009ApJS..184...18G}.
The BRC cores were then isolated from the lower density distribution by clipping
MST branches longer than the critical length found above.
Similarly, we have enclosed all the YSOs associated with the BRC by selecting the point where the 
curved transition segment meets the  shallow-sloped segment at longer spacings.
We have called this region in the BRC as its active region where recent star formation took place or 
contains YSOs moved  out from the cores due to dynamical evolution.
Black dots and black MST connections in Figs. \ref{Fall54} - \ref{Fall79} (bottom left panels)  represent the more closely spaced YSOs
than the critical length. In this way we can very easily pick up the major groupings of the 
YSOs along with some other sub-groupings scattered in the regions.
We have then plotted the respective convex hull \citep[cf.][]{2009ApJS..184...18G} for these cores and for the whole active SFR region in Figs.  \ref{Fall54} - \ref{Fall79}, lower panels
(solid red and solid grey lines, respectively).
The physical details of these sub-groups (cores) and the active regions are given in Tables \ref{Tp1} - \ref{Tp3}. 
The median value of the critical branch lengths for the cores and the active regions are 0.18 pc and 0.30 pc, respectively.

\section{Discussion}

\subsection{General trends in the spatial distributions}

In this Section, we investigate the distribution of the YSOs, their separation, and their relation to the associated molecular clouds as a function of their evolutionary status.

\subsubsection{YSOs and their association with the surrounding molecular cloud}

The identification of YSOs (cf. \S 3)  in  a sample of eight BRCs which are classified  
as triggered by \citet{2009A&A...497..789U}  reveals recent star formation in them.
A newer observational characterization of the relationship
between the spatial distribution of the YSOs
and their associated cloud material is vital to understand the nature
of their spatial distribution, to constrain the model of star formation, 
and to ascertain the underlying physics 
\citep[e.g.,][]{2005MNRAS.356.1201B,2007ApJ...654..304K,2009ApJ...706.1341M,2011ApJ...739...84G}.
We have studied the spatial distribution of the YSOs in and around the BRCs by
superimposing them on the  $\sim10^\prime\times10^\prime$ color-composite image obtained from the 8.0 $\mu$m (red);  3.6 $\mu$m (green)
and  $K$ (blue) band images (Figs. \ref{Fall54} - \ref{Fall79} (top left panels)).
The yellow and red dots are Class I and Class II objects, respectively.
The distribution of gas and dust as seen by the MIR emissions along with the 
several concentrations of YSOs can be easily seen in the images.
The distribution of YSOs reveals that a majority of Class I sources belong generally
to these concentrations, whereas the comparatively older population, i.e. Class II objects, are rather randomly distributed throughout the region.

Star formation usually takes place inside the dense cores of molecular clouds and
the YSOs often follow clumpy structures of their parental molecular clouds
\citep[see e.g.][]{1993AJ....105.1927G, 1996AJ....111.1964L, 1998A&A...336..150M, 2002ApJ...566..993A, 2005ApJ...632..397G, 2008ApJ...674..336G, 2006ApJ...636L..45T, 2007ApJ...669..493W}.
The IR extinction maps can be used to represent the column density distribution of the molecular cloud
associated with BRCs \citep[cf.][]{2009ApJS..184...18G, 2013MNRAS.432.3445J}.
We  have compared the isodensity contours of YSOs with the extinction maps (Figs. \ref{Fall54} - \ref{Fall79}; top right panels), 
and found that most of the YSOs are distributed in groups in the regions of detectable extinction. 
\citet{2011ApJ...739...84G} also found similar trends in eight nearby molecular clouds.
They have reported a power-law correlation between the local surface densities of
YSOs and the column density of gas (as traced by NIR extinction), agreeing with the
prediction of the thermal fragmentation of a sheet-like isothermal layer. 
If we compare the extinction contours with the IRAC  8.0 $\mu$m images (Figs. \ref{Fall54} - \ref{Fall79} (top left panels)), they roughly follow each other's distribution except in one or two BRCs, but the peaks do not match well. 
IRAC  8.0 $\mu$m band includes the PAH emission and
its intensity would be the strongest in the photo-dissociation region (PDR) which faces the ionizing source. 
The extinction contours, on the other hand, would indicate the distribution of the molecular cloud.  
Therefore, there may be instances where the IRAC and extinction maps don't match well. 
Also, the cut-off in the  8.0 $\mu$m intensity and extinction contours can raise some differences in these two distributions.

\citet{2008MNRAS.389.1209S} studied the spatial distribution of different classes of YSOs in embedded
clusters and found that they mostly evolve from a hierarchical to a more centrally concentrated distribution.
\citet{2008ApJ...674..336G} have shown that the  sources in each of the Class I and Class II evolutionary states have very 
different spatial distributions relative to the distribution of the dense gas in their natal cloud.
We have also compared the extinction maps with the positions of YSOs of different evolutionary status and found that
the Class I sources are located towards places with higher extinction (cf. Figs. \ref{Fall54} - \ref{Fall79}, top panels).
These properties agree well with previous findings in the W5 region \citep{2008ApJ...688.1142K,2012A&A...546A..74D}
and with the assumed evolutionary stages of both classes: the
younger Class I sources are more clustered and associated with the
most dense molecular material in which they were born, while the
Class II sources are scattered probably by moving away from their birthplaces.
\citet{2014MNRAS.439.3719C} showed that their samples of embedded clusters are likely gravitationally unbound,
supporting the result that the more evolved members move further away due
to the weak gravitational well of the cluster.

\subsubsection {Scattered YSOs population }

Many ground-based near-IR surveys of molecular clouds 
\citep[e.g.,][]{1991ApJ...368..432L,1993ApJ...412..233S,2000ApJS..130..381C,2003AJ....126.1916P,2003ARAA..41...57L}
have shown that the molecular clouds contain both, a dense
`clustered' and a diffuse `distributed' population.
\citet{2008ApJ...688.1142K} have analyzed the clustering properties of objects classified as young stars across 
the W5 region and found 40-70\% of these sources belong to groups with $\geq$10 members
and remaining were termed as scattered populations.
Although the cluster cores of the BRCs have sizes of the order of a parsec \citep{2006MNRAS.369..143M}, 
the stars in them may have moved off from their formation site in  few Myr. 
\citet{2011MNRAS.410.1861W} using N-body calculations, 
studied the numbers and properties of escaping stars from low number (N=100 and 1000)  young embedded star 
clusters prior to gas expulsion over the first 5 Myr of their existence.  
They have found that these clusters can lose substantial amounts (upto 20\%) 
of stars within 5 Myr. In the present sample of BRCs (except SFO 75), the cores have $<$100 stars as their members.
These stars probably have mean velocity of $\sim$2 km s$^{-1}$ \citep{2011MNRAS.410.1861W} and can
travel the distance of $\sim$2 - 6 pc during the 1-3 Myr of their formation \citep[the typical age of the BRC YSOs:][]{2009MNRAS.396..964C,2011MNRAS.415.1202C,2014MNRAS.443.1614P}.  
Therefore, for the $\sim10\times10$ arcmin square FOV of ISPI, we expect 5-10\% of the stars would have 
escaped from the core region and are not included in our analyses.
We have calculated the fraction of the scattered YSOs population (the YSOs outside the cores, but in the active regions)
and found  they are between $\sim$20-45 \% of the total YSOs in the whole active region.
Similar numbers have also been found in other studies also \citep[30-50 \% 
:][]{2014MNRAS.439.3719C}.
Out of these YSOs in the outer regions, 62 \% 
are Class II objects, which is more or less similar to previous findings \citep[cf. 67 \% 
,][]{2014MNRAS.439.3719C}.
This higher percent of the comparatively older  population in the outer regions is in accordance with 
 their dynamical evolution. \citet{2014MNRAS.439.3719C} have argued that the 10-20 \% of the scattered population in their sample 
were the  members of the cluster and happened to move away because of the dynamical relaxation.
In addition to the above, the explanation of the scattered populations may include: 
dynamical interaction between cluster members, small groups
merging, cluster definition and isolated star formation \citep[for details, cf.][]{2014MNRAS.439.3719C}.

\subsubsection{YSOs spacings}

The complex patterns  (e.g. filaments, bubbles and irregular clumps etc.) found in YSO 
population in SFRs
are the result of the interplay of fragmentation process, turbulence, magnetic
fields, crossing the Galactic arms' potential, activities of massive stars in a region etc.
Fragmentation of the gas with turbulence \citep[e.g.][ and refs. therein]{2007prpl.conf...63B} and
magnetic fields \citep[e.g.][and refs. therein]{2007prpl.conf...33W} have been discussed, leading to detailed
predictions of the distributions of fragment spacings. 
Observations of SFRs 
\citep[see][]{2005ApJ...632..397G, 2006ApJ...636L..45T} suggest a strong peak in their
histograms of NN spacings of the protostars in young embedded clusters. This peak indicates
a significant degree of Jeans fragmentation, since this most frequent spacing agrees with an
estimate of the Jeans length for the dense gas within which YSOs are embedded.  

From the histograms (cf. Fig. \ref{Fnn2}) of NN2/NN6, it is clear that all the BRCs in the current sample have peaks at small spacings 
and they have a relatively long tail of large spacings. Although the peak may be sharp,
or broad, or one of several near equivalent peaks, such a peaked
character is often observed, regardless of the two dimensional
distribution of sources  \citep{2009ApJS..184...18G}.
Peaked NN2 distance distributions typically suggest a significant subregion 
(or subregions) of relatively uniform, elevated surface density  \citep{2009ApJS..184...18G}.
\citet{2009ApJS..184...18G} in their study of 36 star-forming clusters, have also observed that 
short spacings are relatively more frequently than longer ones.
Their sample shows a well defined peak at 0.02-0.05 pc and a tail extending to the spacings of 0.2 pc or greater. 
All the BRCs in the present study also show a well-defined single peak with extended spacing upto 1 pc in their respective histograms of NN2/NN6 distributions.
Since NN2 is more sensitive to the local density fluctuations, it shows the median value of 0.03 pc over all BRCs for the peak of this distribution, whereas,
NN6, which is an indicator of larger scale fluctuation, shows a comparatively larger peak value of 0.19 pc.

\subsection{Physical properties of YSO cores and the active regions in the BRCs}

In this Section, we investigate the physical properties of the identified active regions/cores 
in the present sample of BRCs.

\subsubsection{ Class I versus Class II distribution}

We have calculated the median values of $A_K$ as $\sim$1.3 mag and $\sim$1.0 mag
and the YSOs separation as 0.07 pc and 0.11 pc for the Class I and Class II sources, 
respectively, in all cores of the studied BRCs.
Similar values of these quantities have been found even for the active regions (cf. Table \ref{Tp3}).
Therefore, we can conclude that the Class I sources are located towards the places with 
higher extinction  and are relatively closer to each other than the Class II sources
(for details, see Table \ref{Tp3} and Figs. \ref{Fall54} - \ref{Fall79} (top left panels)).
We have also calculated the fraction of the Class I objects among all the YSOs (cf. Table \ref{Tp1}, last column) as an indicator of the "star formation age" of a region.
The median values for this in the cores and  in the active region are 66\% and 52\%  (cf. Table \ref{Tp3}), respectively.
If we calculate this fraction only for the outer active region, excluding the inner cores, it falls to 38\%.
If we include only the YSOs being categorized by the $Spitzer$ data (cf. Table \ref{Tp3}), we still find a similar trend
indicating higher percentages of younger sources in the inner regions of BRCs having high column densities.  
This is also in agreement with the conclusions of \citep{2009ApJS..184...18G, 2011ApJ...739...84G} that
protostars are found in regions having marginally higher stellar surface densities than the more evolved PMS stars.

\subsubsection{YSO surface densities}

\begin{figure*}
\centering\includegraphics[height=7cm,width=9cm,angle=0]{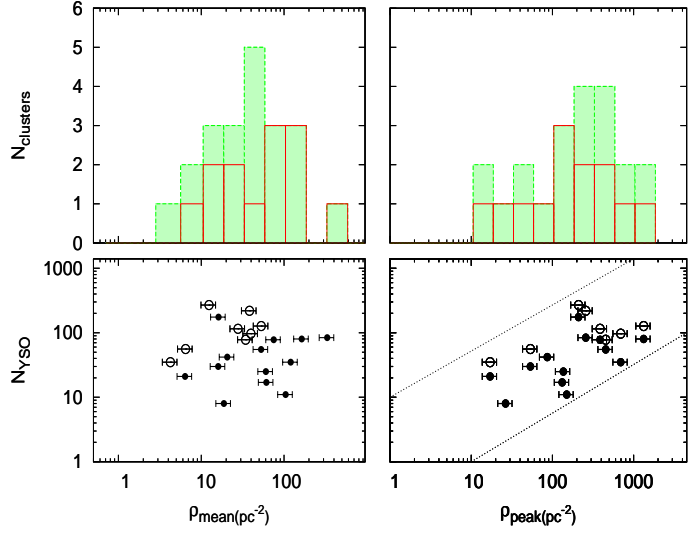}
\caption{\label{Fdensity} 
Histogram showing the  mean YSOs surface density (Upper Left Panel) and
the plot of the mean YSOs surface density versus the number of cluster members (Lower Left Panel). 
The red solid histogram and filled circles are for the cores, and the green dotted histogram
and open circles are for active regions.
(Right Panels): Same as Left, but for the  peak YSOs surface density distribution.
Dotted lines in the lower-right chart enclose all the regions 
with a slope of 0.8 as given  in \citet{2014MNRAS.439.3719C}.
 }
\end{figure*}

The YSOs in our cluster sample have mean surface densities mostly between 10 and 300 pc$^{-2}$ (see Table \ref{Tp2} and Fig. \ref{Fdensity})  
for the cores and the active regions, respectively. 
These values are in agreement with the values given by \citet{2009ApJS..184...18G} for their sample of low-mass embedded clusters (LECs).
The median values for the surface densities for the cores and the active region come out 
to be around 60 pc$^{-2}$ and 28 pc$^{-2}$, respectively.
The peak surface densities vary between 17 -1330 pc$^{-2}$ for our sample (cf. Table \ref{Tp2} and Fig. \ref{Fdensity}).
Cores show a peak in the distribution of peak surface density at around 150 pc$^{-2}$ (cf. Fig. \ref{Fdensity}).
\citet{2014MNRAS.439.3719C} in their sample of embedded clusters found a weak trend between the peak surface density 
and the number of cluster members, suggesting that the clusters are better characterized by their peak YSOs surface density.
In the present sample, the YSOs also follow a similar correlation (cf. Fig. \ref {Fdensity}).

\subsubsection{Core morphology}

\begin{figure*}
\centering\includegraphics[height=7cm,width=9cm,angle=0]{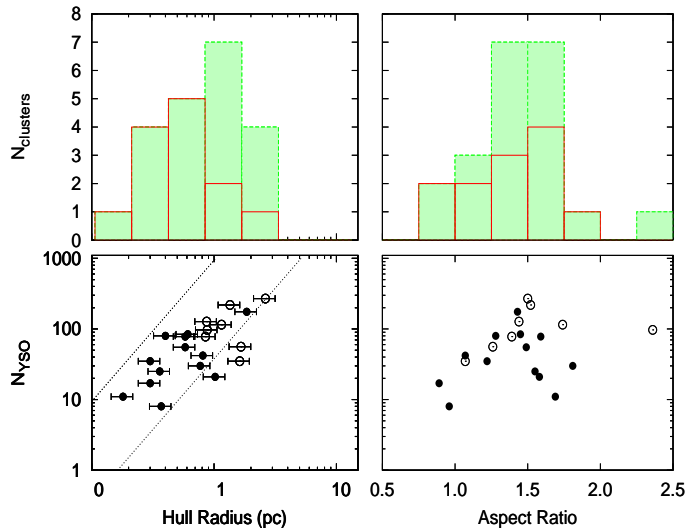}
\caption{\label{Fhull} 
Histogram showing the hull radius distribution (Upper Left Panel) and
the plot of the hull radius versus the number of cluster members (Lower Left Panel). 
The red solid histogram and filled circles are for the cores, and the green dotted histogram
and open circles are for active regions.
The doted lines in the lower-left panel represent the 
constant surface densities at 12 and 300 pc$^{-2}$. Those
correspond to the range spanned by the embedded clusters from \citet{2009ApJS..184...18G}.
(Right Panels): Same as Left, but for the aspect ratio distribution.
 }
\end{figure*}

The groups of young stars in SFRs show a wide range of sizes, 
morphologies and star numbers \citep[cf.][]{2008ApJ...674..336G,2009ApJS..184...18G,2011ApJ...739...84G,2014MNRAS.439.3719C}.
Recently \citet{2014ApJ...787..107K} studied 142 sub-clusters in different SFRs, 
and found their elongated morphologies with the core radius peaking at 0.17 pc.
We use the clusters convex hull radius ($R_H$) and aspect ratio  to
investigate their morphology (see Table \ref{Tp2} and Fig. \ref{Fhull} ). The $R_H$ values of the cores range
between 0.2 and 2.0 pc with a median value of 0.6 pc (cf. Table \ref{Tp3}). 
These values are similar to those reported by \citet{2014MNRAS.439.3719C} for a sample of LEC (0.5 pc).
Most of the cores and active regions in the present sample are also found within a range of
constant surface density of 12 - 300 pc$^{-2}$ (cf. Fig. \ref{Fhull}), as reported by \citet{2009ApJS..184...18G}.
Almost all the cores in the present sample show an elongated morphology with the median value of  the aspect ratios around 1.45.

The median number of YSOs in cores and in active region are 35 and 97 (cf. Table \ref{Tp3}), respectively, for the present sample of BRCs. 
The median MST branch length for these cores is found to be 0.09 pc.
The total sum of YSOs in the active regions for all the BRCs is 997, out of which 602 (60\%) falls in the cores.
These numbers are very similar to those given in literature: 62\% \citep{2009ApJS..184...18G} and 66\% \citep{2014MNRAS.439.3719C}.

\subsubsection{Structural $Q$ parameter}

The spatial distribution of YSOs associated with the BRCs is also investigated
by their structural $Q$ parameter values.
The $Q$ parameter 
\citep{2004MNRAS.348..589C,2006A&A...449..151S}
is used to measure the level of
hierarchical versus radial distributions of a set of points, and it is defined by
the ratio of the MST normalized mean branch length and the normalized mean separation between points \citep[cf.][for details]{2014MNRAS.439.3719C}.
By using the normalized values, the $Q$ parameter 
becomes independent from the cluster size \citep{2006A&A...449..151S}.
According to \citet{2004MNRAS.348..589C}, 
a group of points distributed radially will
have a high $Q$ value ($Q$ $>$ 0.8), while clusters with a more fractal
distribution will have a low $Q$ value ($Q$ $<$ 0.8).

\begin{figure*}
\centering\includegraphics[height=7cm,width=8cm,angle=0]{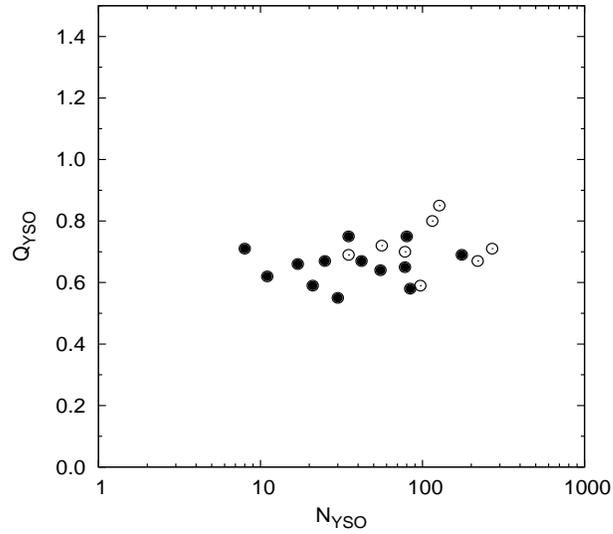}
\caption{\label{Fq} Structural $Q$ parameter ($Q$$_{\rm YSO}$) for the YSOs in the cores (filled circles) 
and in the active regions (open circle).
  }
\end{figure*}

We find that our sample of BRCs has median $Q$ values less than 0.8 (0.66 in cores and 0.70 in active regions, cf. Tables \ref{Tp2} and \ref{Tp3}), 
showing a more fractal distribution especially in the inner regions of BRCs.
\citet{2014MNRAS.439.3719C} have found a weak trend in the distribution of $Q$ values per number of members,
suggesting a higher occurrence of sub-clusters merging
in the most massive clusters, which decreases the value of the $Q$ parameter.
For our sample we didn't find any such correlation (cf. Fig. \ref{Fq}).

We compared the $Q$ parameters for Class I and Class II sources (see Table \ref{Tp3}), and found that the Class I sources 
are distributed more hierarchically than the Class II sources ($Q$$_{\rm Class~I} <$ $Q$$_{\rm Class~II}$). 
A similar result is shown by  \citet{2014MNRAS.439.3719C} and \citet{2008MNRAS.389.1209S}
for low-mass SFRs and is likely a consequence of the cluster's dynamic relaxation.

\subsubsection{Associated molecular material}

\begin{figure*}
\centering\includegraphics[height=7cm,width=9cm,angle=0]{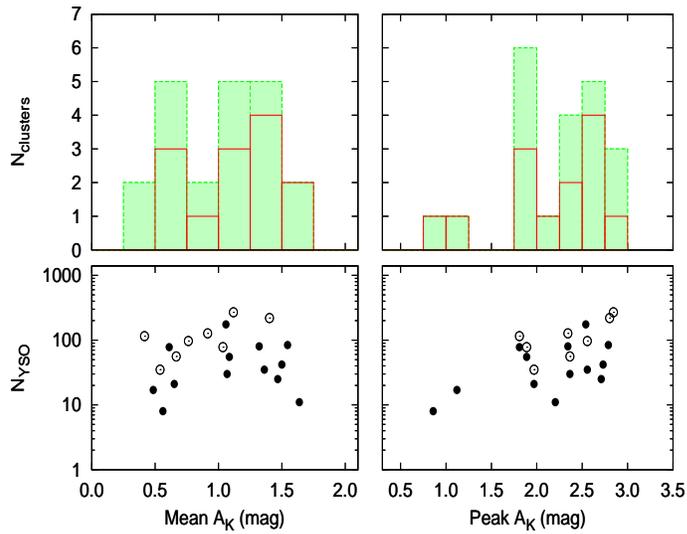}
\caption{\label{Fak} 
Histogram showing the  mean  $K$-band extinction (Upper Left Panel) and
the plot of the mean  $K$-band extinction versus the number of cluster members (Lower Left Panel). 
The red solid histogram and filled circles are for the cores, and the green dotted histogram
and open circles are for active regions.
(Right Panels): Same as Left, but for the  peak $K$-band extinction distribution.
 }
\end{figure*}

The  mean $A_K$ values for the identified cores 
have been found to be in between 0.5 and 1.6 mag, with a median value of  1.1 mag (cf. Table \ref{Tp3} and Fig. \ref{Fak}). 
This is very similar to the values given by \citet{2014MNRAS.439.3719C} for LECs.
However, the peak value of $A_K$ (2.3 mag) for the current sample is higher than the value (1.5 mag) reported for LECs by \citet{2014MNRAS.439.3719C}.
We have found a weak correlation between the peak $A_K$ and the number of cluster members (cf. Fig. \ref{Fak}).
The median $A_K$ value for the whole active region is 0.8 mag, which is lower than the core value, naturally indicating the
distribution of higher density of YSOs towards the high density molecular clouds.

\subsubsection{Jeans Length}

\citet{2009ApJS..184...18G} analyzed  the YSOs spacings of the YSOs in the stellar cores of 36 star-forming clusters and 
suggested that Jeans fragmentation is a starting
point for understanding the primordial structure in SFRs.
We have also calculated the minimum radius required for the gravitational collapse of a 
homogeneous isothermal sphere (Jeans length `$\lambda_J$')
in order to investigate the fragmentation scale by using the formulae given in \citet{2014MNRAS.439.3719C}.

\begin{figure*}
\centering\includegraphics[height=7cm,width=8cm,angle=0]{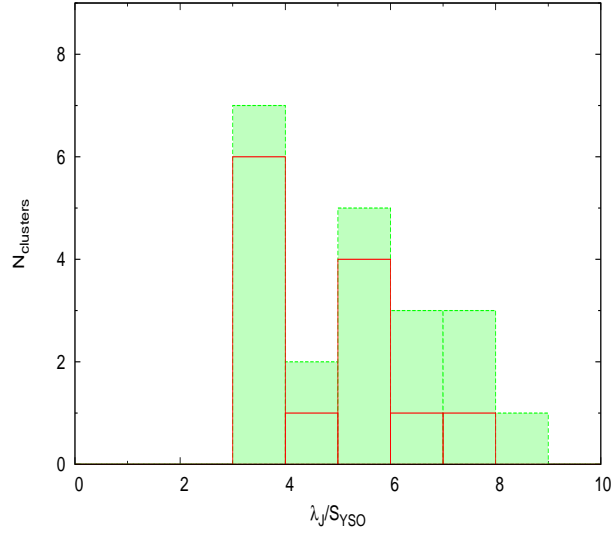}
\caption{\label{Fjeans} Histogram showing the distribution of the ratio between the cluster Jeans 
length ($\lambda_J$) and the mean projected distance between the members of cores and active regions.
The red solid histogram is for the cores, and the green dotted histogram
is for active regions.
}
\end{figure*}

The  Jeans length $\lambda_J$ for the cores in the current study has 
values between 0.2 - 1 pc, with a median value of 0.46 pc (cf. Tables \ref{Tp2} and \ref{Tp3}).
We have also compared $\lambda_J$ and the mean separation `$S_{YSO}$' between cluster members (Fig. \ref{Fjeans})
and found that the ratio $\lambda_J/S_{YSO}$ has an average value of $4.9\pm1.2$.
Similarly,  \citet{2014MNRAS.439.3719C} reported the ratio for their sample of embedded clusters as $4.3 \pm 1.5$.
Present results agree with a non-thermal driven fragmentation since it took place at
scales smaller than the Jeans length \citep{2014MNRAS.439.3719C}. 

\subsubsection{Molecular content}

We have calculated the  molecular mass of the identified cores/active 
regions using the extinction maps generated in the \S 3.3. 
First, we have converted the average $A_V$ value (corrected for the 
foreground extinction) in each grid of our map into $H_2$ column density 
using the relation given by  \citet{1978ApJS...37..407D,1989ApJ...345..245C},  i.e.
$\rm N(H_2) = 1.25\times10^{21} \times A_V ~~cm^{-2}~~ mag^{-1}  $.
Then, this $H_2$ column density has been integrated over the convex hull of each region
and multiplied by the $H_2$ molecule mass to get the molecular mass of the cloud.
The extinction maps used for this are shown in  Figs. \ref{Fall54} - \ref{Fall79}.
The extinction law, $A_K/A_V=0.090$ \citep{1981ApJ...249..481C} has been used to convert $A_K$ values of our maps to $A_V$.
Foreground contributions have been corrected by using the relation: 
$A_{K_{foreground}}$ =0.15$\times$D \citep[][D is distance in kpc]{2005ApJ...619..931I}.  
The properties of the molecular clouds associated with the cores and active regions
are listed in Table \ref{Tp2}.
The cores in the present sample show a wide range in their cloud mass distribution 
($\sim$20 to 2400 M$_\odot$), with a median value around 130  M$_\odot$.

\begin{figure*}
\centering\includegraphics[height=7cm,width=8cm,angle=0]{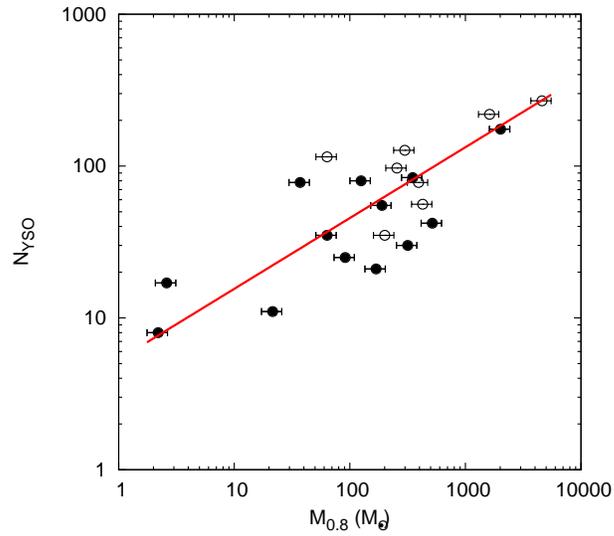}
\caption{\label{Fmass} Relation between the number of YSOs `$N_{YSO}$' and the molecular mass above
$A_K$ = 0.8 mag (M$_{0.8}$) in the cores and the active regions. The solid line shows the best fit to the data points.
  }
\end{figure*}

\citet{2010ApJ...724..687L} found that the number of YSOs in a
cluster are directly proportional to the dense cloud mass M$_{0.8}$ 
(the mass above a column density equivalent to $A_K \sim$ 0.8 mag) with a slope equal to one.
This gives an  empirical relation between the content of YSOs and their parent molecular cloud.
Recently, \citet{2014MNRAS.439.3719C} have checked the same relation
for their sample of embedded clusters and found a similar slope of $0.89\pm0.15$ 
between the mass of the dense cloud and the number of cluster members.
This suggests that the star formation rates depend linearly on the mass of the dense cloud \citep{2010ApJ...724..687L}. 
We have also calculated the M$_{0.8}$  \citep[cf. Table \ref{Tp2}, as explained in][]{2014MNRAS.439.3719C} for our sample of BRCs.
We find that the number of YSOs is proportional to M$_{0.8}$ with a slope of 0.47 (cf. Fig. \ref{Fmass}),
which is very similar to the slope of 0.5, as calculated by \citet{2014MNRAS.439.3719C},
when  they have not corrected their sample for the non-detection of fainter members.
Once they applied the corrections for these non-detections, they found an almost similar slope ($0.89\pm0.15$) as given by  \citet{2010ApJ...724..687L}. 
The data used and the distances of the present sample of BRCs (1 - 2.7 kpc) are more or less similar to 
those embedded clusters (1.6 - 2.7 kpc) studied in \citet{2014MNRAS.439.3719C}.
We therefore expect a similar level of non-detection for our sample.
Hence, it may give a similar relation as that of  \citet{2010ApJ...724..687L}.
In summary, we can at-least say that there is a linear 
relation between the dense cloud mass and the number of YSOs for the present sample of BRCs.

\subsubsection{Star formation efficiency}

The wide range in observed YSO surface densities provides
an opportunity to study how this quantity is related to the
observed star formation efficiency (SFE) and the properties
of the  associated molecular cloud \citep{2011ApJ...739...84G}. Recent works indicate that 
SFE increases with the stellar density. \citet{2009ApJS..181..321E} 
showed that YSO clusterings of higher surface density
tend to exhibit higher SFE (30\%) than their lower
density surroundings (3\%-6\%). 
\citet{2008ApJ...688.1142K} found SFEs of $>$10\%-17\% for high surface
density clusterings and 3\% for lower density regions. 

We have calculated the SFE, defined as the percentage of
gas  mass converted into stars by using the cloud mass derived from $A_K$  
inside the cluster convex hull area and the number of YSOs found in the same area \citep[see also][]{2008ApJ...688.1142K}.
For simplicity, we have assigned 0.5 solar mass \citep[cf.][]{2008ApJ...688.1142K} to each of the 
identified YSOs and found SFEs between 3 and 30 \% with an average of $\sim$14 \%.
\citet{2014MNRAS.439.3719C} have obtained the SFE of a range of 3-45 \% with an average 20 \% 
for the sample of embedded clusters.
Our results are comparable to those of \citet{2008ApJ...682..445C} for the S254 region (4-33 \% range and 10 \% average) 
with the exception of  G192.54-0.15.
Although the cluster G192.54-0.15 is located inside the H II region, S254 having distance of 2.4 kpc, 
these authors have concluded that it is located in the background of the complex 
at a distance 9 kpc, based on its kinematics using the rotational model 
from \citet{1993A&A...275...67B}. The SFE of G192.54-0.15 is 54\% for the distance of 2.4 kpc then
decreases to 1\% if we adopt 9 kpc.

\begin{figure*}
\centering\includegraphics[height=7cm,width=8cm,angle=0]{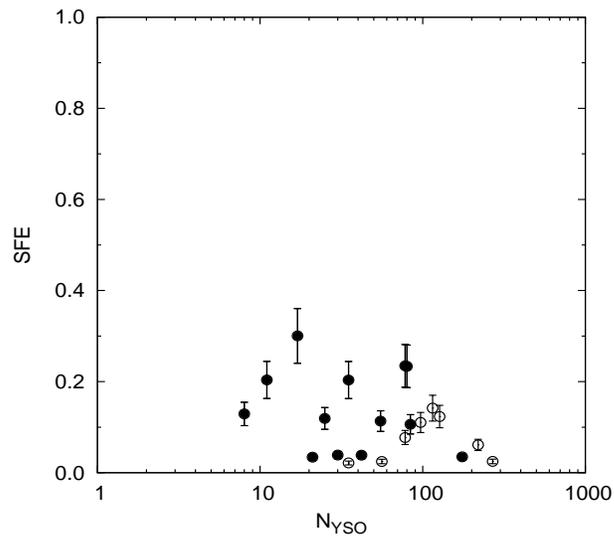}
\caption{\label{Fsfe} Star formation efficiency in the cores (filled circles)
and in the active regions (open circles) with respect to the number of YSOs `$N_{YSO}$'. 
  }
\end{figure*}

The present SFE values are in agreement with the efficiencies needed to go
from the core mass function to the initial mass function
\citep[e.g. 30 per cent in the Pipe nebula and 40 per cent in Aquila, from][respectively]{2007A&A...462L..17A, 2010A&A...518L.102A}.
The SFE distribution as a function of the number of the cluster member of each region is shown in Fig. \ref{Fsfe}. 
There seem to be no correlation between them. This suggests that 
feedback processes may start having an impact only in the later stages of the cluster evolution

\subsection{Biases}

\subsubsection{Effect of uncertainty in distance on the analyses}

Since, the distances of the BRCs  studied in the present survey have 
been taken from the literature (cf. \S 1),
which are essentially the photometric distances of the bright ionizing source(s) with typical uncertainties  
of the order of $\sim$10\%, this will have corresponding effects 
on the various parameters given in Table 5 and Table 6. For example,  
as for $R_{hull}$, $R_{circ}$, $\rho$, MST/NN2 length and $D_{crit.}$, 
the dependence on the distance is simple and linear, therefore the percentage error 
associated with them due to the error in distance would be of the same order. 
The uncertainty due to it would be the largest in the derived mass, 
stellar density and SFE ($\sim$20\%), whereas the Jeans length would be in error by 15\%.
We have also plotted these errors in the Figs. 19, 20, 24 and 25. 
There are no significant changes in the distributions in the figures.

\subsubsection{Effect of non-detection of YSOs on the analyses}

The detection of YSOs is affected by the presence of bright infrared sources, 
bright nebulosity, and as well as by the high crowding in the BRCs, 
and we expect varying degrees of the detection completeness as a function of 
the above factors.
To quantitatively analyze these effects, we followed the same approach as 
demonstrated by \citet{2014MNRAS.439.3719C}. 
Percentage of the corrected number of cluster members
($N^*_{YSO}$) can be estimated from the equation given by  \citet{2014MNRAS.439.3719C} i.e.,

$N^*_{YSO} = {{N_{YSO}/(1 - IR)}\over{1-k}}$,

where $IR$ is the percentage of cluster members without IR-excess and $k$ is the percentage of 
synthetic stars with $K$ magnitudes fainter than 90\% completeness limit of a synthetic 
cluster having similar properties as the observed clusters. 
 \citet{2014MNRAS.439.3719C} found that $k$ varies on average by 10\% for their sample of 
embedded clusters. The percentage of cluster members without IR-excess in the present sample 
of BRC should be of the order of 10\% (cf. \S 3.1). 
Therefore, the percentage of lost YSOs will be around 20\% of the total YSOs  identified in the
present survey in each BRC. This bias will have a considerable effect on the 
reported peak YSO surface densities as well as on the higher density portions of the nearest 
neighbor surface density maps. We expect only a modest effect on the mean YSO surface 
densities and overall member counts. 
As the completeness decreases, we expect to lose preferentially those stars which are deeply embedded in the nebulosity of the BRCs
and most of them are likely of younger evolutionary class (Class I) located in the cores.
Therefore, most probably, adding non-detected stars might complement some of the results in this study.

\section{Conclusion}

We carried out deep and wide field NIR observations of eight southern BRCs
previously categorized as triggered candidates by \citet{2009A&A...497..789U}
 using the ISPI camera on the 4m CTIO Blanco telescope. 
These data along with the $Spitzer$ archive data have been used to identify deeply embedded low mass YSOs.
Spatial distribution of these YSOs has been compared with their associated 
molecular clouds which have been inferred by the extinction analysis. 
Quantitative techniques have been used for analyzing the spatial structure of the YSO aggregates.
The main results of our studies are as follows:

\begin{itemize}

\item
We have identified a large number of YSOs, implying recent star formation in each of these BRCs. 
The number of the YSOs ranges from 44 to 433.
This sample of YSOs is complete down to sub-solar masses.
We have classified these YSOs based on their excess emission in IR.

\item
Isodensity contours of the YSOs 
and NIR extinction maps for the regions studied have been 
generated and compared  with those of other SFRs.
The YSOs are generally distributed in groups in the regions of 
higher extinction, peaking at or near the peak of the molecular column density. 
A majority of younger population, i.e., Class I sources belong 
to these groups, whereas the comparatively 
older population, i.e., Class II objects, are more randomly distributed throughout the regions.
This distribution is in accordance with the notion the star formation usually takes place inside the dense cores of the molecular clouds and
the YSOs often follow their clumpy structures.

\item
19-46 percent of the YSOs  in the present sample of BRCs are a scattered population.
Out of these, 62 percent are comparatively older Class II objects  (cf. 38 percent in the cores).
This higher percent in the outer regions can be attributed 
to their dynamical evolution. 

\item
The concentrations of YSOs in our sample have similar projected NN 
spacings as in  many other embedded cluster cores, having a well defined single peak 
which indicates a significant degree of Jeans fragmentation.
NN2,  being  more sensitive to the local density fluctuations, shows a median 
value of 0.03 pc for the peak of their histogram, whereas,
NN6, which is an indicator of larger scale fluctuation, shows a comparatively larger peak value (0.19 pc).

\item
The Class I sources are associated with
higher extinction  and are located relatively closer to each other than the Class II sources.
Higher numbers of younger sources in the inner regions of 
BRCs having high column densities of molecular cloud confirm that
the young protostars are  usually found in regions having 
marginally higher surface densities than the more evolved PMS stars.

\item
MST analyses is used to isolate star forming stellar cores 
from a diffuse, sparsely distributed, and potentially varying density background.
We have extracted thirteen stellar cores of eight
or more members. The active regions in all the studied BRCs contain 
in total  997 YSOs, out of which 602 (60\%) belong to the cores.
The median MST branch length in these cores is found to be 0.09 pc.
The members of these cores are mostly younger YSOs.

\item
Several basic structural measurements of these
cores have been done, finding that the median core is 0.6 pc in radii
and somewhat elongated (aspect ratio of 1.45), of relatively low
density (60 pc$^2$), small (35 members), young (66\% Class I), 
and partially embedded (median $A_K$ =1.1 mag).

\item
BRCs show fractal distribution of YSOs in their inner regions.
Also, the Class I sources are distributed more hierarchically than the Class II,
which is likely a consequence of the dynamical relaxation.
The Jeans lengths for the cores have values between 0.2 - 1 pc, with a median value of 0.46 pc.
Longer Jeans lengths in comparison to YSOs separations support the non-thermally driven fragmentation in the BRCs.

\item

The cores in the present sample show a wide range in their mass distribution 
($\sim$20 to 2400 M$_\odot$) with a median value around 130  M$_\odot$.
We found a linear  relation between the density of the clouds and the number of YSOs 
for the present sample of BRCs.
We also found the SFEs for the identified cores to be between 3 and 30 \% with an average of $\sim$14 \%.
These values are in support of the previous findings agreeing with the efficiencies needed to go
from the core mass function to the initial stellar mass function.

\end{itemize}

\section*{Acknowledgments}
The observations reported in this paper were carried out by using the BLANCO telescope at CTIO.  We thank the staff members for their assistance during the  observations. 
Financial supports for J.B and R.K. are provided by the Ministry of Economy, Development, and Tourism's Millennium Science Initiative through grant 
IC120009, awarded to The Millennium Institute of Astrophysics, MAS. and Fondecyt Reg. No. 1130140 and 1120601.
This work is based in part on data obtained by the $Spitzer$ Space Telescope, which is operated by the Jet Propulsion Laboratory, 
Caltech, under a contract with NASA. Support to this work was provided by NASA through a contract issued by JPL/Caltech.
This publication also makes use of data from the Two Micron All Sky Survey, which is a joint project of the University of 
Massachusetts and the Infrared Processing and Analysis Center/California Institute of Technology, 
funded by the National Aeronautics and Space Administration and the National Science Foundation.

\bibliography{ms2}{}
\bibliographystyle{aj}

\newpage

\begin{table*}
\centering
\caption{\label{Tlog} Regions observed in the present study. The last column gives the stellar mass at
the completeness limit of 
YSO detection after correcting for foreground and peak (in parenthesis) extinction values (cf. \S 3.2).}
\begin{tabular}{@{}lrrccccc@{}}
\hline
Name& $\alpha_{(2000)}$&$\delta_{(2000)}$& $l$ &$b$ & Distance  & Pixel & 90\% Completeness \\
 & {\rm $(^h:^m:^s)$} & {\rm $(^o:^\prime:^{\prime\prime)} $} && (degrees) & (kpc) & size (pc) & Limit (M$_\odot$)\\
\hline
SFO 54 & 08:35:31.7&  -40:38:28   &  259.941383 & -0.040549 & 0.95 &0.0014  & 0.04(0.15)\\
SFO 55 & 08:41:13.0&  -40:52:03   &  260.775002 &  0.678340 & 1.15 &0.0017  & 0.03(0.10)\\
SFO 64 & 11:12:18.0&  -58:46:20   &  290.374065 &  1.661209 & 2.70 &0.0040  & 0.40(1.40)\\
SFO 65 & 11:33:00.0&  -63:27:20   &  294.301328 & -1.905817 & 1.70 &0.0025  & 0.08(0.20)\\
SFO 68 & 11:35:31.9&  -63:14:51   &  294.511530 & -1.623458 & 1.70 &0.0025  & 0.08(0.20)\\
SFO 75 & 15:55:50.4&  -54:38:58   &  327.573745 & -0.851722 & 2.80 &0.0041  & 0.80(3.00)\\
SFO 76 & 16:10:38.6&  -49:05:52   &  332.956319 &  1.803776 & 1.80 &0.0027  & 0.15(1.40)\\
SFO 79 & 16:40:00.1&  -48:51:45   &  336.491077 & -1.475060 & 1.35 &0.0020  & 0.10(2.00)\\
Field 1& 08:34:42.4&  -40:40:45   &  259.878445 & -0.188092 & $-$   $-$    & \\
Field 2& 16:39:03.7&  -48:51:34   &  336.390516 & -1.357565 & $-$   $-$    & \\
\hline
\end{tabular}
\end{table*}

\begin{table*}
\centering
\caption{\label{Tslopes} Color coefficients and constants derived by using 2MASS data for the calibration of ISPI data. 
`N' represents the total number of the common stars used in the fitting.}
\begin{tabular}{@{}r@{ }c@{ }r@{ }r@{ }r@{ }r@{ }r@{ }r@{ }r@{}}
\hline
   ID  &N &$       M1    $&$     C1      $&$        M2    $&$     C2     $&$          M3      $&$        C3      $  \\
\hline
SFO 54 & 220  &$ 1.013\pm0.013$ &$-0.077\pm0.020 $&$      0.963\pm0.026$&$ 0.420\pm0.005$&$       -0.079\pm0.025$&$ 0.522\pm0.012$\\
SFO 55 & 223  &$ 0.997\pm0.013$ &$-0.038\pm0.020 $&$      0.996\pm0.027$&$ 0.460\pm0.005$&$       -0.070\pm0.025$&$ 0.532\pm0.011$\\
SFO 64 & 617  &$ 1.003\pm0.008$ &$-0.086\pm0.008 $&$      0.952\pm0.017$&$ 0.427\pm0.005$&$       -0.078\pm0.021$&$ 0.472\pm0.005$\\
SFO 65 & 671  &$ 0.993\pm0.007$ &$-0.008\pm0.009 $&$      0.943\pm0.026$&$ 0.449\pm0.004$&$       -0.078\pm0.015$&$ 0.516\pm0.006$\\
SFO 68 & 672  &$ 1.015\pm0.006$ &$ 0.012\pm0.008 $&$      0.967\pm0.014$&$ 0.461\pm0.003$&$       -0.065\pm0.016$&$ 0.501\pm0.005$\\
SFO 75 & 955  &$ 0.986\pm0.006$ &$ 0.079\pm0.011 $&$      0.946\pm0.010$&$ 0.596\pm0.002$&$       -0.064\pm0.014$&$ 0.374\pm0.009$\\
SFO 76 & 993  &$ 0.981\pm0.006$ &$ 0.058\pm0.010 $&$      0.952\pm0.011$&$ 0.559\pm0.002$&$       -0.069\pm0.014$&$ 0.554\pm0.008$\\
SFO 79 & 775  &$ 0.991\pm0.005$ &$ 0.002\pm0.010 $&$      0.970\pm0.010$&$ 0.458\pm0.003$&$       -0.048\pm0.011$&$ 0.499\pm0.006$\\
Field 1& 263  &$ 1.009\pm0.013$ &$-0.044\pm0.014 $&$      0.985\pm0.036$&$ 0.439\pm0.008$&$       -0.089\pm0.034$&$ 0.490\pm0.010$\\
Field 2& 1366 &$ 0.990\pm0.004$ &$-0.065\pm0.008 $&$      0.928\pm0.010$&$ 0.452\pm0.002$&$       -0.052\pm0.007$&$ 0.502\pm0.007$\\
\hline
\end{tabular}
\end{table*}

\begin{table*}
\centering
\tiny
\caption{\label{Tyso} Sample of YSOs identified in the present study on the basis of their excess IR-emission using CTIO/$Spitzer$ data. Their respective magnitudes and photometric errors are also given. The last column gives the information about the scheme used in their classification. The complete table is available in the electronic form only.}
\begin{tabular}{@{}l@{ }c@{ }c@{ }r@{ }r@{ }r@{ }r@{ }r@{ }r@{ }r@{ }r@{}}
\hline
Name& $\alpha_{(2000)}$&$\delta_{(2000)}$& $J\pm\sigma~~~~~$&$H\pm\sigma~~~~~$&$K\pm\sigma~~~~~$&$3.6\pm\sigma~~~~~$&$4.5\pm\sigma~~~~~$&$5.8\pm\sigma~~~~~$&$8.0\pm\sigma~~~~~$&comment$^*$\\
(BRC\_ID) & {\rm $(^h:^m:^s)$} & {\rm $(^o:^\prime:^{\prime\prime)} $} & (mag)$~~~~~$&  (mag)$~~~~~$& (mag)$~~~~~$& (mag)$~~~~~$& (mag)$~~~~~$& (mag)$~~~~~$& (mag)$~~~~~$& \\
\hline
SFO54\_1  & 08:35:40.30& 40:40:07.2&$  9.704 \pm 0.017$&$  8.880 \pm 0.019$&$  8.128 \pm 0.015$&$  7.191 \pm 0.110$&$  6.818 \pm 0.070$&$  6.135 \pm 0.018$&$  5.430 \pm 0.046$& 2   \\
SFO54\_2  & 08:35:32.82& 40:38:36.2&$                -$&$ 11.049 \pm 0.033$&$  9.421 \pm 0.022$&$  7.890 \pm 0.061$&$  7.385 \pm 0.036$&$  6.926 \pm 0.019$&$  6.436 \pm 0.023$& 2,6 \\ 
SFO54\_3  & 08:35:30.95& 40:38:26.7&$                -$&$ 13.023 \pm 0.067$&$  9.865 \pm 0.033$&$                -$&$                -$&$                -$&$                -$&   6 \\ 
SFO54\_4  & 08:35:22.56& 40:38:50.2&$ 12.684 \pm 0.024$&$ 11.081 \pm 0.051$&$  9.927 \pm 0.039$&$  8.357 \pm 0.079$&$  7.842 \pm 0.035$&$  7.365 \pm 0.022$&$  6.574 \pm 0.022$& 2,5 \\ 
SFO54\_5  & 08:35:31.03& 40:38:22.3&$ 14.586 \pm 0.004$&$ 12.171 \pm 0.005$&$ 10.523 \pm 0.016$&$                -$&$                -$&$                -$&$                -$&   5 \\ 
SFO54\_6  & 08:35:17.76& 40:38:23.4&$ 12.290 \pm 0.006$&$ 11.420 \pm 0.006$&$ 10.932 \pm 0.007$&$ 10.499 \pm 0.068$&$ 10.097 \pm 0.051$&$  9.795 \pm 0.031$&$  9.087 \pm 0.042$& 2   \\ 
SFO54\_7  & 08:35:32.80& 40:38:15.5&$ 13.628 \pm 0.002$&$ 11.964 \pm 0.002$&$ 11.095 \pm 0.004$&$  9.773 \pm 0.086$&$  9.222 \pm 0.050$&$  8.507 \pm 0.036$&$  7.732 \pm 0.023$& 2   \\ 
SFO54\_8  & 08:35:34.65& 40:37:27.3&$ 15.692 \pm 0.009$&$ 12.856 \pm 0.003$&$ 11.156 \pm 0.006$&$  9.573 \pm 0.068$&$  9.006 \pm 0.020$&$  8.486 \pm 0.034$&$  7.597 \pm 0.017$& 2,5 \\ 
SFO54\_9  & 08:35:42.10& 40:38:05.1&$ 13.125 \pm 0.002$&$ 12.003 \pm 0.002$&$ 11.227 \pm 0.006$&$ 10.383 \pm 0.070$&$ 10.400 \pm 0.039$&$ 10.311 \pm 0.042$&$ 10.220 \pm 0.031$&   5 \\ 
SFO54\_10 & 08:35:25.92& 40:38:58.3&$ 13.164 \pm 0.003$&$ 12.040 \pm 0.002$&$ 11.419 \pm 0.002$&$ 10.385 \pm 0.042$&$  9.887 \pm 0.035$&$  9.362 \pm 0.038$&$  8.462 \pm 0.025$& 2   \\ 
\hline
\end{tabular}

$^*$ 1=Class I (Through $Spitzer$ data), 2=Class II (Through $Spitzer$ data) , 3=Class I (Through CTIO/$Spitzer$ data), 4=Class II (Through CTIO/$Spitzer$ data), 5=Class II (Through CTIO data), 6=Class I (Through CTIO data)\\

\end{table*}

\begin{table*}
\centering
\caption{\label{Tp1} Center coordinates of the identified cores and active regions along with the total number of the YSOs, 
and their distribution as a function of their evolutionary status. 
The numbers in bracket represent the YSOs classified by using the $Spitzer$ IRAC data.}
\begin{tabular}{@{}lccrrcc@{}}
\hline
Name& $\alpha_{(2000)}$&$\delta_{(2000)}$&N & Class& Class& Frac$^a$   \\
 & {\rm $(^h:^m:^s)$} & {\rm $(^o:^\prime:^{\prime\prime)} $} &  &  I& II& (\%)\\
\hline
&&&&&&\\                                                              
Cores\\
 SFO 54   C1   &08:35:28.6& -40:38:30 &   81   &  36( 3)  &  45(22) & 44(12) \\
 SFO 55   C1   &08:41:13.5& -40:52:11 &   55   &  37( 4)  &  18(12) & 67(25) \\
 SFO 64   C1   &11:12:19.1& -58:46:23 &   78   &  41( 1)  &  37(12) & 53( 8) \\
 SFO 65   C1   &11:33:00.4& -63:28:16 &   21   &  15( 0)  &   6( 2) & 71( 0) \\
 SFO 68   C1   &11:35:31.7& -63:14:58 &   35   &  23( 2)  &  12( 2) & 66(50) \\
 SFO 68   C2   &11:34:58.5& -63:16:45 &   17   &  10( 0)  &   7( 0) & 59( -) \\
 SFO 75   C1   &15:55:49.2& -54:39:15 &  175   & 114( 4)  &  61( 8) & 65(33) \\
 SFO 75   C2   &15:56:12.3& -54:39:57 &   42   &  34( 0)  &   8( 0) & 81( -) \\
 SFO 76   C1   &16:10:39.1& -49:06:53 &   30   &  21( 0)  &   9( 0) & 70( -) \\
 SFO 76   C1   &16:10:25.5& -49:02:55 &    8   &   0( 0)  &   8( 0) &  -( -) \\
 SFO 79   C1   &16:40:01.3& -48:51:43 &   84   &  53( 8)  &  31(15) & 63(35) \\
 SFO 79   C2   &16:39:46.3& -48:51:05 &   11   &  10( 0)  &   1( 1) & 91( 0) \\
 SFO 79   C3   &16:40:11.3& -48:48:59 &   25   &  17( 2)  &   8( 4) & 68(33) \\
&&&&&&\\                                                              
Active regions\\
SFO 54  A  &08:35:31.1& -40:38:43 &  128   &  48( 4)  &  80(45) & 38( 8) \\
SFO 55  A  &08:41:13.0& -40:52:27 &   78   &  49( 6)  &  29(19) & 63(24) \\
SFO 64  A  &11:12:21.1& -58:46:20 &  115   &  45( 2)  &  70(39) & 39( 5) \\
SFO 65  A  &11:32:49.4& -63:28:23 &   35   &  19( 0)  &  16( 5) & 54( 0) \\
SFO 68  A  &11:35:30.8& -63:13:59 &   97   &  50( 3)  &  47(11) & 52(21) \\
SFO 75  A  &15:55:54.4& -54:39:16 &  269   & 181( 4)  &  88(10) & 67(29) \\
SFO 76  A  &16:10:33.9& -49:05:56 &   56   &  26( 0)  &  30( 0) & 46( -) \\
SFO 79  A  &16:40:00.6& -48:51:31 &  219   & 155(12)  &  64(35) & 71(26) \\\\
\hline
\end{tabular}

a: Class 1/(Class 1 + Class II)
\end{table*}

\begin{table*}
\centering
\scriptsize
\caption{\label{Tp2} Properties of the identified cores and active regions. The hull and circle radius along 
with the aspect ratio are given in columns 2, 3 and 4, respectively. 
Columns 5 and 6 represent the mean and peak stellar density obtained using the isodensity contours. 
Columns 7  and 8 are the mean MST branch length and NN distances, respectively.
The mean and peak extinction values are given in Columns 9 and 10, respectively. Column 11 represents the cloud mass in the convex hull derived using the extinction maps.
Column 12 represents the mass of the dense cloud having $A_K$ greater than 0.8 mag. 
Columns 13, 14 and 15 represent the $Q$ value, Jeans length and critical branch length for MST, respectively.}
\begin{tabular}{@{}l@{ }c@{  }c@{ }c@{ }c@{ }c@{ }c@{ }c@{ }c@{ }c@{ }c@{ }c@{ }c@{ }c@{ }c@{ }c@{}}
\hline
Name& $R_{\rm hull}$& $R_{\rm cir}$& Aspect & $\rho_{\rm mean}$ & $\rho_{\rm peak}$ & MST & NN2 & $A_{K_{mean}}$ & $A_{K_{peak}}$& M$_{A_K}$& M$_{A_K}$ & $Q$ & J & D$_{crit.}$\\
 &  (pc)& (pc)& Ratio & (pc$^{-2}$) & (pc$^{-2}$) & (pc) &  (pc) & (mag) &(mag) & (M$_\odot$) & (M$_\odot$)  & & (pc) & (pc)\\
\hline
&&&&&&&&&&&&&&\\
Cores\\
SFO 54   C1   & 0.40& 0.45& 1.28& 163.21&1330.4& 0.04& 0.03&  1.32  &  2.34  &  131.3&  125.3& 0.75& 0.30& 0.12\\ 
SFO 55   C1   & 0.58& 0.70& 1.49&  52.83& 454.1& 0.07& 0.06&  1.09  &  1.89  &  215.3&  189.5& 0.64& 0.46& 0.18\\
SFO 64   C1   & 0.58& 0.73& 1.59&  75.11& 386.9& 0.06& 0.05&  0.61  &  1.81  &  127.2&   37.2& 0.65& 0.52& 0.18\\
SFO 65   C1   & 1.02& 1.29& 1.58&   6.38&  17.1& 0.20& 0.12&  0.65  &  1.97  &  296.9&  168.8& 0.59& 0.98& 0.56\\
SFO 68   C1   & 0.30& 0.34& 1.22& 120.36& 693.6& 0.06& 0.04&  1.36  &  2.56  &   68.5&   63.5& 0.75& 0.27& 0.15\\
SFO 68   C2   & 0.30& 0.28& 0.89&  61.15& 132.0& 0.09& 0.03&  0.49  &  1.12  &   19.8&    2.6& 0.66& 0.46& 0.15\\
SFO 75   C1   & 1.85& 2.22& 1.43&  16.20& 209.7& 0.13& 0.10&  1.06  &  2.54  & 2416.1& 2015.7& 0.69& 0.84& 0.30\\
SFO 75   C2   & 0.81& 0.83& 1.07&  20.61&  86.4& 0.11& 0.09&  1.50  &  2.73  &  525.9&  517.5& 0.67& 0.52& 0.30\\
SFO 76   C1   & 0.77& 1.04& 1.81&  16.06&  53.6& 0.16& 0.10&  1.07  &  2.36  &  372.7&  316.8& 0.55& 0.57& 0.37\\
SFO 76   C1   & 0.37& 0.36& 0.96&  18.76&  26.5& 0.19& 0.15&  0.56  &  0.86  &   27.0&    2.2& 0.71& 0.65& 0.37\\
SFO 79   C1   & 0.61& 0.73& 1.45& 331.86& 258.0& 0.07& 0.04&  1.55  &  2.79  &  353.6&  349.8& 0.58& 0.39& 0.16\\
SFO 79   C2   & 0.18& 0.24& 1.69& 104.62& 150.0& 0.07& 0.06&  1.64  &  2.20  &   21.5&   21.5& 0.62& 0.23& 0.16\\
SFO 79   C3   & 0.36& 0.45& 1.55&  60.39& 136.2& 0.11& 0.05&  1.47  &  2.71  &   92.4&   91.2& 0.67& 0.34& 0.16\\

&&&&&&&&&&&&&&\\
Active regions\\

SFO 54  A  & 0.87& 1.05& 1.44&  53.06&1330.4& 0.06& 0.05&  0.91  &  2.34  &  451.8&  299.0& 0.85& 0.57&0.26  \\  
SFO 55  A  & 0.85& 1.00& 1.39&  34.37& 454.1& 0.09& 0.06&  1.03  &  1.89  &  465.3&  393.1& 0.70& 0.57&0.29  \\  
SFO 64  A  & 1.15& 1.52& 1.74&  27.75& 386.9& 0.09& 0.06&  0.42  &  1.81  &  348.3&   63.6& 0.80& 0.97&0.40  \\  
SFO 65  A  & 1.62& 1.68& 1.07&   4.24&  17.1& 0.32& 0.15&  0.54  &  1.97  &  800.1&  200.3& 0.69& 1.21&0.84  \\  
SFO 68  A  & 0.88& 1.35& 2.36&  39.72& 693.6& 0.07& 0.05&  0.76  &  2.56  &  391.3&  255.6& 0.59& 0.63&0.30  \\  
SFO 75  A  & 2.63& 3.22& 1.50&  12.42& 209.7& 0.16& 0.11&  1.12  &  2.84  & 5286.7& 4601.4& 0.71& 0.97&0.40  \\  
SFO 76  A  & 1.66& 1.87& 1.26&   6.47&  53.6& 0.19& 0.15&  0.67  &  2.36  & 1108.5&  427.3& 0.72& 1.06&0.56  \\  
SFO 79  A  & 1.35& 1.66& 1.52&  38.30& 258.0& 0.09& 0.06&  1.40  &  2.80  & 1683.8& 1619.9& 0.67& 0.61&0.24  \\\\
\hline
\end{tabular}
\end{table*}
\clearpage

\begin{table*}
\centering
\caption{\label{Tp3} Median averaged parameters of all the  cores and active region. 
The figure in brackets represent the numbers of the YSOs classified by using the $Spitzer$ IRAC data.}
\begin{tabular}{@{}lll@{}}
\hline
Properties  & Core   &  Active region\\
\hline
Fraction of Class I sources (\%)&  66(33)  &  52(21) \\
NN2$_{(Class~I)}$ (pc)            &    0.07 &   0.07  \\
NN2$_{(Class~II)}$ (pc)           &    0.10 &   0.11  \\
A$_K{_{(Class~I)}}$ (mag)         &    1.3  &   1.1  \\
A$_K{_{(Class~II)}}$ (mag)        &    1.0  &   0.8  \\
Number of YSOs                 &  35    & 97     \\
R$_{\rm hull}$  (pc)           &    0.58 &   1.15  \\
Aspect Ratio                   &    1.45 &   1.44  \\
Mean number density (pc$^{-2}$)&   60    &  28     \\
Peak number density (pc$^{-2}$)&  150    & 258     \\
A$_K$ (mag)               &    1.1  &   0.8  \\
Peak  A$_K$(mag)               &    2.3  &   2.3  \\
Cloud mass (M$_\odot$)         &  131    & 465     \\
Dense cloud mass (M$_\odot$)   &  125    & 299     \\
MST branch length (pc)         &    0.09 &   0.09  \\
Structural $Q$ parameter         &    0.66 &   0.70  \\
Structural $Q$ parameter$_{(Class~I)}$ &    0.68 &   0.64  \\
Structural $Q$ parameter$_{(Class~II)}$ &    0.72 &   0.81  \\
Jeans Length (pc)              &    0.46 &   0.63  \\
Star formation efficiency (\%) &     12  &    6   \\
\hline
\end{tabular}
\end{table*}


\clearpage

\clearpage

\clearpage

\clearpage

\clearpage

\label{lastpage}

\end{document}